\documentclass[a4paper]{amsproc}
\usepackage{amssymb}
\usepackage{amsfonts}
\usepackage{amsmath,amsthm}
\usepackage{graphicx}
\usepackage{epstopdf}

\def\text#1{\hbox{#1}}
\newcommand\beq{\begin{equation}}
\newcommand\eeq{\end{equation}}

\newtheorem{theorem}{Theorem}
\newtheorem{prop}[theorem]{Proposition}

\newtheorem{remark}{Remark}
\newtheorem{example}{Example}

\newcommand{\vague}{\stackrel{\lower0.2ex\hbox{$\scriptscriptstyle
                    \it{v} $}}{\rightarrow}}
\newcommand{\weak}{\stackrel{\lower0.2ex\hbox{$\scriptscriptstyle
                    \it{w} $}}{\rightarrow}}
\newcommand{\distr}{\stackrel{\lower0.2ex\hbox{$\scriptscriptstyle
                    \it{d} $}}{\rightarrow}}
\newcommand{\eqdis}{\stackrel{\lower0.2ex\hbox{$\scriptscriptstyle
                    \it{d} $}}{=}}

\def\text#1{\hbox{#1}}

\def\build #1_#2{\mathrel{\mathop{\kern 0pt #1}
\limits_{#2}}} 
\newcommand{\zs}[1]{{\mathchoice{#1}{#1}
{\lower.25ex\hbox{$\scriptstyle#1$}}
{\lower0.25ex\hbox{$\scriptscriptstyle#1$}}}}

\newcommand{\R}{\mathbf{R}}

\newcommand{\Prob}{\operatorname{P}}

\newcommand{\trans}{{\mathrm{T}}}

\newcommand{\E}{\operatorname{E}}

\newcommand{\Normal}{{\operatorname{N}}}

\begin{document}

\title[A model for the forward density process]{A simple time-consistent model for the forward density process}
\author[H.~Hult \and F.~Lindskog \and J.~Nykvist]{Henrik Hult \and Filip Lindskog \and Johan Nykvist*}
\address{Department of Mathematics, KTH Royal Institute of Technology, SE 100 44, Stockholm, Sweden. Email: hult@kth.se; lindskog@kth.se; jnykvist@kth.se}

\thanks{H.~Hult acknowledges support from the G\"oran Gustafsson Foundation.}
\thanks{{}* Corresponding author}




\date{\today}




\begin{abstract}
In this paper a simple model for the evolution of the forward density of the future value of an asset is proposed. The model allows for a straightforward initial calibration to option prices and has dynamics that are consistent with empirical findings from option price data. The model is constructed with the aim of being both simple and realistic, and avoid the need for frequent re-calibration. The model prices of $n$ options and a forward contract are expressed as time-varying functions of an $(n+1)$-dimensional Brownian motion and it is investigated how the Brownian trajectory can be determined from the trajectories of the price processes.
An approach based on particle filtering is presented for determining the location of the driving Brownian motion from option prices observed in discrete time. A simulation study and an empirical study of call options on the S\&P 500 index illustrates that the model provides a good fit to option price data.
%
\end{abstract}

\maketitle




\pagenumbering{arabic}

\section{Introduction}

Consider a financial market consisting of a collection of European options with maturity $T > 0$, written on the value $S_T$ of an asset at time $T$.  Suppose that the option prices at any time $t \in [0,T]$ can be expressed as discounted expected option payoffs, where the expectations are computed with respect to a density $f_t$ of $S_{T}$. The density $f_{t}$ is often called the forward density of $S_T$. 
This paper addresses the modeling of the initial density $f_0$ and the evolution of the density $f_t$ over time, $\{f_t\}_{t\in [0,T]}$.  

The model is constructed on a filtered probability space $(\Omega,\mathcal{F},\{\mathcal{F}_t\}_{t\in [0,T]},\Prob)$ with expectation operator $\E$. For each $t \in [0,T]$, the forward price of a 
derivative payoff $g(S_T)$ is the expected payoff computed with respect to the density $f_t$:
\begin{align}\label{eq:intro1}
\E[g(S_T)\mid \mathcal{F}_t]=\int g(x)f_t(x)dx.
\end{align} 
In statements such as the above, to avoid technicalities, the functions mentioned are assumed to satisfy measurability and integrability conditions necessary for the statements to be meaningful.
The market is assumed to consist of $n+1$ forward contracts on European call options with payoffs $(S_T-K_j)_+$ for $0=K_0<K_1<\dots<K_n$. If the original market consists of a mix of European puts and calls, then the put-call parity may be used to define an equivalent market consisting entirely of forward contracts on call option payoffs. From \eqref{eq:intro1} it follows that the forward price processes $\{G^j_t\}_{t\in [0,T]}$ are martingales satisfying the initial condition
\begin{align}\label{eq:intro2}
G^j_0=\int (x-K_j)_+f_0(x)dx \quad \text{for } j=0,1,\dots,n.
\end{align}
A parametric form for $f_0$ will be selected that allows its parameters to be set in a straightforward manner from the $n+1$ equations in \eqref{eq:intro2} and internally consistent  forward prices $G^0_0,G^1_0,\dots,G^n_0$.

The filtration $\{\mathcal{F}_t\}_{t\in [0,T]}$ is assumed to be generated by a standard $(n+1)$-dimensional Brownian motion $(V^1,V^2,\dots,V^{n+1})$ and we take, for all $t$, $f_t$ to be a function with parameters $t,V^1_t,V^2_t,\dots,V^{n+1}_t$ that vary over time and other parameters that are set in the initial calibration of $f_0$ to the current price data.
The choice of $f_t$ allows the $\R^{n+1}$-valued forward price process $(G^0,G^1,\dots,G^{n})$ to be expressed in terms of the Brownian motion $(V^1,V^2,\dots,V^{n+1})$ as
\begin{align*}
(G^0_t,G^1_t,\dots,G^{n}_t)=h_t(V^1_t,V^2_t,\dots,V^{n+1}_t)
\end{align*}
for functions $h_t:\R^{n+1}\to\R^{n+1}$, $t \in [0,T]$. It is desirable that the functions $h_t$ are locally invertible so that the filtration $\{\mathcal{G}_t\}_{t\in [0,T]}$ generated by the prices, the filtration with an economic interpretation, equals the Brownian filtration. For the model to be relevant the functions $h_t$ must give rise to price processes with joint dynamics that are in line with empirically observed stylized facts for option price processes. Moreover, the range of option prices that the model can produce must be large enough to capture the fluctuations of observed option prices and avoid the need for frequent recalibration.  Frequent recalibration of a model's parameters is unattractive from a theoretical point of view and limits its practical utility. 

The model for $\{f_t\}_{t\in [0,T]}$ set up at time $0$ is intended to be relevant also at time $t>0$. Therefore, it makes sense to require that the realized forward prices at time $t>0$ should be possible realizations of the model prices $G^0_t,G^1_t,\dots,G^n_t$. The following example illustrates that a simple model such as Black's model does not satisfy this requirement.

\begin{example}[Black's model]\label{ex:blacksmodel}
  Consider the case $n=1$ (one forward contract and one call option on $S_T$) and let $(W_t)_{t\in [0,T]}$ be standard Brownian motion with respect to $\Prob$. Black's model, see \cite{B76}, says that 
  \begin{align*}
    S_T&=G^0_0\exp\Big\{\sigma_0 W_T-\frac{\sigma_0^2}{2}T\Big\}
  \end{align*}
  which implies that the forward price $G_0^1$ for the call option payoff  $(S_T-K)_+$ is given by
  \begin{align*}
    &G_0^1=G^0_0\Phi(d_1)-K\Phi(d_2), \\ 
    &d_1=\frac{\log(G^0_0/K)}{\sigma_0\sqrt{T}}+\frac{\sigma_0\sqrt{T}}{2},
    \quad d_2=d_1-\sigma_0\sqrt{T}. 
  \end{align*}
  The parameter $\sigma_0$ solving this equation is the option's implied volatility (implied from $G_0^0$ and $G_0^1$). 
Writing 
  \begin{align*}
    S_T=G^0_0\exp\Big\{\sigma_0W_t-\frac{\sigma_0^2}{2}t\Big\}
    \exp\Big\{\sigma_0(W_T-W_t)-\frac{\sigma_0^2}{2}(T-t)\Big\}
  \end{align*}
  and $\mathcal{F}_t=\sigma(\{W_s\}_{s\in [0,t]})$ we notice that the model allows stochastic fluctuations in the forward price of $S_T$: 
  \begin{align*}
    G^0_t=G^0_0\exp\Big\{\sigma_0W_t-\frac{\sigma_0^2}{2}t\Big\}.
  \end{align*}
However, the option's implied volatility is required to stay constant over time.
  In particular, the future realized prices are practically guaranteed to violate the model which therefore has to be frequently recalibrated to fit the price data.
\end{example}

One reason for the inability of the dynamic version of Black's model in Example \ref{ex:blacksmodel} to generate future option prices is that the filtration $\{\mathcal{F}_t\}_{t\in [0,T]}$ is generated by a one-dimensional Brownian motion. After the initial calibration the range of possible forward prices that the model produces is very limited: it is likely that, after a short period of time, the observed option prices lie outside the range of the model. Similar problems occur for instance for the local volatility model by Dupire \cite{D94} and for many stochastic volatility models. In addition, the initial calibration for these models is non-trivial. 

In the model we will consider below we want $\{\mathcal{F}_t\}_{t\in [0,T]}$ to be equivalent to the filtration generated by the price processes and consider the situation when no price process can be determined from the other price processes.

We do not consider the spot price process for the underlying asset, only its value at time $T$
and forward and other derivative contracts written on that value. If the asset is a non-dividend paying stock, then the spot price must equal the discounted forward price in order to rule out arbitrage opportunities. 

We do not pay attention to the subjective probability views of market participants. Therefore it does not make much sense here to discuss equivalent martingale measures. However, by requiring that the conditional density process is a martingale and that it produces realistic dynamics for the price processes we are implicitly saying that the model could be a natural candidate for an equivalent martingale measure for informed market participants. 

The paper \cite{SW08} has a similar objective as ours. However, whereas in \cite{SW08} the authors set up a system of stochastic differential equations (diffusion processes) for the evolution of the spot price and the implied volatilities and address the difficult mathematical problem of determining conditions for the absence of arbitrage opportunities, we consider a more explicit but less general class of models for the conditional density process $\{f_t\}_{t\in [0,T]}$. A more general problem is investigated in \cite{CN09} and \cite{KK10}, where characterizations are provided of arbitrage-free dynamics for markets with call options available for all strikes and all maturities.  Conditional density models, which are studied in this paper, are also studied in \cite{FHM11}, where the authors characterize the ``volatility processes" $\{\sigma^f_t(x)\}_{t\in [0,T]}$ in the stochastic exponential representation 
\begin{align*}
f_t(x)=f_0(x)+\int_0^t \sigma^f_s(x) f_s(x)dV_s
\end{align*} 
that generate proper conditional density processes. In contrast, we take a particular model for $\{f_t\}_{t\in [0,T]}$ as the starting point whereas in \cite{FHM11} the conditional density model is implied from the model for $\{\sigma^f_t(x)\}_{t\in [0,T]}$.

In \cite{B06}, \cite{DH07}, \cite{C05}, and \cite{LL00} the authors consider a setting with a finite number of traded options for a finite set of maturities on one underlying asset, and characterize absence of arbitrage in this setting. Both static arbitrage and arbitrage when dynamic trading in the options is allowed are considered.  In \cite{B06}, \cite{C05}, and \cite{LL00} explicit Markov martingales are constructed that give perfect initial calibration to the observed option prices.

The outline of this paper is as follows. In Section \ref{sec:staticmodel} we consider a rather naive model for $f_0$, a distribution of $S_T$ that reproduces the given option prices, and present a straightforward calibration procedure for the model parameters. The model is the starting point for the conditional density model for $\{f_t\}_{t\in [0,T]}$ that is presented in Section \ref{sec:dynamicmodel}.  The theoretical properties of the model and a discussion on  how the model can be set up to meet the natural requirements for a good derivative pricing model are also included in Section \ref{sec:dynamicmodel}. 
Section \ref{sec:evaluationofthemodel} contains further theoretical and numerical investigations of the properties of the conditional density model and it is evaluated on S\&P 500 index option data and through simulation studies.

Our contributions can be summarized as follows. We propose a simple model for the evolution of the forward density. At each time the forward density is a mixture of lognormal distributions which makes it easy to make the initial calibration of its parameters and price European type derivatives. On a market with $n$ liquidly traded call options and a forward contract, the model is driven by an $(n+1)$-dimensional Brownian motion, making it flexible enough to capture realized option price fluctuations in a satisfactory way and avoids the need for frequent recalibration. 
The model is set up so that the filtration generated by the $n+1$ price processes is essentially, see Section \ref{sec:filtrations} for details, equal to the $(n+1)$-dimensional Brownian filtration. 
Moreover, the model can easily be set up to capture stylized features of option prices, such as a negative correlation between changes in the forward price and changes in implied volatility. A simulation study and an empirical study of call options on the S\&P 500 index illustrates that the model provides a good fit to option data.

\section{The spot price at maturity}\label{sec:staticmodel}

We start by investigating a very simple model for $S_T$, which will be refined later, that reproduces 
the $n+1$ observed forward prices. The random variable $S_T$ is assumed to be discrete and  takes one of the values $0\leq x_1<\dots<x_{n+2}<\infty$. Let $p_0^k$ be the forward probability of the event $\{S_T=x_k\}$. The initial calibration requires solving a linear system of equations  of the form $Ap=b$, where $p$ is 
the vector of forward probabilities of the events $\{S_T=x_k\}$:
\begin{align}\label{eq:lineqsyst}
  \left(\begin{array}{llll}
    1 & 1 & \dots & 1\\
    x_1 & x_2 & \dots & x_{n+2}\\
    (x_1-K_1)_+ & (x_2-K_1)_+ & \dots & (x_{n+2}-K_1)_+\\
    \vdots & & & \\
    (x_1-K_n)_+ & (x_2-K_n)_+ & \dots & (x_{n+2}-K_n)_+
  \end{array}\right)
  \left(\begin{array}{l}
    p_0^1 \\ p_0^2 \\ \vdots \\ p_0^{n+2}
  \end{array}\right)
  = 
  \left(\begin{array}{l}
    1 \\ G^0_0 \\ G^1_0 \\ \vdots \\ G^n_0
  \end{array}\right).
\end{align}
If further $x_2\leq K_1$, $x_{n+2}>K_n$, 
and $x_k\in (K_{k-2},K_{k-1}]$ for $k=3,\dots,n+1$, then 
the matrix on the left-hand side in \eqref{eq:lineqsyst} is one row operation 
away from an invertible triangular matrix. 
In particular, the matrix equation $Ap=b$ can be solved explicitly for $p$
by backward substitution and then it only remains to verify that $p$ is a probability
vector. In order to ensure the existence of a probability vector solving \eqref{eq:lineqsyst} it must be assumed that 
\begin{align}\label{eq:noarbcon1}
  \frac{G_0^{j-1}-G_0^j}{K_j-K_{j-1}} \in [0,1], \quad j\geq 1,
\end{align}
where we set $K_0=0$, and
\begin{align}\label{eq:noarbcon2}
  G_0^{j-1}-\frac{K_{j+1}-K_{j-1}}{K_{j+1}-K_{j}}G_0^j
  +\frac{K_{j}-K_{j-1}}{K_{j+1}-K_{j}}G_0^{j+1}\geq 0, \quad j\geq 1.
\end{align}
The conditions \eqref{eq:noarbcon1} and \eqref{eq:noarbcon2} were considered
in \cite{CM05} and ensure that the market of linear combinations of forward contracts together with a linear pricing rule is free of static arbitrage opportunities. 
The following result, which is proved at the end of the paper, is used as a starting point in the initial calibration of the model for the forward price processes presented in Section \ref{sec:dynamicmodel}. The result gives (necessary and) sufficient conditions for the existence of a discrete distribution of $S_T$ that is consistent with the forward prices on $S_T$. 
The statement of Proposition \ref{prop:simplemodel} below is a slight generalization of Proposition 
3.1 in \cite{B06}.

 
\begin{prop}\label{prop:simplemodel}
  Suppose that the non-negative forward prices $G^0_0$ and $G_0^1,\dots,G_0^n$ 
  on the values $S_T$ and $(S_T-K_j)_+$, for $j=1,\dots,n$, at time $T>0$,
  are ordered so that $K_1<\dots<K_n$ and satisfy 
  \eqref{eq:noarbcon1} and \eqref{eq:noarbcon2}.
  If $x_k=K_{k-1}$ for $k=2,\dots,n+1$,
  \begin{align}
    x_1 &\leq \frac{G^0_0(K_2-K_1)+G_0^2K_1-G_0^1K_2}{(K_2-K_1)-(G_0^1-G_0^2)}, \label{eq:noarbcon3}
    \quad\text{and}\\
    x_{n+2} &\geq \frac{G_0^{n-1}K_n-G_0^nK_{n-1}}{G_0^{n-1}-G_0^n},\label{eq:noarbcon4}
  \end{align} 
  then there exist a unique probability vector $(p_1,\dots,p_{n+2})$ such that
  \begin{align*}
    G^0_0=\sum_{k=1}^{n+2}p_kx_k 
    \quad\text{and}\quad
    G_0^j=\sum_{k=j+2}^{n+2}p_k(K_{k-1}-K_j)
    \quad \text{for } j=1,\dots,n.
  \end{align*}
  The $p_k$s are given by
  \begin{align}
    p_1&=\frac{K_1+G_0^1-G^0_0}{K_1-x_1}, \nonumber \\
    p_2&=\frac{x_1[G_0^1-G_0^2-(K_2-K_1)]+G^0_0(K_2-K_1)-G_0^1K_2+G_0^2K_1}{(K_1-x_1)(K_2-K_1)}, \nonumber \\
    p_k &= \frac{G_0^{k-2}}{K_{k-1}-K_{k-2}} - \frac{G_0^{k-1}(K_k-K_{k-2})}{(K_{k-1}-K_{k-2})(K_k-K_{k-1})} + \frac{G_0^k}{K_k-K_{k-1}}, \label{pk} \\
    & \text{for } k=3,\dots,n, \nonumber \\ 
    p_{n+1} &= \frac{G_0^{n-1}}{K_n-K_{n-1}}
    -\frac{G_0^n(x_{n+2}-K_{n-1})}{(K_n-K_{n-1})(x_{n+2}-K_n)}, \nonumber \\
    p_{n+2} &= \frac{G_0^n}{x_{n+2}-K_n}. \nonumber
  \end{align}
\end{prop}
\begin{remark}
Notice that \eqref{eq:noarbcon1} and \eqref{eq:noarbcon3} imply that $x_1<K_1$ and that \eqref{eq:noarbcon4} implies that $x_{n+2}>K_n$.
Notice also that Proposition \ref{prop:simplemodel} says that there exist indicators $I_k\in \{0,1\}$ satisfying $I_1+\dots+I_{n+2}=1$ and $p_k=\E[I_k]$ such that 
  \begin{align*}
    G^0_0=\E\Big[\sum_{k}I_kx_k\Big]
    \quad\text{and}\quad
    G_0^j=\E\Big[\Big(\sum_{k}I_kx_k-K_j\Big)_+\Big]
    \quad \text{for } j=1,\dots,n.
  \end{align*}
  The conditions \eqref{eq:noarbcon1}-\eqref{eq:noarbcon4} are sharp: it can be seen from the proof that if any of them is violated, then the conclusion of Proposition \ref{prop:simplemodel} does not hold.
\end{remark}

Although the model in Proposition \ref{prop:simplemodel} for $S_T$ under the forward probability
provides explicit expressions for the model parameters in terms of the prices and reproduces any set of observed prices satisfying \eqref{eq:noarbcon1} and \eqref{eq:noarbcon2} it is not a good model. If we want to use the model for pricing new derivative contracts, then we should feel uncomfortable with having a finite grid of points as the only possible values for $S_T$. For instance, the contract that pays $1$ if $S_T$ takes a value other than one of the grid points would be assigned a zero price and this would be viewed as an arbitrage opportunity by most (all) market participants.

A simple extension is to model $S_T$ as the random variable
\begin{align}\label{eq:STrandomgrid}
  S_T=\sum_{k=1}^{n+2} I_kx_kZ_k
\end{align}
which corresponds to replacing the fixed values $x_1,\dots,x_{n+2}$ by 
random values $x_1Z_1,\dots,x_{n+2}Z_{n+2}$ for some suitably chosen random 
variables $Z_1,\dots,Z_{n+2}$ that are independent of $I_1,\dots,I_{n+2}$. 
Take 
\begin{align}\label{eq:Zks}
  Z_k=\exp\Big\{-\frac{\sigma_k^2}{2}T+\sigma_kB_T\Big\},
\end{align}
where $B_T$ is $\Normal(0,T)$-distributed. Then $\E[(x_kZ_k-K_j)_+]=G^{{\small\text{B}}}(x_k,\sigma_k,K_j,T)$, where 
\begin{align}
  &G^{{\small\text{B}}}(x,\sigma,K,T)=x\Phi(d_1)-K\Phi(d_2), \label{eq:BformulaFP}\\
  &d_1=\frac{\log(x/K)}{\sigma\sqrt{T}}+\frac{\sigma\sqrt{T}}{2},
  \quad d_2=d_1-\sigma\sqrt{T},\nonumber
\end{align}
is Black's formula for the forward price of a European call option 
maturing at time $T$, where $x$ is the forward price of $S_T$,
$\sigma$ is the volatility, and $K$ is the strike price.

The initial calibration problem for the modified model amounts to finding a probability vector $p$ solving the linear equation $Ap=b$, where $A$ is a square matrix with $n+2$ rows and columns with 
\begin{align*}
A_{1,k}=1 \text{ for all } k 
\quad \text{and } A_{j,k}=G^{{\small\text{B}}}(x_k,\sigma_k,K_{j-2},T) 
\text{ for } j\geq 2 \text{ and all } k, 
\end{align*}
and where $b=(1,G_0^0,\dots,G_0^n)^{\trans}$.
%
%
The solution $p$ to $Ap=b$ can, as before, be expressed as $p=A^{-1}b$
as long as we specify the $x_k$s and $\sigma_k$s so that $A$ is invertible. 
In general $A$ will not be close to a diagonal matrix and therefore $p=A^{-1}b$
has to be computed numerically.
Notice that for a vector $b$ of internally consistent forward prices and an invertible 
matrix $A$ we may find that $A^{-1}b$ has negative components. In that case the price vector $b$ is outside the range of price vectors that the model can generate.
Fortunately it is not hard to determine the range of forward price vectors 
that the model can produce. 
The simplex $\mathcal{S}=\{p \in \R^{n+2} : p\geq 0, 1^\trans p=1\}$, where $1^{\trans} = (1,\dots,1)$, is a convex set and a linear transformation $A$ of a convex set is a convex 
set. Moreover, the extreme points of $\mathcal{S}$ are mapped to the extreme
points of $A\mathcal{S}$. Therefore it is sufficient to determine the points 
$b_k=Ae_k$ for $k=1,\dots,n+2$, where $e_k$ is the $k$th basis vector in 
the standard basis for $\R^{n+2}$, and investigate the convex hull of
$\{b_1,\dots,b_{n+2}\}$. This is the set of price vectors that the model
can produce.
 
\section{The forward price processes}\label{sec:dynamicmodel}

A choice of the initial forward distribution, $F_0(x)=\Prob(S_T\leq x)$ and forward density $f_0(x)=F_0'(x)$ has been proposed implicitly from \eqref{eq:STrandomgrid} and \eqref{eq:Zks}. In this section, the evolution of the forward distribution and density  will be treated as a stochastic process $\{f_t\}_{t\in [0,T]}$, where $F_t(x)=\Prob(S_T\leq x \mid \mathcal{F}_t)$ and $f_t(x)=F_t'(x)$. The filtration $\{\mathcal{F}_t\}_{t\in [0,T]}$ is taken to be generated by an $(n+1)$-dimensional standard Brownian motion. The $n$-dimensional Brownian motion corresponding to the first $n$ components is denoted by $W$ ($W^k=V^k$ for $k=1,\dots,n$) and is used to model the indicators $I_{1}, \dots, I_{n+2}$, whereas the $1$-dimensional Brownian motion corresponding to the last component is denoted by $B$ ($B=V^{n+1}$) and is used to model the variables $Z_{1}, \dots, Z_{n+2}$ as in \eqref{eq:Zks}.  

The forward price at time $t$ of a derivative contract on $S_T$ with payoff function $g$ is given by
\begin{align*}
\E[g(S_T)\mid \mathcal{F}_t]&=\E\Big[g\Big(\sum_{k=1}^{n+2}I_kx_kZ_k\Big)\mid \mathcal{F}_t\Big]\\
&=\sum_{k=1}^{n+2}\Prob(I_k=1\mid \mathcal{F}_t)\E[g(x_kZ_k)\mid \mathcal{F}_t].
\end{align*}
We consider a partition $\{D_1,\dots,D_{n+2}\}$ of $\R^n$ and set $I_k=I\{W_T\in D_k\}$. The factors $\Prob(I_k=1\mid \mathcal{F}_t)$ and $\E[g(x_kZ_k)\mid \mathcal{F}_t]$ can  be computed as follows:
\begin{align*}
	\Prob(I_k=1\mid \mathcal{F}_t)&=\Prob(W_T \in D_k\mid W_t),\\
	\E[g(x_kZ_k)\mid \mathcal{F}_t]
	&=\E\Big[g\Big(x_k\exp\Big\{-\frac{\sigma_k^2}{2}T+\sigma_kB_T\Big\}\Big)\mid B_t\Big].
\end{align*}
We write
\begin{align*}
p_t^k=\Prob(I_k=1\mid \mathcal{F}_t) = \Prob(W_t+W_T-W_t\in D_k\mid W_t)=\Phi_n\Big(\frac{D_k-W_t}{\sqrt{T-t}}\Big),
\end{align*}
where $\Phi_n$ is the standard Gaussian distribution in $\R^n$.
Note that the stochastic process $\{p_t\}_{t\in [0,T]}$, where $p_t=(p_t^1,\dots,p_t^{n+2})$, is a martingale on the simplex $\mathcal{S}=\{p \in \R^{n+2} : p\geq 0, 1^\trans p=1\}$ with the property that $p_T\in \{e_1,\dots,e_{n+2}\}$, where the $e_k$s are the basis vectors of the standard Euclidean basis in $\R^{n+2}$.
The forward prices at time $t$ are given by
\begin{align*}
  G^0_t=\sum_k p_t^k x_t^k \quad \text{and} \quad 
  G_t^j=\sum_k p_t^k G^{{\small\text{B}}}(x_t^k,\sigma_k,K_j,T-t) 
  \quad \text{for } j=1,\dots,n,
\end{align*}
where $G^{{\small\text{B}}}$ denotes Black's formula \eqref{eq:BformulaFP} for the forward price of a European call option and
\begin{align*}
  x_t^k=x_k\E\Big[\exp\Big\{-\frac{\sigma_k^2}{2}T+\sigma_k B_T\Big\}
    \mid B_t\Big]
  =x_k\exp\Big\{-\frac{\sigma_k^2}{2}t+\sigma_k B_t\Big\}.
\end{align*}

\subsection{Tracking the Brownian particle in continuous time}
\label{sec:filtrations}
In order to use the model at time $t \in (0,T)$ for pricing a European derivative with payoff function $g$, it is necessary to know the location of the Brownian particle $(W^{1}_{t}, \dots, W^{n}_{t}, B_{t})$. That is, given the observed forward prices $(G^{0}_{t}, \dots, G^{n}_{t})$ we need to infer the location of $(W^{1}_{t}, \dots, W^{n}_{t}, B_{t})$. 
We may express $(G^0_t,G_t^1,\dots,G_t^n)$ as the value of a function $h_t$ evaluated at $(W_t^1,\dots,W_t^n,B_t)$.
The filtration $\{\mathcal{G}_t\}_{t\in [0,T]}$ generated by the vector $(G^0,G^1,\dots,G^n)$ of price processes is therefore smaller than or equal to the Brownian filtration $\{\mathcal{F}_t\}_{t\in [0,T]}$ generated by $(W^1,\dots,W^n,B)$. We now investigate the functions $h_t$ in order to compare the two filtrations and to determine the dynamics of the price processes.

The mixture probabilities can be written as $p_t^k=p_t^k(W_t)$, where 
\begin{align}\label{eq:pkexpression}
  p_t^k(w)=\int_{D_k}(2\pi(T-t))^{-n/2}\exp\Big\{-\frac{(x-w)^{\trans}(x-w)}{2(T-t)}\Big\}dx.
\end{align}
Write $h_t=(h^{0}_{t}, \dots, h^{n}_{t})$. Then the forward price $G^0_t$ can be expressed as $G^0_t=h^0_t(W_t,B_t)$, where
\begin{align*}
  h^0_t(w,b)=\sum_{k=1}^{n+2}p_t^k(w)
  x_k\exp\Big\{-\frac{\sigma_k^2}{2}t+\sigma_k b\Big\}.
\end{align*}
Similarly, $G_t^j=h^j_t(W_t,B_t)$, $j=1,\dots, n$, where
\begin{align*}
  h^j_t(w,b)&=\sum_{k=1}^{n+2}p_t^k(w)
  G^{{\small\text{B}}}\Big(x_k\exp\Big\{-\frac{\sigma_k^2}{2}t+\sigma_k b\Big\},\sigma_k,K_j,T-t\Big)\\
  &=\sum_{k=1}^{n+2}p_t^k(w)\Big(x_k\exp\Big\{-\frac{\sigma_k^2}{2}t+\sigma_k b\Big\}\Phi(d_1)-K_j\Phi(d_2)\Big)
\end{align*}
with $d_1=d_1(j,k,b,T-t)$ and $d_2=d_2(j,k,b,T-t)$ given by
\begin{align*}
  d_1&=\frac{1}{\sigma_k\sqrt{T-t}}\Big(-\frac{\sigma_k^2}{2}t+\sigma_k b+\log(x_k/K_j)\Big)+\frac{1}{2}\sigma_k\sqrt{T-t},\\
  d_2&=d_1-\sigma_k\sqrt{T-t}.
\end{align*}
In particular,  
\begin{align}
  (G^0_t,G_t^1,\dots,G_t^n)&=h_t(W_t,B_t)  \label{eq:f_t} \\
  &=(h^0_t(W_t,B_t),h^1_t(W_t,B_t),\dots,h^n_t(W_t,B_t)). \nonumber
\end{align}
If, for every $t\in [0,T)$, $h_t:\R^{n+1}\to\R^{n+1}$ is locally one-to-one everywhere, then an $(n+1)$-dimensional trajectory for the forward prices can be transformed into a unique $(n+1)$-dimensional trajectory for the $(n+1)$-dimensional standard Brownian motion $(W,B)$. 
From the inverse function theorem (Theorem 9.24 in \cite{R76}) we know that if the Jacobian matrix 
\begin{align}\label{eq:jacobian}
h_t'(w,b)=\left(\begin{array}{lclr}
\frac{\partial h^0_t}{\partial w_1}(w,b) & \dots & \frac{\partial h^0_t}{\partial w_n}(w,b) & \frac{\partial h^0_t}{\partial b}(w,b) \\
\vdots & & & \vdots \\
\frac{\partial h^n_t}{\partial w_1}(w,b) & \dots & \frac{\partial h^n_t}{\partial w_n}(w,b) & \frac{\partial h^n_t}{\partial b}(w,b)
\end{array}\right)
\end{align}
of the continuously differentiable function $h_t$ is invertible at the point $(w,b)$, then $h_t$ is one-to-one in a neighborhood of $(w,b)$ and has a continuously differentiable inverse in a neighborhood of $h_t(w,b)$. The set
\begin{align*}
\Gamma_t = \{(w,b) \in \R^{n+1} : \det h_t'(w,b)=0\}
\end{align*}
is the subset of $\R^{n+1}$ where $h_t$ is not locally one-to-one. 

\begin{figure}[!ht]
	\centering
	\includegraphics[scale=.35]{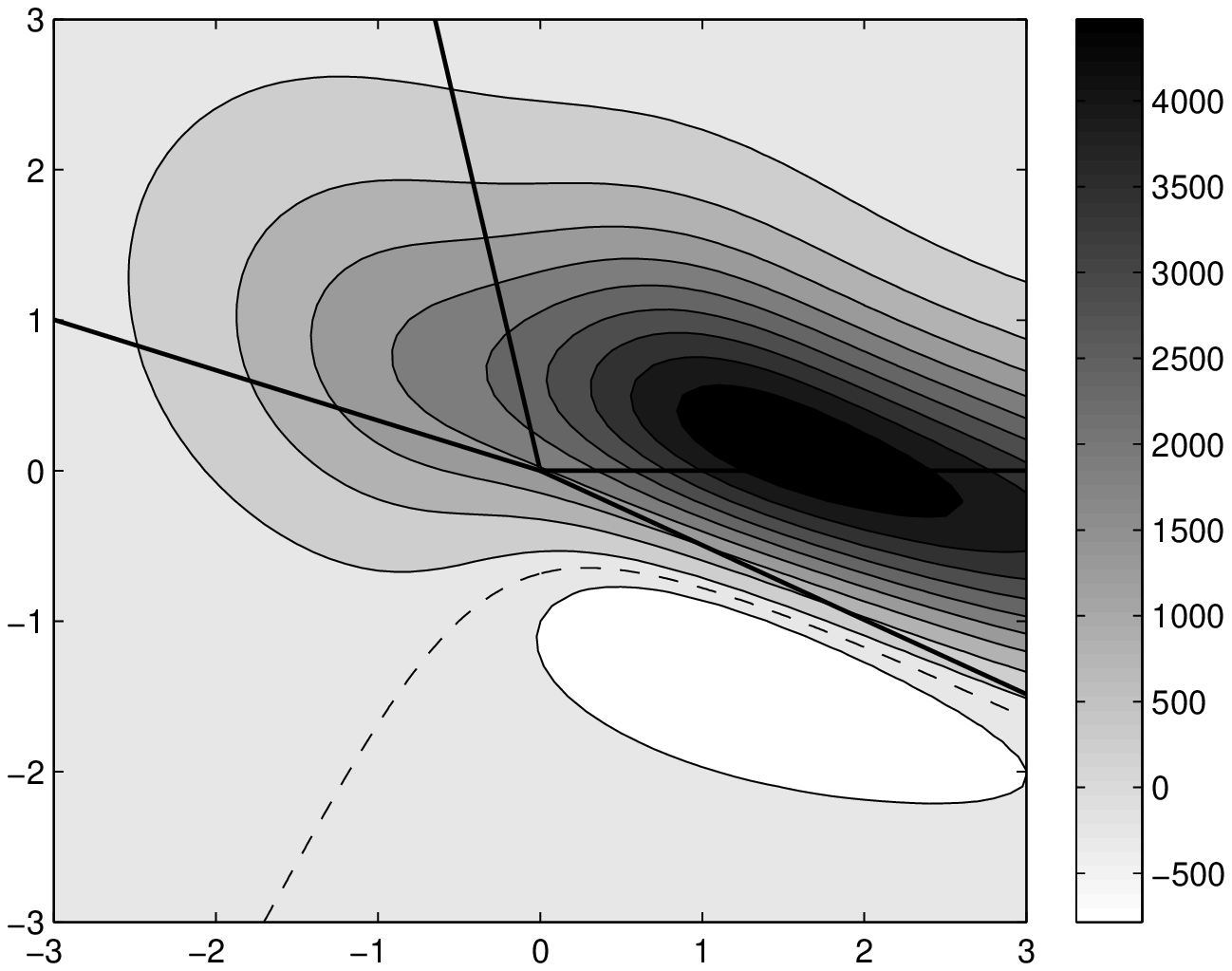}
	\includegraphics[scale=.35]{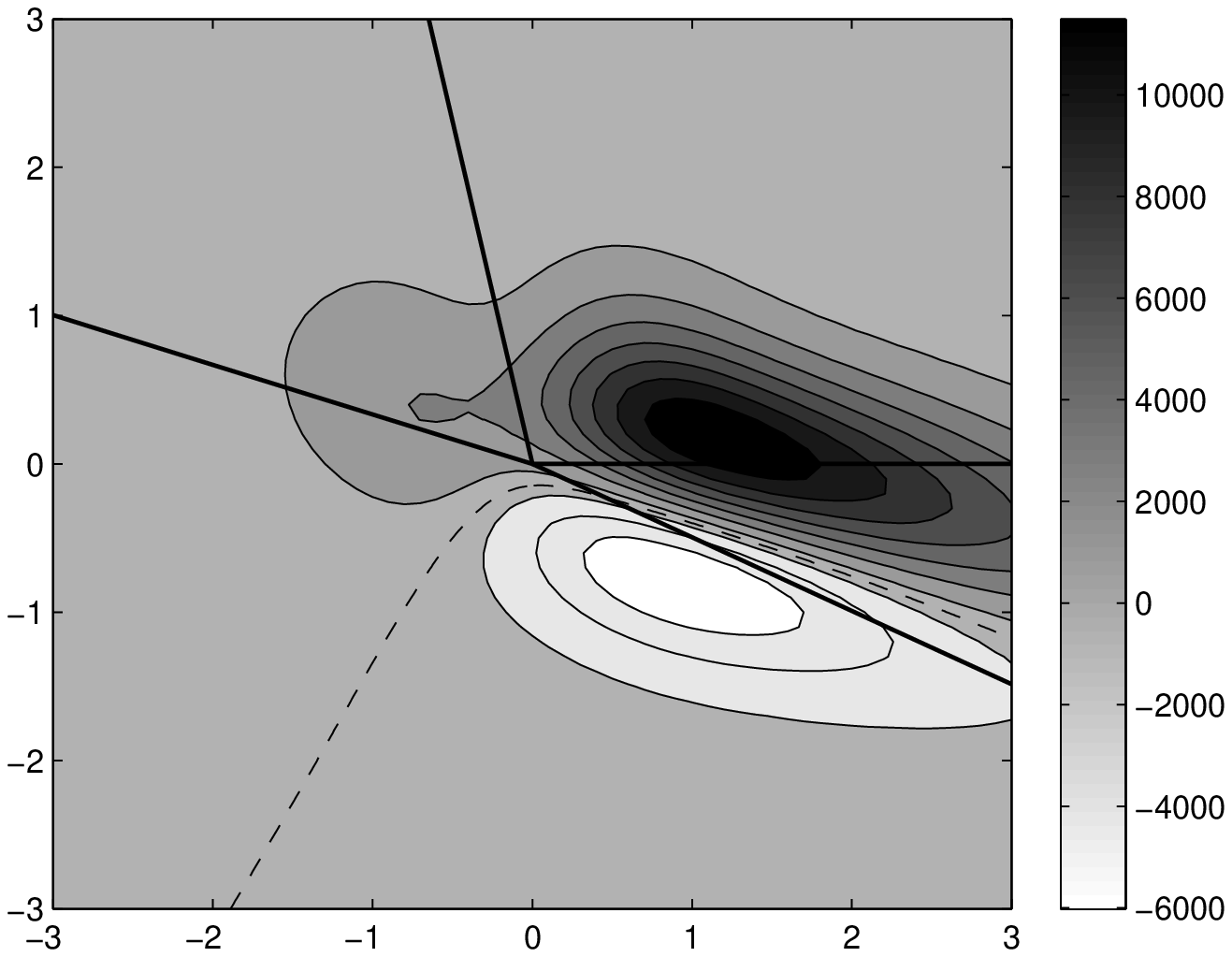}
	\includegraphics[scale=.35]{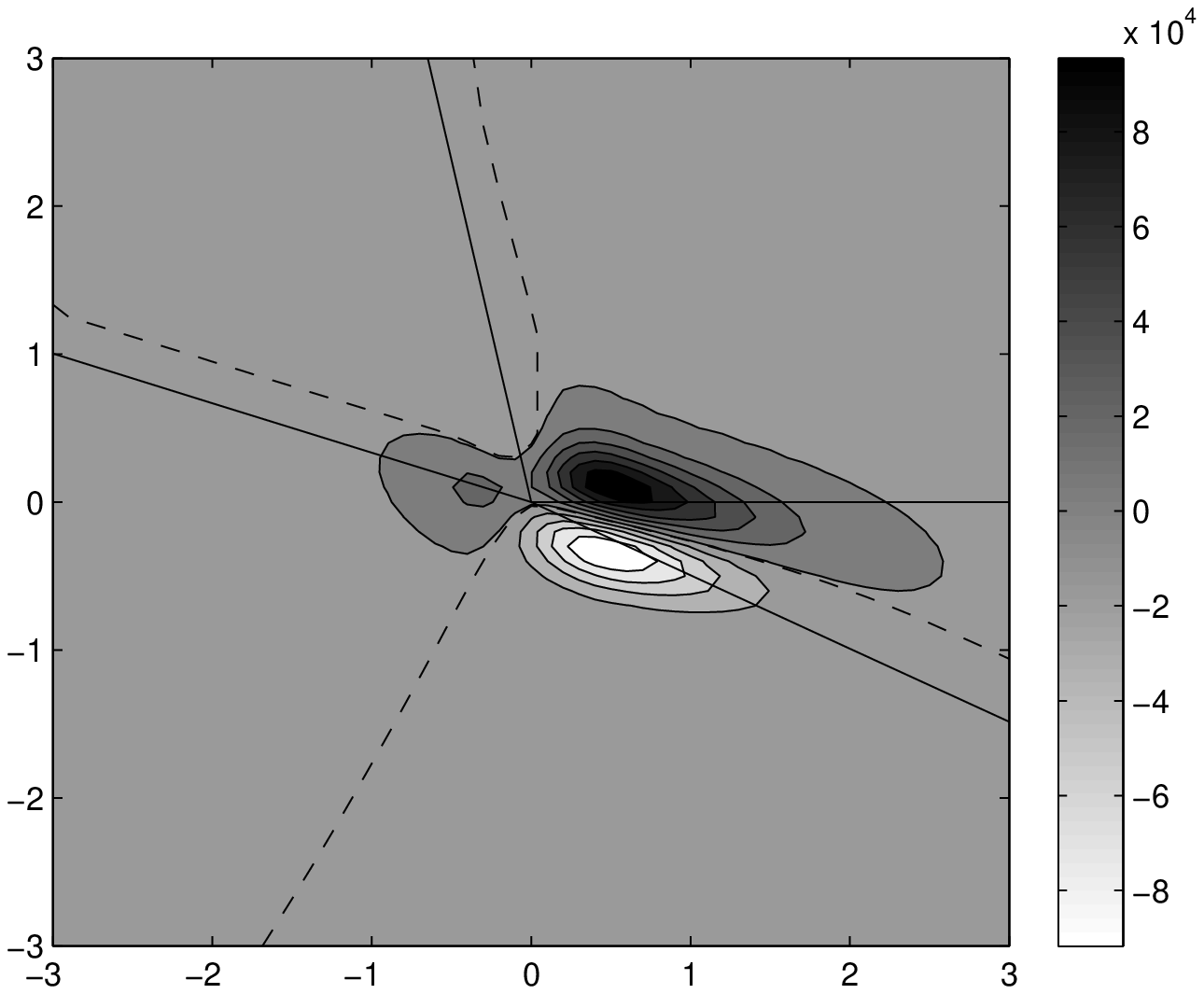}
	\caption{Contour plots of the Jacobian determinant $\det h'_t$, for $n=2$ and $b=0$, as a function of $(w_1,w_2)$ at times $t=0$, $t=0.5$, and $t=0.9$. The zeros of the determinant are 
displayed along the dotted curves. The functions $h_t$ correspond to a forward density process calibrated to S\&P 500 option data presented in Section \ref{sec:evaluationofthemodel} and parameterized as in \eqref{eq:paramsneq2}.}
 	\label{fig:jacdetcolbar}
\end{figure}

In order to investigate the sets $\Gamma_t$ and in order to express the dynamics of the price processes using It\^o's formula the  partial derivatives of the functions $h_t$ must be computed. We find that 
\begin{align*}
  \frac{\partial h^0_t}{\partial b}(w,b)=\sum_{k=1}^{n+2}p_t^k(w)
  \sigma_k x_k\exp\Big\{-\frac{\sigma_k^2}{2}t+\sigma_k b\Big\}
\end{align*}
and
\begin{align*}
  \frac{\partial h^j_t}{\partial b}(w,b)=\sum_{k=1}^{n+2}p_t^k(w)
  \Phi(d_1)\sigma_k x_k\exp\Big\{-\frac{\sigma_k^2}{2}t+\sigma_k b\Big\},
\end{align*}
where $d_1=d_1(j,k)$ depends on $j$ and $k$ through $K_j$ and $\sigma_k$.
Similarly,
\begin{align*}
  \frac{\partial h^0_t}{\partial w_i}(w,b)=\sum_{k=1}^{n+2}
  \frac{\partial p_t^k}{\partial w_i}(w) x_k\exp\Big\{-\frac{\sigma_k^2}{2}t+\sigma_k b\Big\}
\end{align*}
and
\begin{align*}
  \frac{\partial h^j_t}{\partial w_i}(w,b)=\sum_{k=1}^{n+2}
  \frac{\partial p_t^k}{\partial w_i}(w)
  G^{{\small\text{B}}}\Big(x_k\exp\Big\{-\frac{\sigma_k^2}{2}t+\sigma_k b\Big\},\sigma_k,K_j,T-t\Big).
\end{align*}
Finally,
\begin{align*}
\frac{\partial p_t^k}{\partial w_i}(w)=\int_{D_k}(2\pi(T-t))^{-n/2}\frac{(x_i-w_i)}{T-t}\exp\Big\{-\frac{(x-w)^{\trans}(x-w)}{2(T-t)}\Big\}dx.
\end{align*}
Numerical investigations,  illustrated in Figure \ref{fig:jacdetcolbar}, indicate that $\Gamma_t$ is a smooth surface of dimension $n$ that varies continuously with $t$. If the function $\det h_t' : \R^{n+1}\to\R$ has a nonzero gradient almost everywhere in $\Gamma_t=\{(w,b)\in \R^{n+1}: \det h_t' =0\}$, then the implicit function theorem (Theorem 9.28 in \cite{R76}) implies that $\Gamma_t$ is a continuously differentiable hypersurface in $\R^{n+1}$.  Similarly, if the function $(t,w,b)\mapsto \det h_t'(w,b)$ has a nonzero gradient almost everywhere in $\Gamma=\{(t,w,b)\in \R^{n+2}: \det h_t' =0\}$, then $\Gamma$ is a continuously differentiable hypersurface in $\R^{n+2}$ from which we conclude that the $\Gamma_t$s vary continuously with $t$.
If the gradients are nonzero almost everywhere, then we conclude that
$\Prob((W_t,B_t)\in \Gamma_t)=0$ for all $t$ but that 
$\Prob((W_t,B_t)\in \Gamma_t \text{ for some } t)>0$. In particular, if 
\begin{align*}
\tau=\inf\{t>0:(W_t,B_t)\in \Gamma_t\},
\end{align*}
the first time that the $(n+1)$-dimensional Brownian motion $(W,B)$ arrives at a point where $h_{\tau}$ is not locally invertible, then the trajectory of $\{(W_t,B_t)\}_{t\in [0,\tau]}$ is uniquely determined by the trajectory of $\{h_t(W_t,B_t)\}_{t\in [0,\tau]}$. Therefore, $\tau$ is a stopping time with respect to $\{\mathcal{G}_t\}_{t\in [0,T]}$ and $\mathcal{G}_{t\wedge \tau}=\mathcal{F}_{t\wedge \tau}$. 
However, whether the trajectory of $\{(W_t,B_t)\}_{t\in [0,T]}$ is uniquely determined by the trajectory of $\{h_t(W_t,B_t)\}_{t\in [0,T]}$ or not depends on the function $h_{\tau}$ in a neighborhood of $(W_{\tau},B_{\tau})\in \Gamma_{\tau}$.

In practice, only discrete observations of the forward prices are available, so it will be impossible to track the Brownian motion exactly based on the discretely observed forward prices. This issue is treated in some detail in Section \ref{sec:evaluationofthemodel} where both a local linear approximation and a particle filtering method is applied to track the location of the Brownian particle. 

\subsection{The forward price dynamics}
Many popular models for derivative pricing are based on modeling the dynamics of the underlying spot price or forward price directly. Examples are Black's model, Dupire's model, and stochastic volatility models. Our starting point is a model for the dynamics of the forward density. From the model for the forward density process, the dynamics of the forward price process $\{G^{0}_t\}_{t\in [0,T]}$ are derived from the expressions for the partial derivatives of $h_{t}$ and It\^{o}'s formula (Theorem 33, p.~81, in \cite{P04}):
\begin{align*}
G^0_t&=G^0_0+\sum_{i=1}^{n}\int_0^t  \frac{\partial h^0_s}{\partial w_i}(W_s,B_s)dW^i_s+\int_0^t \frac{\partial h^0_s}{\partial b}(W_s,B_s)dB_s\\
&\quad +\frac{1}{2}\int_0^t\Big(\sum_{i=1}^{n} \frac{\partial^2 h^0_s}{\partial w_i^2}(W_s,B_s)+2\frac{\partial h^0_s}{\partial s}(W_s,B_s)+\frac{\partial^2 h^0_s}{\partial b^2}(W_s,B_s)\Big)ds\\
&=G^0_0+\sum_{i=1}^{n}\sum_{k=1}^{n+2}\int_0^t 
  \frac{\partial p_s^k}{\partial w_i}(W_s) x_k\exp\Big\{-\frac{\sigma_k^2}{2}s+\sigma_k B_s\Big\}dW^i_s\\
&\quad +  
  \sum_{k=1}^{n+2}\int_0^t p_s^k(W_s)
  \sigma_k x_k\exp\Big\{-\frac{\sigma_k^2}{2}s+\sigma_k B_s\Big\}dB_s\\
 &\quad + \sum_{k=1}^{n+2}\int_0^t \Big(\frac{1}{2}
 \sum_{i=1}^{n}\frac{\partial^2 p_s^k}{\partial w_i^2}(W_s)+\frac{\partial p_s^k}{\partial s}(W_s)\Big)x_k\exp\Big\{-\frac{\sigma_k^2}{2}s+\sigma_k B_s\Big\}ds.
\end{align*}
From e.g.~the martingale representation theorem (Theorem 43, p.~186, in \cite{P04}) it follows that the last sum above vanishes so that 
\begin{align*}
G^0_t&=G^0_0+\sum_{i=1}^{n}\sum_{k=1}^{n+2}\int_0^t 
  \frac{\partial p_s^k}{\partial w_i}(W_s) x_k\exp\Big\{-\frac{\sigma_k^2}{2}s+\sigma_k B_s\Big\}dW^i_s\\
&\quad +  
  \sum_{k=1}^{n+2}\int_0^t p_s^k(W_s)
  \sigma_k x_k\exp\Big\{-\frac{\sigma_k^2}{2}s+\sigma_k B_s\Big\}dB_s.
\end{align*}

The derivatives computed so far can also be used to study the conditional density process $\{f_t(x)\}_{t\in [0,T]}$. The conditional density 
\begin{align*}
f_t(x)=\sum_{k=1}^{n+2}p_t^k f_t^k(x)
\end{align*}
is a convex combination, with random probability weights $p_t^k$ as above, of lognormal densities $f_t^k(x)$, where   
\begin{align*}
f_t^k(x)=\frac{1}{x\sigma_k\sqrt{2\pi(T-t)}}\exp\Big\{-\frac{1}{2}\Big(\frac{\log(x/x_k)+\sigma_k^2T/2-\sigma_kB_t}{\sigma_k\sqrt{T-t}}\Big)^2\Big\}.
\end{align*}
It\^o's formula and the martingale representation theorem yield, where the dependence of $f_t(x)$ on $W_t$ through the $p_t^k$s and on $B_t$ through the $f_t^k$s has been suppressed, 
\begin{align*}
f_t(x)&=f_0(x)+\sum_{i=1}^{n}\int_0^t  \frac{\partial f_{s}(x)}{\partial w_i}dW^i_s+\int_0^t \frac{\partial f_{s}(x)}{\partial b}dB_s\\
&\quad +\frac{1}{2}\int_0^t\Big(\sum_{i=1}^{n} \frac{\partial^2 f_{s}(x)}{\partial w_i^2}+2\frac{\partial f_{s}(x)}{\partial s}+\frac{\partial^2 f_{s}(x)}{\partial b^2}\Big)ds\\
&=f_0(x)+\sum_{i=1}^{n}\sum_{k=1}^{n+2}\int_0^t 
  \frac{\partial p_s^k}{\partial w_i} f_s^k(x)dW^i_s\\
  &\quad+ \sum_{k=1}^{n+2}\int_0^t p_s^k f_s^k(x)\Big(\frac{\log(x/x_k)+\sigma_k^2T/2-\sigma_kB_s}{\sigma_k(T-s)}\Big) dB_s.
%
\end{align*}
Since $f_t(x)>0$ everywhere we may write $f_t(x)$ as a stochastic exponential
\begin{align}\label{eq:dyncondens}
f_t(x)=f_0(x)+\sum_{i=1}^{n+1} \int_0^t (\sigma_s^f)^i(x)f_s(x)dV_s^i,
\end{align}
where $(V_1,\dots,V_{n+1})=(W_1,\dots,W_n,B)$ and 
\begin{align*}
(\sigma_s^f)^i(x)&=f_s(x)^{-1}\sum_{k=1}^{n+2}\frac{\partial p_s^k}{\partial w_i} f_s^k(x), \quad i=1,\dots,n,\\
(\sigma_s^f)^{n+1}(x)&=f_s^k(x)^{-1}\sum_{k=1}^{n+2}p_s^kf_s^k(x)\Big(\frac{\log(x/x_k)+\sigma_k^2T/2-\sigma_kB_s}{\sigma_k(T-s)}\Big).
\end{align*}
Conditional density models of the form \eqref{eq:dyncondens} are studied in \cite{FHM11}. In \cite{FHM11}, the authors characterize the processes $\sigma^f_t(x)$ that imply that $\{f_t(x)\}$ in \eqref{eq:dyncondens} is a conditional density process
(a forward density process with the choice of numeraire considered here), and provide several explicit examples. 
Here, we consider a particular forward density process and determine the corresponding volatility process $\{\sigma^f_t(x)\}$.

\subsection{Explicit computations in the case $n=2$}

The expression for $p^{k}_{t}$, $k=1,\dots, n+2$ in \eqref{eq:pkexpression} is an integral of a Gaussian density over a set $D_{k}$. In this section explicit evaluation of the partial derivatives of $p^{k}_{t}$ will be performed in the case where $n = 2$ and $D_{k}$ is a cone. 

Recall that the case $n=2$ corresponds to one forward contract on $S_T$ and two call options on $S_T$. In this case $W=(W_x,W_y)$ is a Brownian motion in $\R^2$. We choose the sets $D_1,D_2,D_3,D_4$ as cones or unions of cones because that gives a convenient parameterization for numerical computations and because the configuration of the number of cones and their placement can be rather easily modified to produce dynamics for the price processes that we find reasonable.

Let $D$ be the cone in the first quadrant expressed in polar coordinates as 
$\{(r,v):r\geq 0,v \in [0,\theta]\}$ for $\theta \in [0,\pi/2]$.
With $Z$ denoting a random vector with the standard two-dimensional Normal distribution we write
\begin{align}
  &\Prob(w+\sqrt{T-t}Z\in D)\label{eq:pktprob}\\
  &\quad=\int_{D}\frac{1}{2\pi(T-t)}
  \exp\Big\{-\frac{1}{2}\frac{(x-w_x)^2+(y-w_y)^2}{T-t}\Big\}dxdy\nonumber\\
  &\quad=\int_0^{\infty}\int_{y/\tan\theta}^{\infty}
  \frac{1}{2\pi(T-t)}
  \exp\Big\{-\frac{1}{2}\frac{(x-w_x)^2+(y-w_y)^2}{T-t}\Big\}dxdy\nonumber.
\end{align}
The derivative of \eqref{eq:pktprob} with respect to $w_x$ is
\begin{align}
  &\int_0^{\infty}\int_{y/\tan\theta}^{\infty}
  \frac{1}{2\pi(T-t)}\frac{\partial}{\partial w_x}
  \exp\Big\{-\frac{1}{2}\frac{(x-w_x)^2+(y-w_y)^2}{T-t}\Big\}dxdy\nonumber\\
  &\quad=\int_0^{\infty}\int_{y/\tan\theta}^{\infty}
  \frac{1}{2\pi(T-t)}\frac{x-w_x}{T-t}
  \exp\Big\{-\frac{1}{2}\frac{(x-w_x)^2+(y-w_y)^2}{T-t}\Big\}dxdy\nonumber\\
  &\quad=\frac{1}{2\pi(T-t)}\int_0^{\infty}\Big[-\exp\Big\{-\frac{1}{2}\frac{(x-w_x)^2+(y-w_y)^2}{T-t}\Big\}\Big]_{y/\tan\theta}^{\infty}dy\nonumber\\
  &\quad=\frac{1}{2\pi(T-t)}\int_0^\infty \exp\Big\{-\frac{1}{2}\frac{(y/\tan\theta-w_x)^2+(y-w_y)^2}{T-t}\Big\}dy\label{eq:pktprobder}.
\end{align}
The identity
\begin{align*}
  (ay-b)^2+(y-c)^2
  =\frac{\left(y-\frac{ab+c}{1+a^2}\right)^2}{\frac{1}{1+a^2}} 
  + \frac{(ac-b)^2}{1+a^2}
\end{align*}
with $a = 1/\tan\theta$, $b = w_x$ and $c = w_y$ can be used to write the 
integral in \eqref{eq:pktprobder} as
\begin{align*}
  \frac{\exp\Big\{-\frac{1}{2}\frac{(w_y/\tan\theta-w_x)^2}{(T-t)(1+1/\tan^2\theta)}\Big\}}{\sqrt{2\pi(T-t)(1+1/\tan^2\theta)}}\int_0^\infty \frac{\exp\Big\{-\frac{1}{2}\frac{\left(y-\frac{w_x/\tan\theta+w_y}{1+1/\tan^2\theta}\right)^2}{\frac{T-t}{1+1/\tan^2\theta}}\Big\}}{\sqrt{2\pi\frac{T-t}{1+1/\tan^2\theta}}}dy.
\end{align*}
The integral expression may not look pretty but can be written explicitly as
\begin{align*}
  &\frac{\partial}{\partial w_x}\Prob(w+\sqrt{T-t}Z\in D)\\
  &\quad=\frac{\exp\Big\{-\frac{1}{2}\frac{(w_y/\tan\theta-w_x)^2}{(T-t)(1+1/\tan^2\theta)}\Big\}}{\sqrt{2\pi(T-t)(1+1/\tan^2\theta)}}\Phi\Big(\frac{w_x/\tan\theta+w_y}{\sqrt{(T-t)(1+1/\tan^2\theta)}}\Big)
\end{align*}
in terms of the univariate standard Normal distribution function $\Phi$.
Similar computations for the derivative of \eqref{eq:pktprob} with respect
to $w_y$ give
\begin{align*}
  &\frac{\partial}{\partial w_y}\Prob(w+\sqrt{T-t}Z\in D)
  =\frac{\exp\Big\{-\frac{w_y^2}{2(T-t)}\Big\}}{\sqrt{2\pi(T-t)}}
  \Phi\Big(\frac{w_x}{\sqrt{T-t}}\Big)\\
  &\quad-\frac{\exp\Big\{-\frac{1}{2}\frac{(w_x\tan\theta-w_y)^2}{(T-t)(1+\tan^2\theta)}\Big\}}{\sqrt{2\pi(T-t)(1+\tan^2\theta)}}\Phi\Big(\frac{w_x+w_y\tan\theta}{\sqrt{(T-t)(1+\tan^2\theta)}}\Big).
\end{align*}
Let $D$ be a cone that can be expressed, in polar coordinates, as
$\{(r,v):r\geq 0,v \in [\phi,\phi+\theta]\}$, where $\theta \in (0,\pi/2]$ and $\phi\in [0,2\pi-\theta]$.
Let further $O_{\phi}$ be the matrix corresponding to a clock-wise rotation of angle $\phi$ around the origin so that  $O_{\phi}D$ is of the form, in polar coordinates, 
$\{(r,v):r\geq 0,v \in [0,\theta]\}$.
Then 
\begin{align*}
  \Prob(w+\sqrt{T-t}Z\in D)=\Prob(O_{\phi}w+\sqrt{T-t}Z\in O_{\phi}D)
\end{align*}
and the above computation, with $\tilde{w}=O_{\phi}w$ instead of $w$, can be used to compute the partial derivatives of $\Prob(w+\sqrt{T-t}Z\in D)$ with respect to $w_x$ and $w_y$. With $u=O_{\phi}e_1$ and $v=O_{\phi}e_2$, where $e_1$ and $e_2$ are the standard basis vectors in $\R^2$, we get
\begin{align*}
  \frac{\partial}{\partial w_x}\Prob(w+\sqrt{T-t}Z\in D)
  &=\cos\phi\frac{\partial}{\partial \tilde{w}_x}
  \Prob(\tilde{w}+\sqrt{T-t}Z\in O_{\phi}D)\\
  &\quad-\sin\phi\frac{\partial}{\partial \tilde{w}_y}
  \Prob(\tilde{w}+\sqrt{T-t}Z\in O_{\phi}D)
\end{align*}
and similarly 
\begin{align*}
  \frac{\partial}{\partial w_y}\Prob(w+\sqrt{T-t}Z\in D)
  &=\sin\phi\frac{\partial}{\partial \tilde{w}_x}
  \Prob(\tilde{w}+\sqrt{T-t}Z\in O_{\phi}D)\\
  &\quad+\cos\phi\frac{\partial}{\partial \tilde{w}_y}
  \Prob(\tilde{w}+\sqrt{T-t}Z\in O_{\phi}D).
\end{align*}

\section{Calibration and evaluation of the model}\label{sec:evaluationofthemodel}


To calibrate and evaluate the model we use 41 daily adjusted closing prices over a 59 day period from September 22nd 2011 to November 19th 2011 of European put and call options with strike prices $1150, 1175, 1200, 1225$, and $1250$ on the S\&P 500 index value.  The options mature on November 19th 2011. 
A rather short time series of option prices is selected in order to have price data corresponding to sufficiently large traded volumes so that the option prices can be considered to be relevant market prices at the end of each trading day. The risk-free interest rate is set to 0.5\% (corresponding approximately to the three-month LIBOR rate) and  the put-call parity 
\begin{align*}
C_t(K)-P_t(K)=e^{-0.005(T-t)}(G^0_t-K), \quad K \in \{1150,\dots,1250\},
\end{align*}
is used, for the most traded pair of put and call options on each trading day, to calculate the forward prices $G^0_t$ of the underlying asset. For example, on September 22nd 2011, the put and call options with strike $1150$  were the most traded options and their option prices were used to calculate the initial forward prices for delivery of the value of the S\&P 500 index on November 19th. The initial forward price was calculated to $G^0_0=1128.12$. During the analyzed time period, September 22 - November 19,  the forward price increased. Simultaneously the largest trading volumes shifted from the options with strike $1150$ to the options with strike $1200$. 

\subsection{The initial calibration and model specification}

In this section the initial calibration and model specification for the S\&P 500 options will be explained in some detail. The first step is to select the grid parameters $x_{k}$, the volatilities $\sigma_{k}$, and the partitions $D_{k}, k=1,\dots,n+2$. The parameters will be selected to get a reasonable shape of the initial density $f_{0}$ and such that the evolution of the prices have features that are present in real data. 

\begin{figure}[!ht]
	\centering
	\includegraphics[scale=.35]{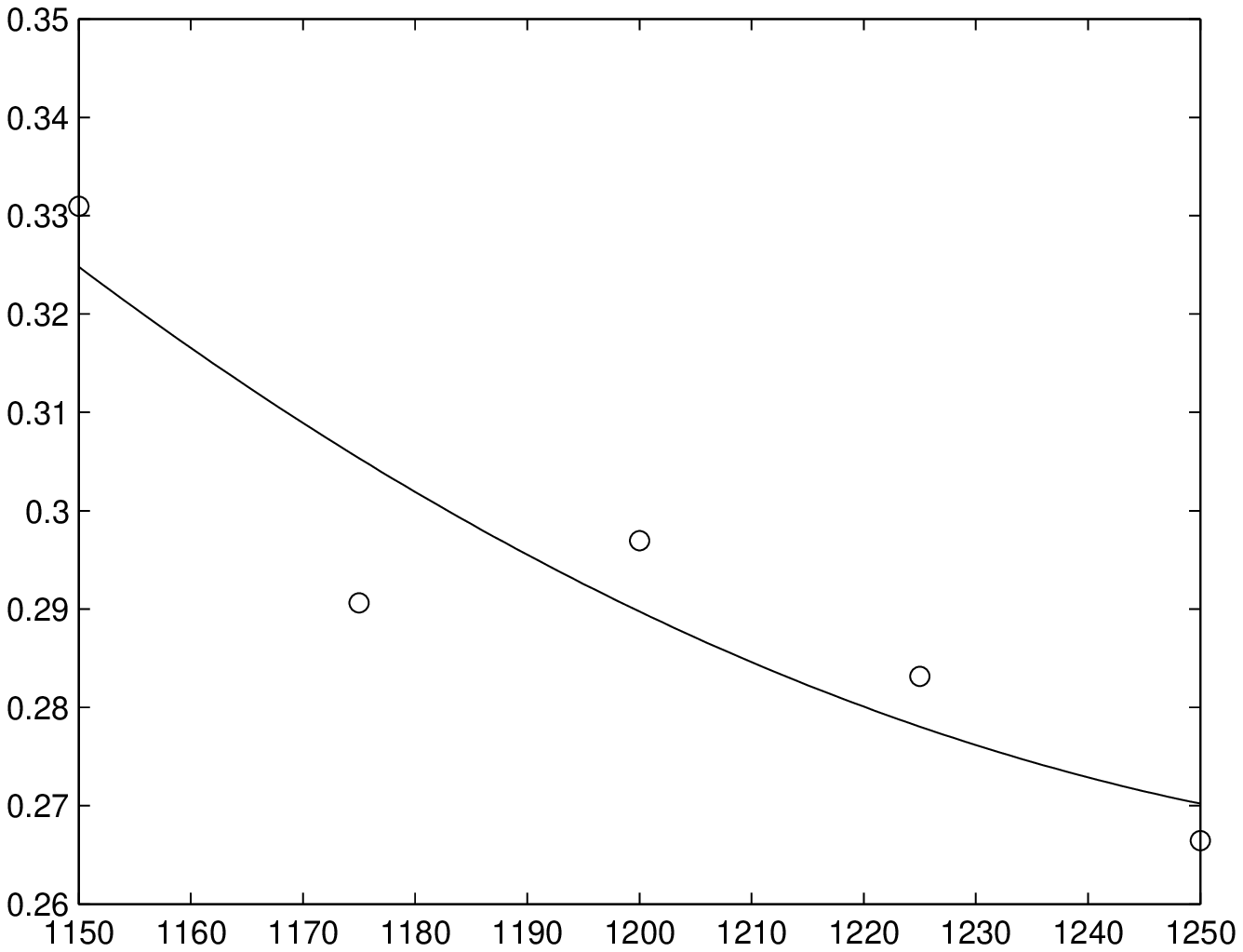}
	\includegraphics[scale=.35]{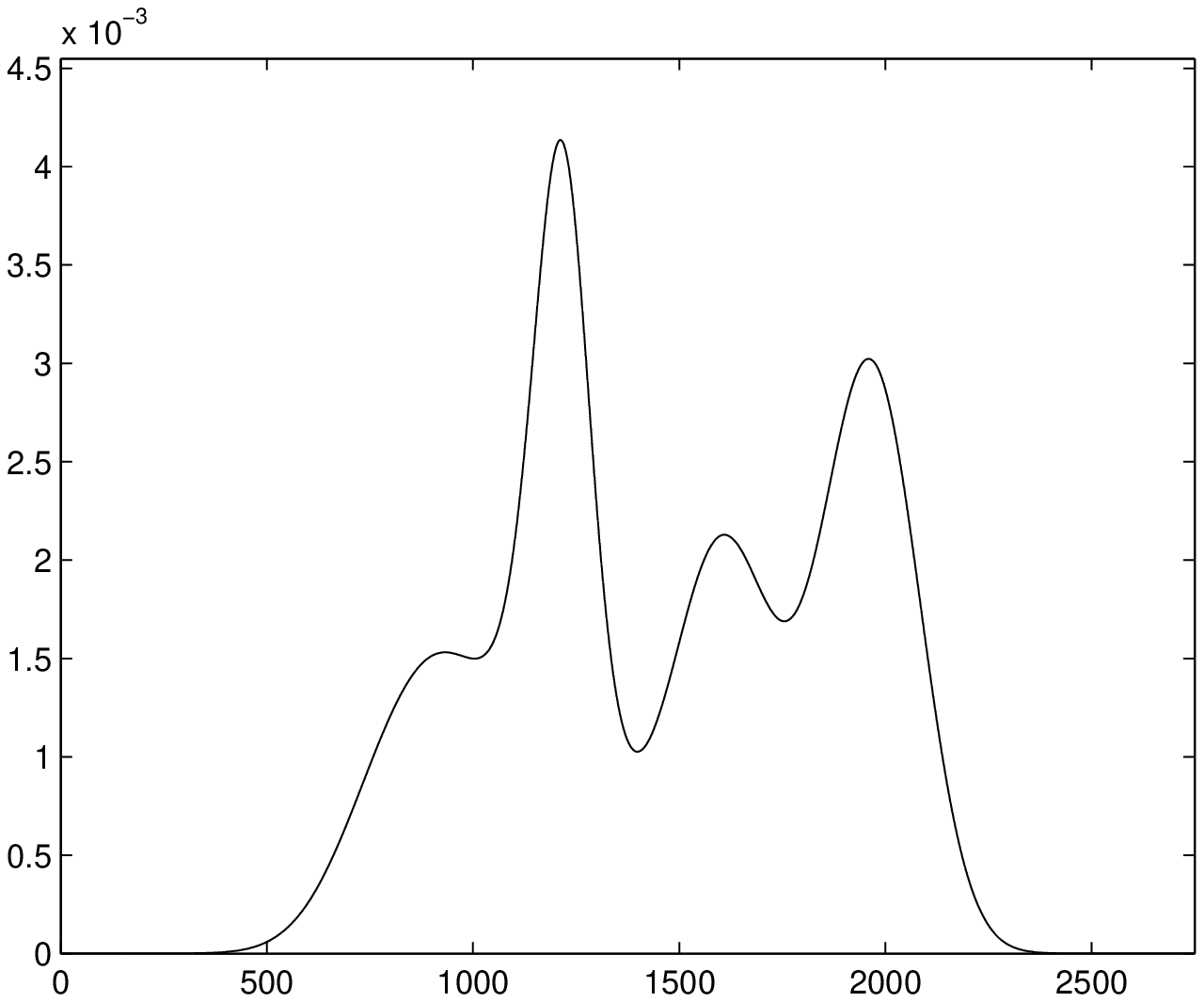}
	\includegraphics[scale=.35]{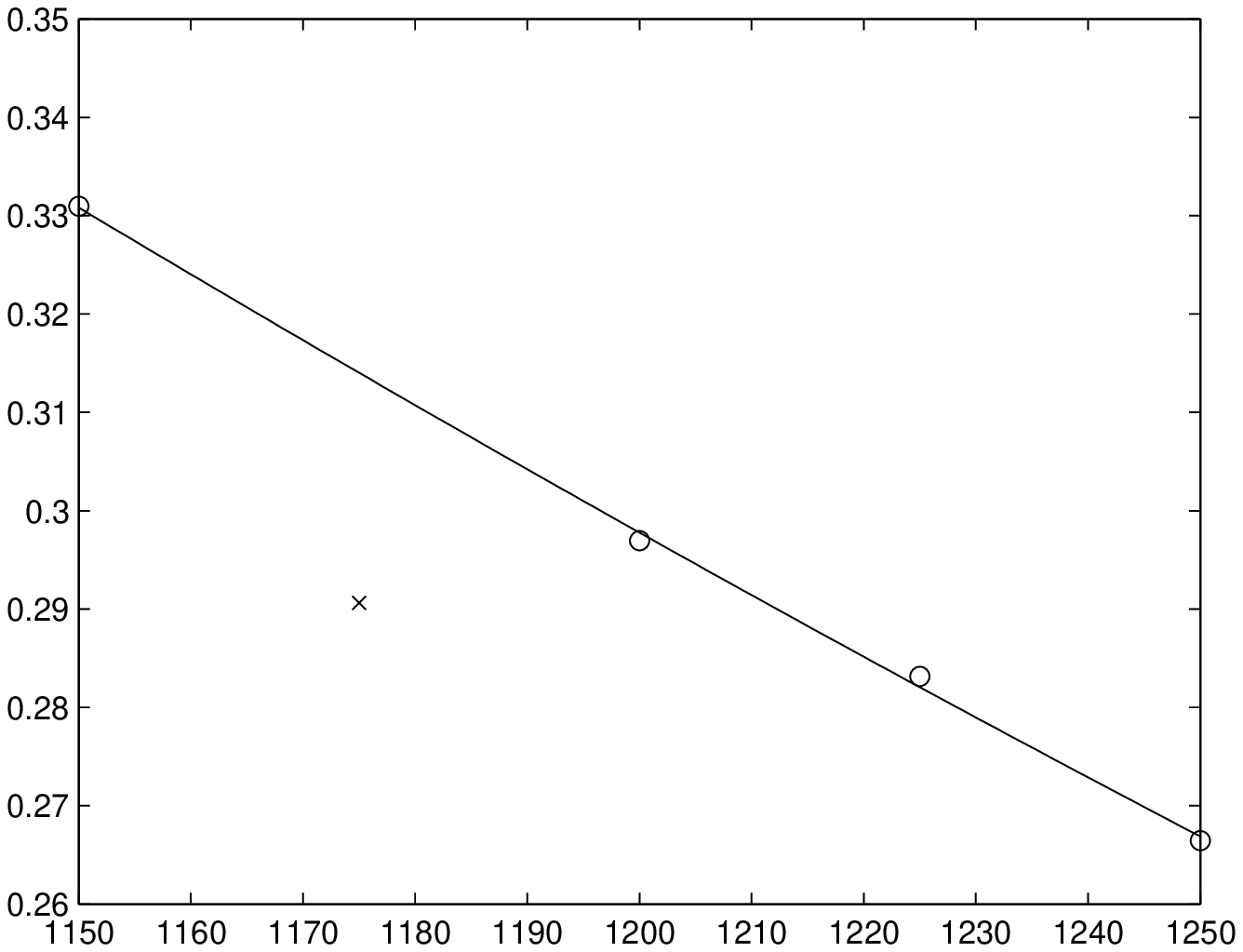}
	\includegraphics[scale=.35]{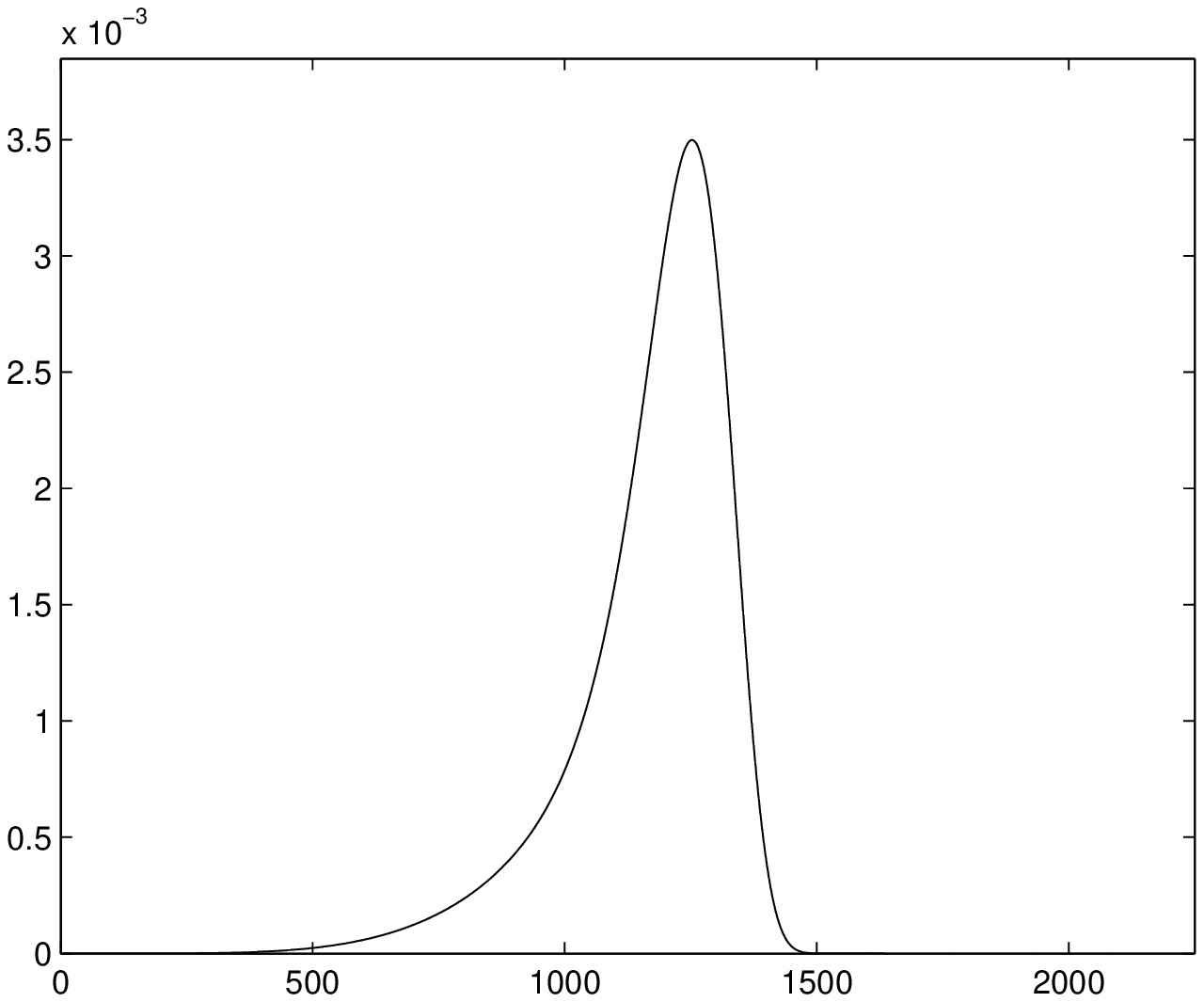}	
 	\caption{
	The upper left plot shows the implied volatilities for the five call options, $\circ$, and the fitted second-degree polynomial (solid curve). The upper right plot shows the probability density for $S_T$ derived from the volatility smile in the upper left plot.
	The lower left plot shows the implied volatilities for the four call options, $\circ$, and the fitted second-degree polynomial (solid curve). The lower right plot shows the probability density for $S_T$ derived from the volatility smile in the lower left plot.}
 	\label{fig:bssmiledens1}
\end{figure}

Let us start by considering the initial density $f_{0}$. As a reference density we will consider the density, $q_{0}^{{\small\text{B}}}$, resulting from Black's formula with a fitted volatility smile. At time $0$ (September 22nd 2011) the implied volatilities, $\sigma^{B}_{i}$, $i=1, \dots,n$, corresponding to the strikes $K_{1}, \dots, K_{n}$, are computed using Black's formula for European call option prices,
\begin{align*}
C^{\small{\text{B}}}_0(K)&=e^{-0.005T}(G^0_0\Phi(d_1)-K\Phi(d_2)),\\ 
d_1&=\frac{\log(G^0_0/K)}{\sigma(K)\sqrt{T}}+\frac{\sigma(K)\sqrt{T}}{2},
\quad d_2=d_1-\sigma(K)\sqrt{T}. 
\end{align*}
A second-degree polynomial (volatility smile or volatility skew) is fitted to the implied volatilities and the formula 
\begin{align*}
\frac{d^2C^{{\small\text{B}}}_0(K)}{dK^2}=e^{-0.005\cdot 59/365}q^{{\small\text{B}}}_0(K),
\end{align*}
see e.g.~\cite{BL78} or \cite{G06}, gives the probability density $q^{{\small\text{B}}}_0$ for $S_T$ implied by the volatility smile and Black's call option price formula (see e.g.~\cite{S93} for details). The prices of the call options with the strike prices 1150, 1175, 1200, 1225, and 1250 produce the volatility smile (second degree polynomial) shown in the upper left plot in Figure \ref{fig:bssmiledens1}. Notice that the fit is rather poor. The corresponding implied density $q^{{\small\text{B}}}_0$ is shown in the upper right plot in Figure \ref{fig:bssmiledens1}. A closer look at the data reveals that the call option with strike price 1175 only has 82 registered trades, so the price of that contract may be unreliable. If that implied volatility is omitted, then the volatility smile in the lower left plot in Figure \ref{fig:bssmiledens1} and the implied probability density in the lower right plot in Figure \ref{fig:bssmiledens1} are obtained. Note that the resulting probability density is smooth, unimodal and left-skewed. 
\emph{We will assume that the volatility smile in the lower left plot in Figure \ref{fig:bssmiledens1}, the graph of the second-degree polynomial $K\mapsto \sigma^{\small\text{B}}(K)$ fitted to the implied volatilities, corresponds to correct market prices which will be used in the calibration of the model in Sections \ref{sec:calibrn2} and \ref{sec:calibrn5}.}

As can be seen in the lower right plot in Figure \ref{fig:bssmiledens1}, implied volatilities that are decreasing with the strike price correspond to a left-skewed implied density function for $S_T$.  This observation is in line with much of the empirical analysis on option price data, see e.g.~\cite{G06}. 

To select the parameters in our model for $f_{0}$,  the parameters $\sigma_k$ will be chosen to produce a left-skewed implied density. 
A commonly held view is that the changes in the implied Black's model volatility and log-returns of the forward (or spot) price are negatively correlated, corresponding to different market responses to good and bad stock market information, see e.g.~\cite{G05}. This behavior is in line with our findings based on the small option price sample used here: for the options with strike prices $K_1=1150$ and $K_2=1200$ the sample correlations between the daily log-returns $\log(G^0_{t+1}/G^0_t)$ and the implied volatility changes $\sigma^{\small\text{B}}_{i,t+1}-\sigma^{\small\text{B}}_{i,t}$, $i=1,2$, are both $-0.35$. For this reason it makes sense to choose the parameters $\sigma_k$ to be decreasing in $k$. If the probability mass of the $p_t^k$s is shifted towards lower indices $k$, then the forward price $G^{0}_t$ decreases and the implied volatility increases.

In Sections \ref{sec:calibrn2} and \ref{sec:calibrn5} below we present the initial calibration of the model to $n=2$ and $n=5$ option prices, respectively. In both cases it is assumed that the volatility smile in the lower left plot in Figure \ref{fig:bssmiledens1} corresponds to correct market prices.

\subsubsection{Initial calibration with two options, $n = 2$}\label{sec:calibrn2}

Let us first consider two options, $n=2$, with strikes $K_1 = 1150$ and $K_2 = 1200$.  The corresponding implied volatilities are $\sigma^{\small\text{B}}(K_1) = 0.33082$ and $\sigma^{\small\text{B}}(K_2) = 0.29777$. 
Since the current forward price of the underlying asset and the risk-free interest rate are known we can use Black's formula to calculate the option prices. We get $C^1_0 = 49.575$ and $C^2_0 = 26.434$, and the corresponding forward prices $G^1_0 = 49.615$ and $G^2_0 = 26.455$, respectively, of the options.
To calibrate the model to these prices, the parameters $x_1$, $x_4$ and $\sigma_k$, $k = 1,\dots, 4$ need to be specified so that the vector $p_0=(p_0^1,\dots,p_0^4)^{\trans}$ is a probability vector, i.e.~has non-negative components that sum up to one.  Proposition \ref{prop:simplemodel} is used to obtain $x_1^{\max} = 1016.81$ and $x_4^{\min} = 1257.11$. We choose $x_1 = 950$ and $x_4 = 1300$ and initially choose $\sigma_k = 0.01$ for all $k$. The resulting density function $f_0(x)=f_0(x;\sigma_1,\dots,\sigma_4)$ is displayed in the upper left plot in Figure \ref{fig:mdens123}.
Even though these parameters are consistent with the observed prices, we are not comfortable with the appearance of the resulting probability density for $S_T$. To get a smoother density we need to increase the $\sigma_k$s. First we increase the $\sigma_k$s simultaneously as long as $p_0$ stays a probability vector. It turns out that $\sigma_k = 0.0542$ is the largest possible value, but the corresponding density $f_0(x)=f_0(x;\sigma_1,\dots,\sigma_4)$ is not left-skewed.  Since the lognormal density is right-skewed, the natural approach is to increase the $\sigma_k$s for small $k$s and decrease the $\sigma_k$s for large $k$s. The parameters $\sigma_1,\dots,\sigma_4=0.18,0.08,0.06,0.03$ give the density  $f_0(x)=f_0(x;\sigma_1,\dots,\sigma_4)$ in the upper right plot in Figure \ref{fig:mdens123}, which is rather similar to the implied density $q_{0}^{{\small\text{B}}}$. We summaries the chosen model parameters:
\begin{align}\label{eq:paramsneq2}
n=2:\quad \left\lbrace
\begin{array}{l}
x_1=950,\; x_2=K_1=1150,\;x_3=K_2=1200,\;x_4=1300, \\
\sigma_1,\dots,\sigma_4=0.18,0.08,0.06,0.03, \\
p_0^1,\dots,p_0^4\approx 0.29, 0.14, 0.51, 0.07.
\end{array}\right.
\end{align}

\subsubsection{Initial calibration with five options, $n=5$}\label{sec:calibrn5}
\begin{figure}
	\centering
	\includegraphics[scale=.35]{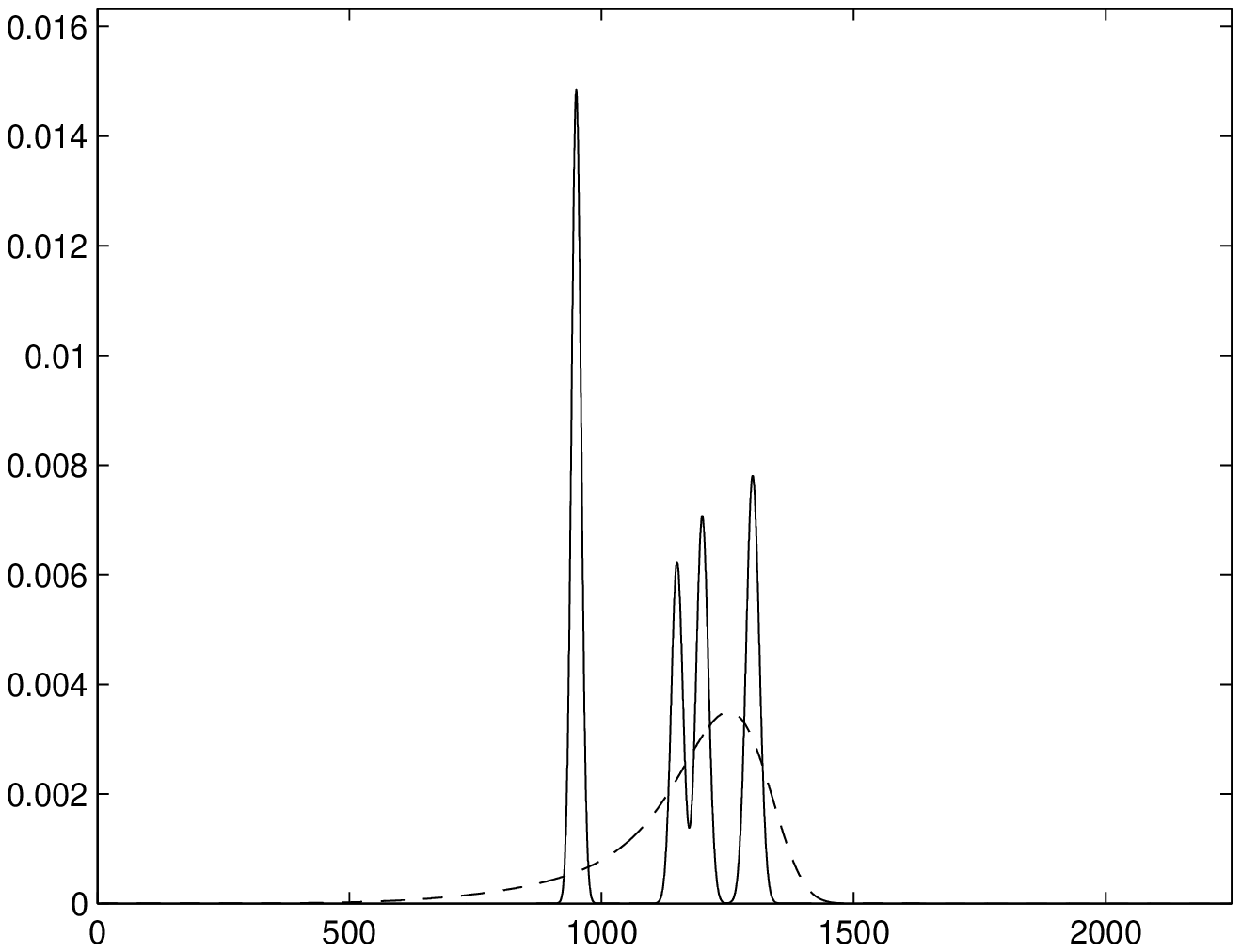}
	\includegraphics[scale=.35]{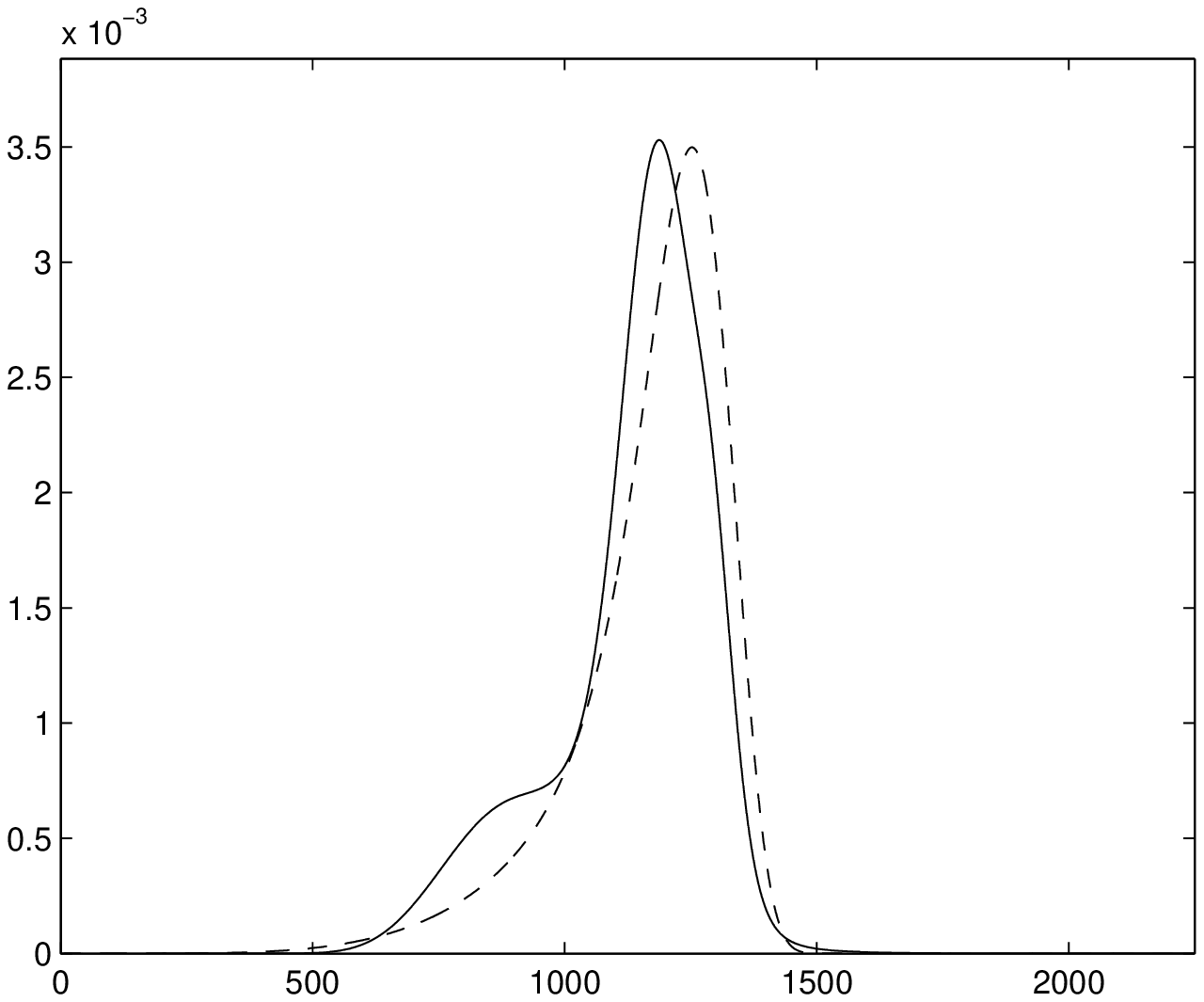}	
	\includegraphics[scale=.35]{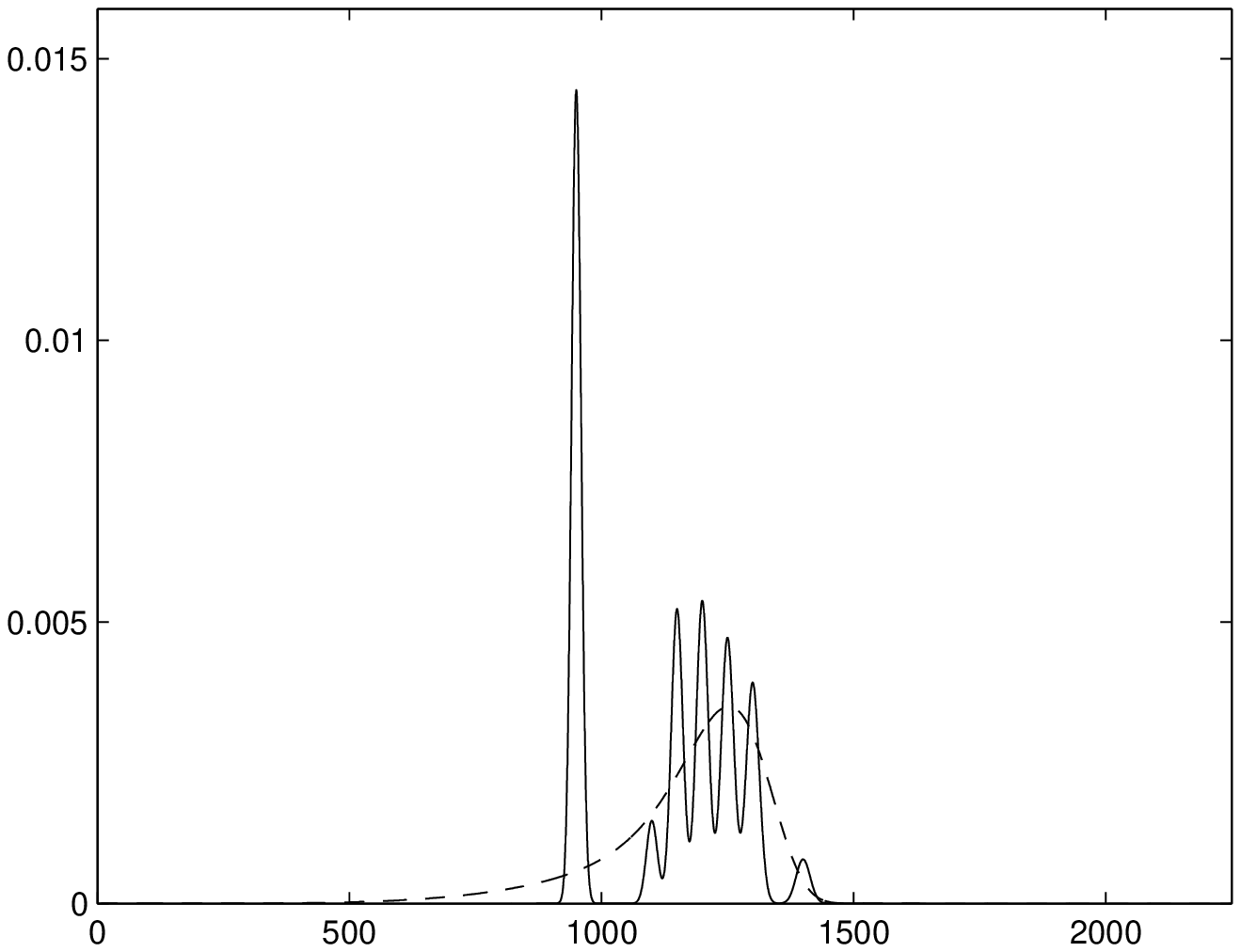}
	\includegraphics[scale=.35]{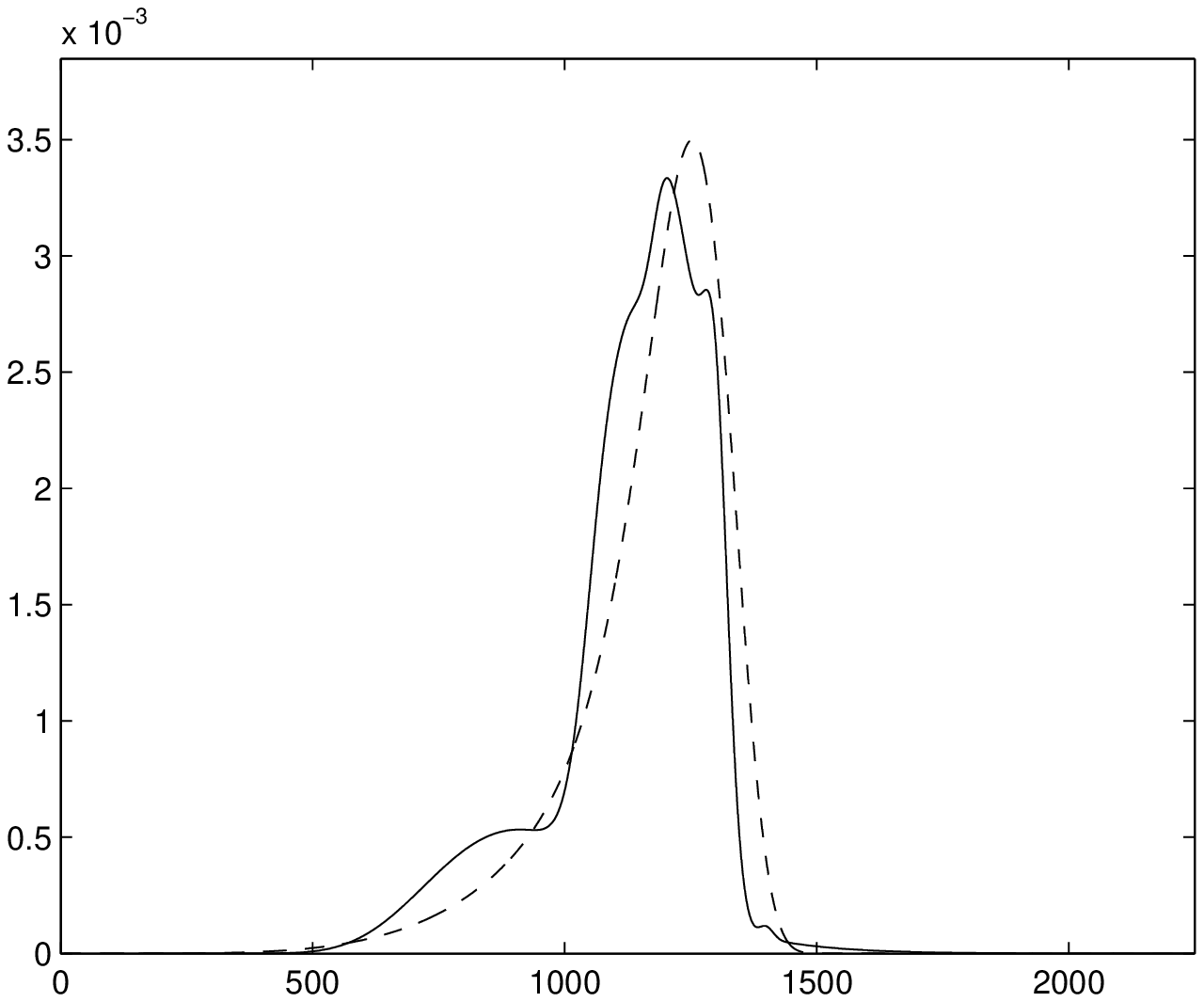}	
 	\caption{The upper plots show the model density $f_0(x)=f_0(x;\sigma_1,\dots,\sigma_4)$ (solid curves) and the implied Black's model density $q^{\small\text{B}}_0(x)$ (dashed curves).
The upper left plot corresponds to $\sigma_k=0.01$ for all $k$ and the upper right plot corresponds to $\sigma_1,\dots,\sigma_4=0.18, 0.08, 0.06, 0.03$.
The lower plots show the model density $f_0(x)=f_0(x;\sigma_1,\dots,\sigma_7)$ (solid curves) and the implied Black's model density $q^{\small\text{B}}_0(x)$ (dashed curves).
The lower left plot corresponds to $\sigma_k=0.01$ for all $k$ and the lower right plot corresponds to $\sigma_1,\dots,\sigma_7=0.21, 0.045, 0.028, 0.025, 0.025, 0.02, 0.01$.}
%
 	\label{fig:mdens123}
\end{figure}

Here $n=5$ options, with strikes $1100,1150,1200,1250,1300$, are considered to illustrate that the calibration procedure easily handles more than two option contracts.
Similar to the setting in Section \ref{sec:calibrn5} the parameters $x_1$, $x_7$ and $\sigma_k$, $k = 1,\dots, 7$ are specified so that $p_0$ is a probability vector.
Using Proposition \ref{prop:simplemodel} to obtain $x_1^{\max} = 968.86$ and $x_7^{\min}= 1321.8$. We choose $x_1 = 950$ and $x_7 = 1400$ and begin
by choosing $\sigma_k = 0.01$ for all $k$. The resulting probability density is displayed in the lower left plot in Figure \ref{fig:mdens123}. We would like the density to spread out the probability mass more evenly and therefore we increase the $\sigma_k$s simultaneously as long as $p_0$ stays a probability vector. It turns out that $\sigma_k = 0.027685$ is the maximum
possible value.
Increasing the $\sigma_k$s for small $k$s and decreasing the $\sigma_k$s for large $k$s will produce a left-skewed density. The density shown in the lower right plot in Figure \ref{fig:mdens123} corresponds to $\sigma_1,\dots,\sigma_7=0.21, 0.045, 0.028, 0.025, 0.025, 0.02, 0.01$. We summaries the chosen model parameters:
\begin{align}\label{eq:paramsneq5}
n=5:\quad \left\lbrace
\begin{array}{l}
x_1=950,\; x_2=K_1=1100,\;x_3=K_2=1150,\;x_4=K_3=1200,\\ x_5=K_4=1250,\;x_6=K_5=1300,\;x_7=1400, \\
\sigma_1,\dots,\sigma_7=0.21, 0.045, 0.028, 0.025, 0.025, 0.02, 0.01, \\
p_0^1,\dots,p_0^7\approx 0.26,0.23,0.08,0.15,0.15,0.13,0.002.
\end{array}\right.
\end{align}
In principle the model can be set up and calibrated to an arbitrarily large number of option contracts. In practice, however, it is difficult to find a large number of reliable option prices for a wide range of strikes. For example, we notice that in our data the options that are actively traded all have strikes close to the current spot price of the underlying asset. Options with strike prices that are far from the current spot price have none or very few trades, so their daily closing prices are unreliable.

\subsubsection{Model dynamics and selection of sets $D_{k}$}
\begin{figure}[!ht]
	\centering
	\includegraphics[scale=.35]{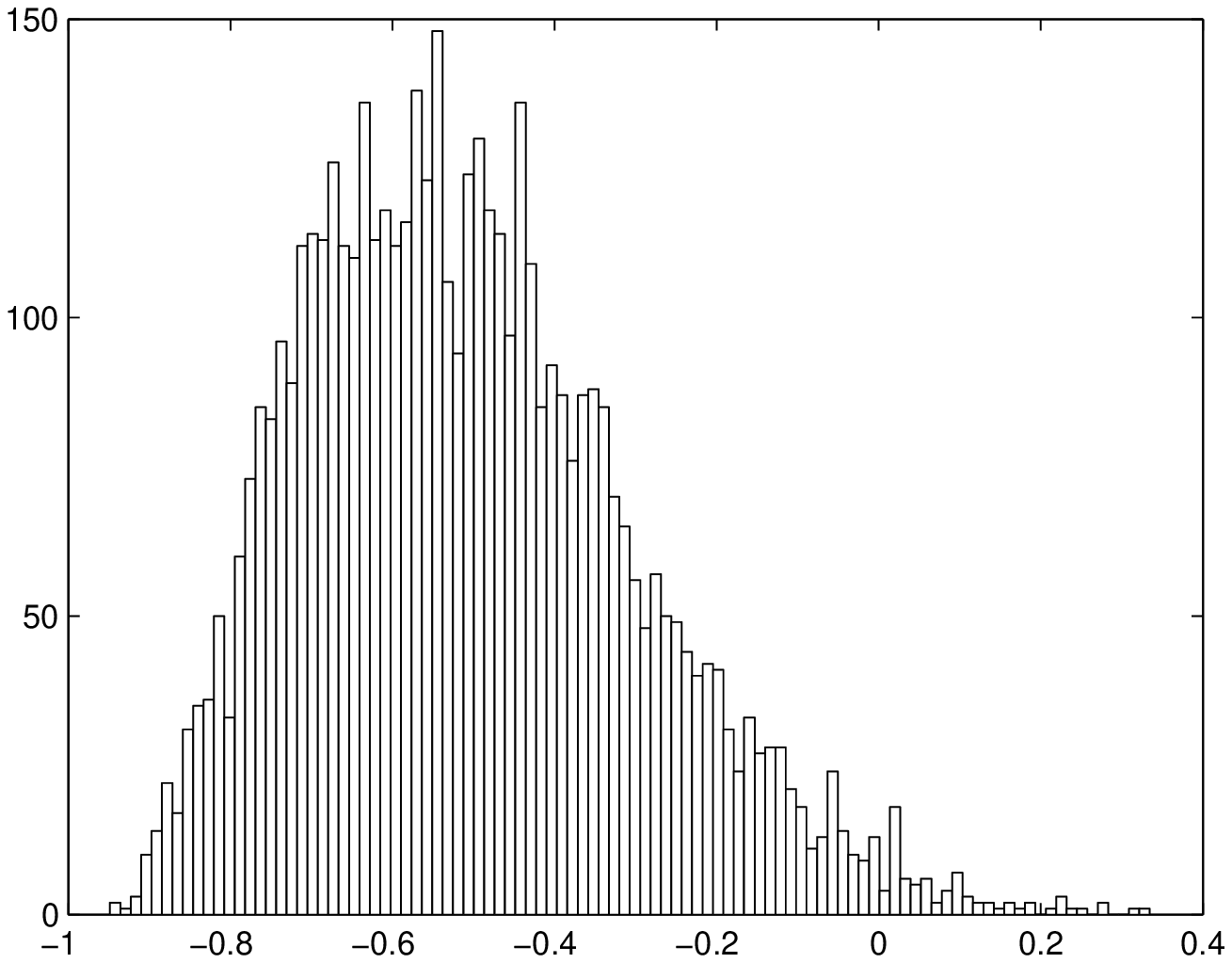}
 	\includegraphics[scale=.35]{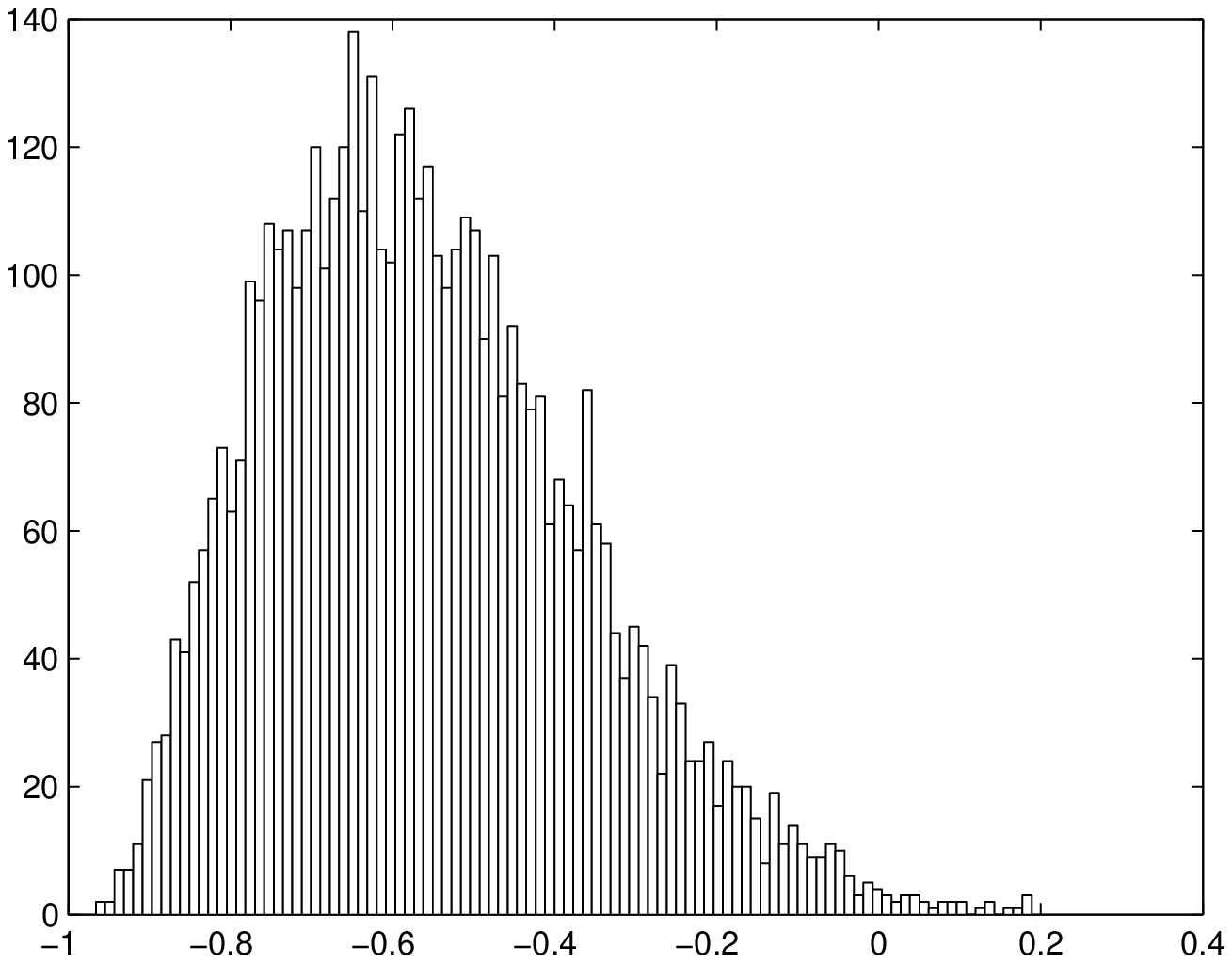}
	\caption{The plots show histograms for the sample correlations between log-returns of the forward price of the underlying asset and the implied volatility changes, based on $5000$ simulated price trajectories. The left plot corresponds to implied volatilities for the call option with strike price $1150$ and the right plot for the strike price $1200$.} 
 	\label{fig:histsim1}
\end{figure}

To examine the dynamics of the model we consider the model parameterized as in \eqref{eq:paramsneq2}. The sets $D_1,\dots, D_4$ are chosen as cones and placed in increasing order starting at the $x$-axis. The cones are illustrated in Figure \ref{fig:jacdetcolbar}.
Then, $N = 5000$ trajectories are simulated of the 3-dimensional Brownian motion (Gaussian random walk) $(W, B)$ with $50$ time steps for each trajectory, corresponding roughly to the number of days of the sample of option prices. For each trajectory and each
time step the corresponding forward price of the underlying asset and the forward prices of the two call options with strike prices $K_1= 1150$ and $K_2 = 1200$, respectively, are calculated.
Next, for each trajectory and each time step Black's formula is used to calculate the implied volatilities for the two options. Finally, the correlation between price changes of the underlying asset and changes in the implied volatilities is calculated. The histograms in Figure
\ref{fig:histsim1} reveal that the correlation is negative in most simulations with mean values
$-0.51$ and $-0.56$. These values appear to be in line with empirical studies, e.g.~\cite{G05}. 

\subsection{Tracking the Brownian particle in discrete time}

As observed in Section \ref{sec:filtrations} the vector of forward prices at every time $t \in [0,T]$ can be expressed as 
$(G^0_t,G^1_t,\dots,G^{n}_t)^{\trans}=h_t(W^1_t,\dots,W^{n}_t,B_t)$,
where $W^1,\dots,W^{n}, B$ are independent one-dimensional Brownian motions. If the vectors $y_t$ of observed prices are within the range of the model, then 
$y_t=h_t(x_t)$ for all $t \in [0,T]$, where the function $h_t$ is given by \eqref{eq:f_t} and 
$\{x_t\}_{t\in [0,T]}$ is an observation of the trajectory of the $(n+1)$-dimensional Brownian motion. Recall that $h_t$ is locally invertible for $t$ smaller than a stopping time $\tau$ measurable with respect to the filtration generated by the price processes. In principle it is possible to uniquely determine the trajectory $\{x_t\}_{t\in [0,\tau]}$ from that of the price process $\{y_t\}_{t\in [0,\tau]}$.

In practice, the situation is more complicated because the price data consist of daily closing prices $\{y_{k\Delta t}:k=1,\dots,T/\Delta t\}$. In particular, the local one-to-one property of the functions $h_t$ do not guarantee that the trajectories of the driving Brownian motions can be well estimated. In this section the aim is to estimate the Brownian motion at the observation times, $\{x_{k\Delta t}:k=1,\dots,T/\Delta t\}$, from the observed option prices. 

\subsubsection{Local linear approximations}\label{sec:loclinapp}

Since the function $h_t$ in \eqref{eq:f_t} is continuously differentiable we may approximate $h_t(x)$ in a neighborhood of a point $x_{0}$ by the best linear approximation $h_t(x_{0})+h_t'(x_{0})(x-x_{0})$. 
Since the $h_t$s are, up to time $\tau$, locally invertible we may use the linear approximations of the $h_t$s together with observations $y_{k\Delta t}$ of the prices to obtain, iteratively, estimates $\widehat{x}_{k\Delta t}$ of the Gaussian random walk $x_{k\Delta t}$:
\begin{align}\label{eq:linearization}
\widehat{x}_{(k+1)\Delta t}&= \widehat{x}_{k\Delta t}+[h'_{(k+1)\Delta t}(\widehat{x}_{k\Delta t})]^{-1}(y_{(k+1)\Delta t}-y_{k\Delta t}),\quad 
 \widehat{x}_{0}=0. 
\end{align}
However, the time step $\Delta t$ corresponding to daily prices is rather large which implies that the linear approximation may be inaccurate. Moreover, the Jacobian matrices $h'_{(k+1)\Delta t}(\widehat{x}_{k\Delta t})$ may be too close to singular leading to poor estimates of the $x_{k\Delta t}$s.


\subsubsection{Particle filtering}


An alternative approach to the local linear approximation is to use an auxiliary particle filter to estimate the $x_{k\Delta t}$s, or rather the posterior distribution of the $x_{k\Delta t}$s. The particle filtering approach considered here works as follows.
\begin{enumerate}
\item At time $k\Delta t$ we have $R$ particles at locations $\alpha_{k\Delta t}^1,...,\alpha_{k\Delta t}^R$, where $\alpha_0^j = 0$ for all $j$.
\item To each particle $\alpha_{k\Delta t}^j$  a first-stage weight $\lambda_j$ is assigned, given by 
\begin{align*}
\lambda_j = \frac{\theta_j}{\sum_{i=1}^R\theta_i}, 
\quad\text{where }\theta_j = \phi(y_{(k+1)\Delta t};h_{(k+1)\Delta t}(\alpha_{k\Delta t}^j),\Sigma_1)
\end{align*}
and $\phi(y; \mu,\Sigma)$ denotes the density at $y$ of the Normal distribution with mean $\mu$ and covariance matrix $\Sigma$. 
\item Draw with replacement from the index set $\{1,\dots,R\}$ according to the weights $\lambda_j$ to produce $R$ indices $n_1,...,n_R$.
\item For each $j$ set $\widetilde{\alpha}_{(k+1)\Delta t}^j=\alpha_{k\Delta t}^{n_j}+\sqrt{\Delta t}Z_j$, where the $Z_j$s are independent and standard Normally distributed vectors.
\item To each particle $\widetilde{\alpha}_{(k+1)\Delta t}^j$  a second-stage weight $\pi_j$ is assigned, given by 
\begin{align*}
\pi_j = \frac{w_j}{\sum_{i=1}^{R} w_i}, \quad \text{where }
w_j = \frac{\phi(y_{(k+1)\Delta t};h_{(k+1)\Delta t}(\widetilde{\alpha}_{(k+1)\Delta t}^j),\Sigma_2)}
{\phi(y_{(k+1)\Delta t};h_{(k+1)\Delta t}(\alpha_{k\Delta t}^{n_j}),\Sigma_1)}.
\end{align*}
\item Draw with replacement from the set $\{\widetilde{\alpha}_{(k+1)\Delta t}^1,\dots,\widetilde{\alpha}_{(k+1)\Delta t}^R\}$ according to the weights $\pi_j$ to produce the set 
of particles $\{\alpha_{(k+1)\Delta t}^1,\dots,\alpha_{(k+1)\Delta t}^R\}$.
\end{enumerate}
In order to use the particle filter the filter parameters $R$, $\Sigma_1$, and $\Sigma_2$ must be specified. Notice that for each $k$ the sample $\{\alpha_{k\Delta t}^1,\dots,\alpha_{k\Delta t}^R\}$ forms an empirical distribution that approximates the conditional distribution of $x_{k\Delta t}$ given $y_0,y_{\Delta t},\dots,y_{k\Delta t}$.

\begin{figure}[!ht]
	\centering
	\includegraphics[scale=.35]{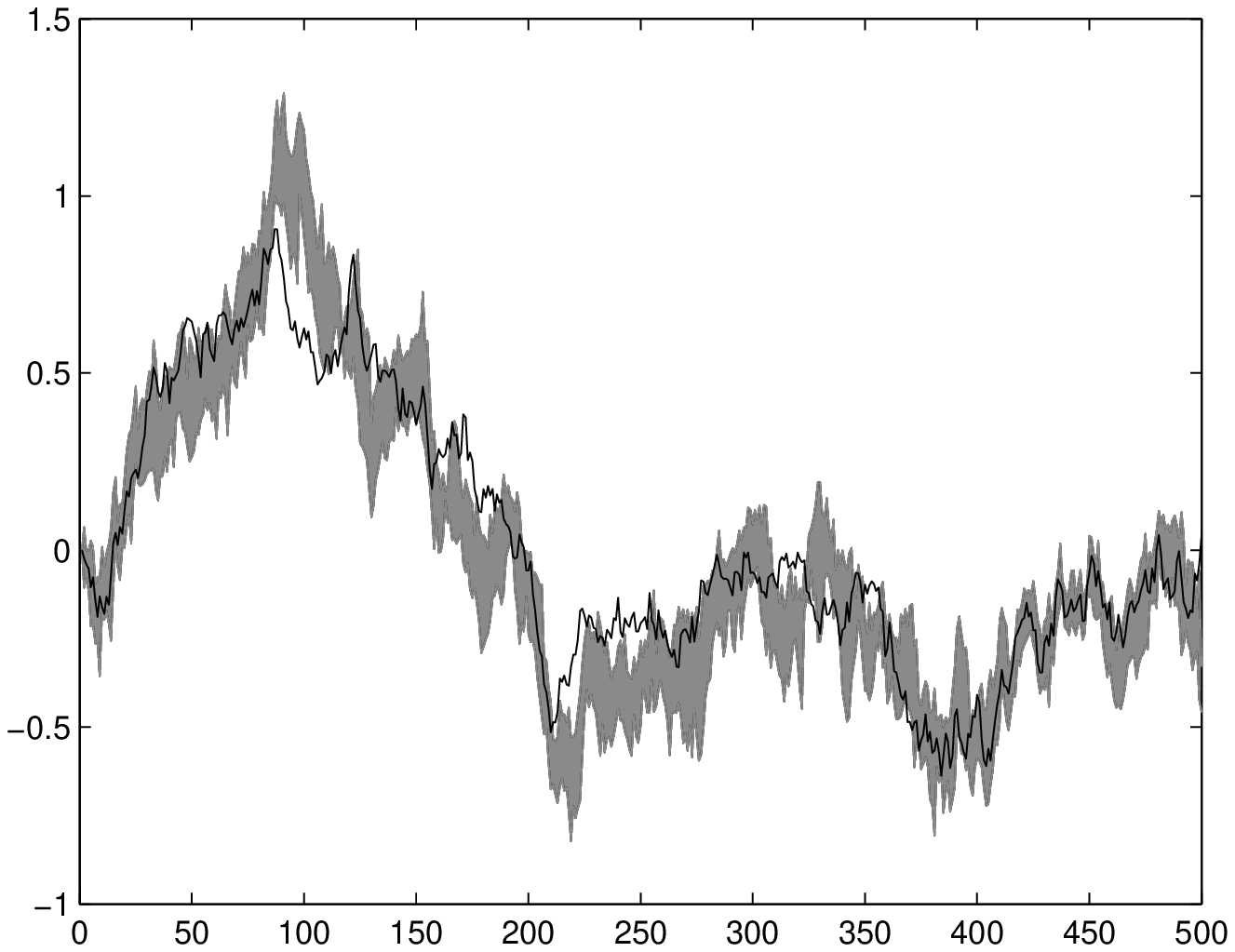}
	\includegraphics[scale=.35]{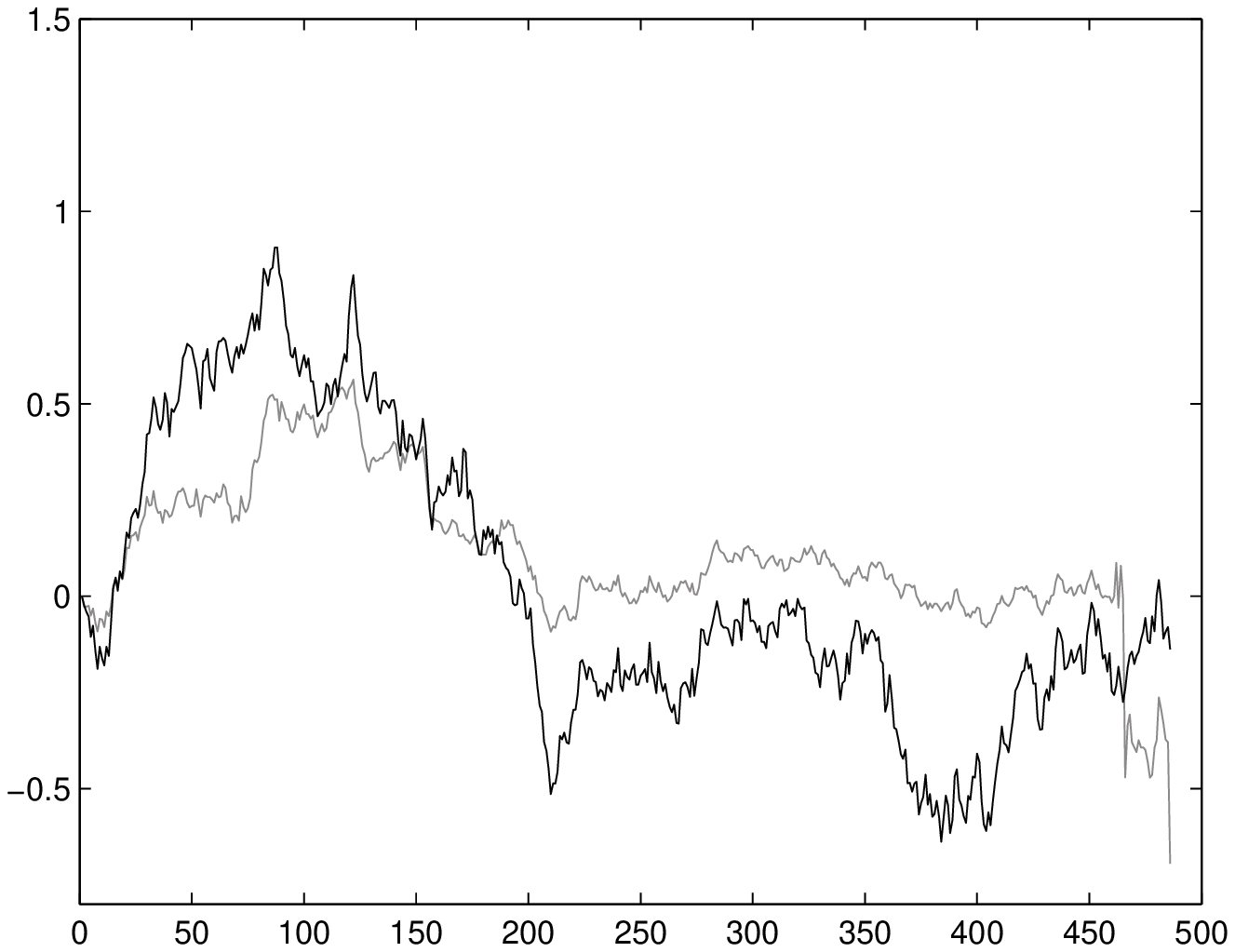}
	\includegraphics[scale=.35]{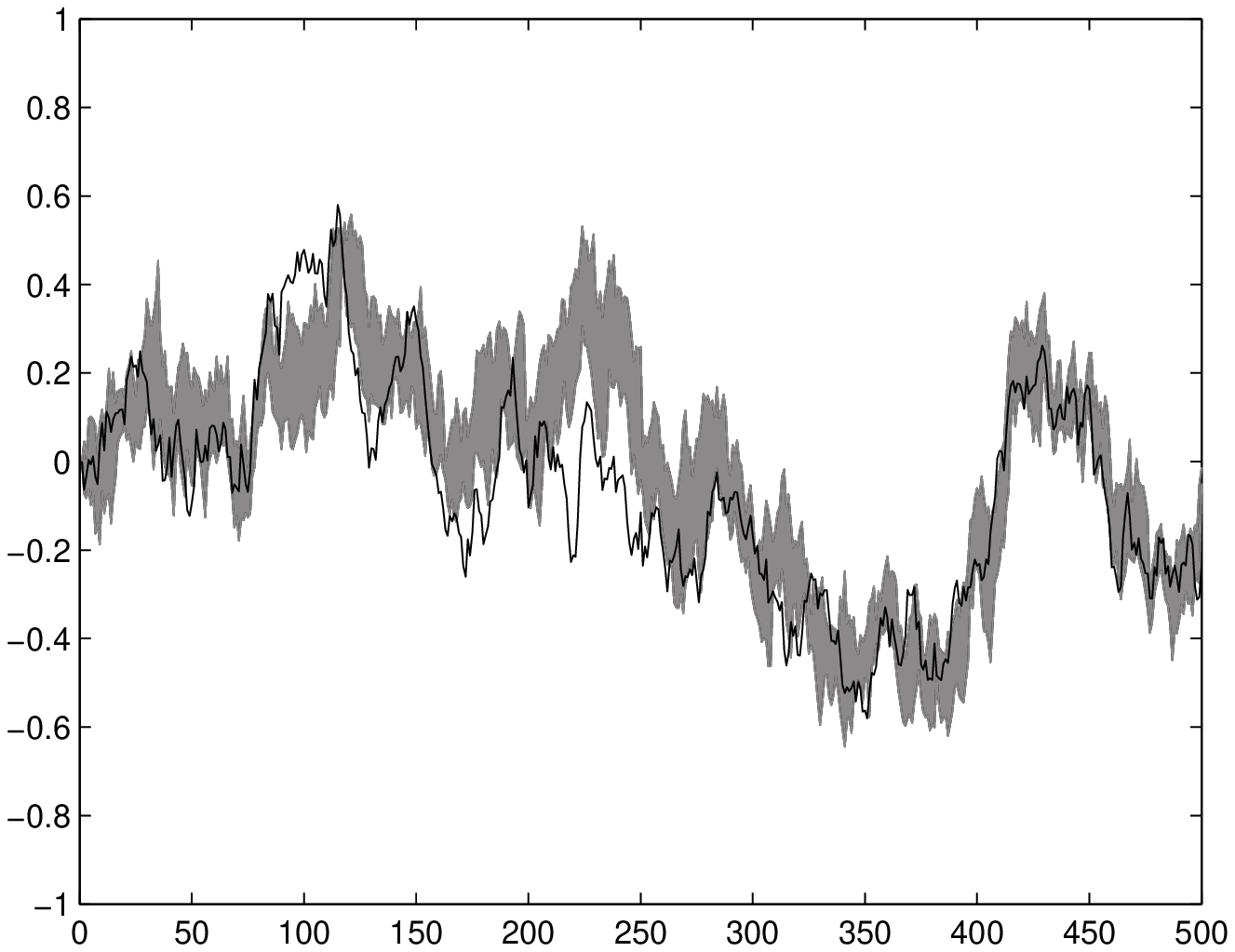}
	\includegraphics[scale=.35]{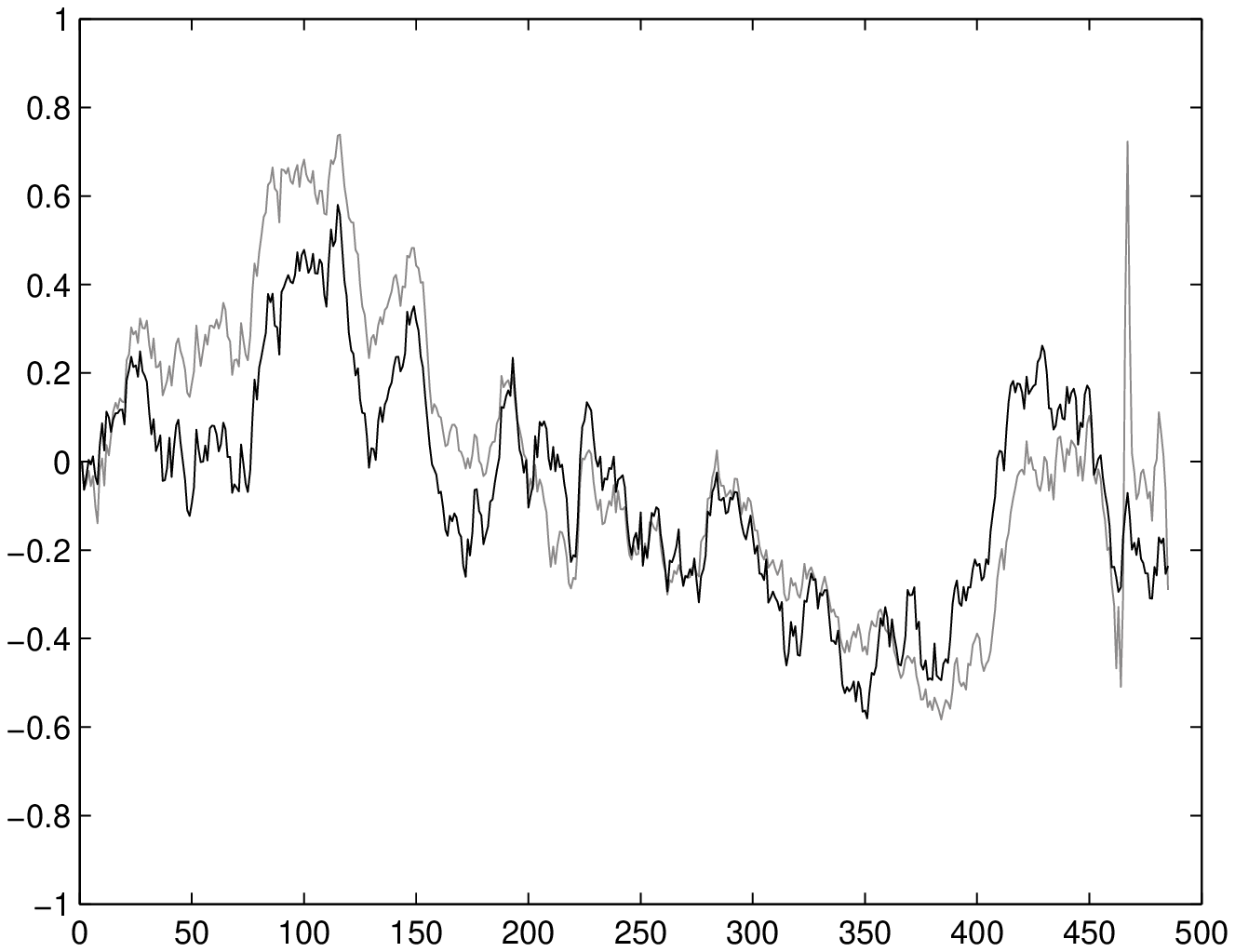}
	\includegraphics[scale=.35]{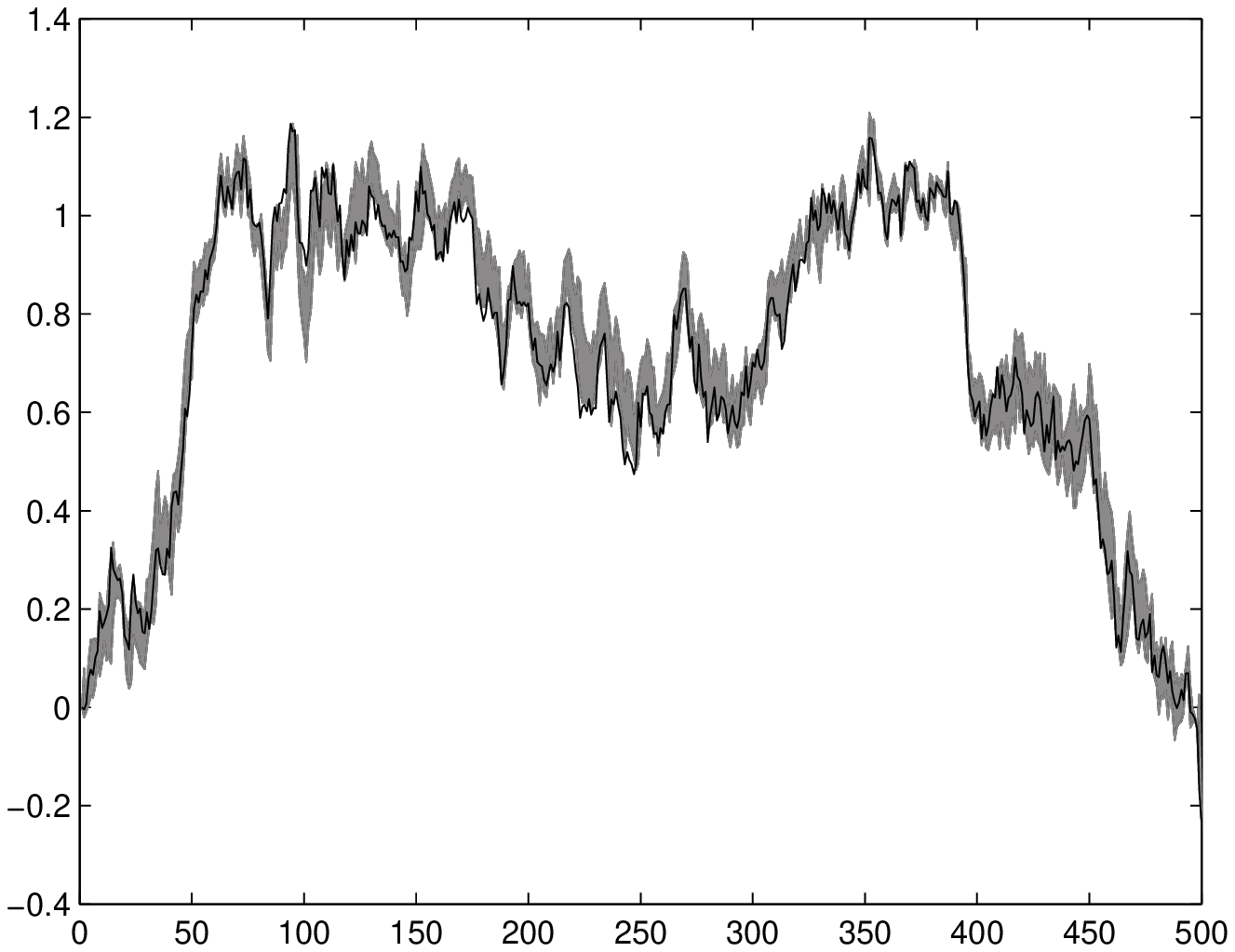}	
	\includegraphics[scale=.35]{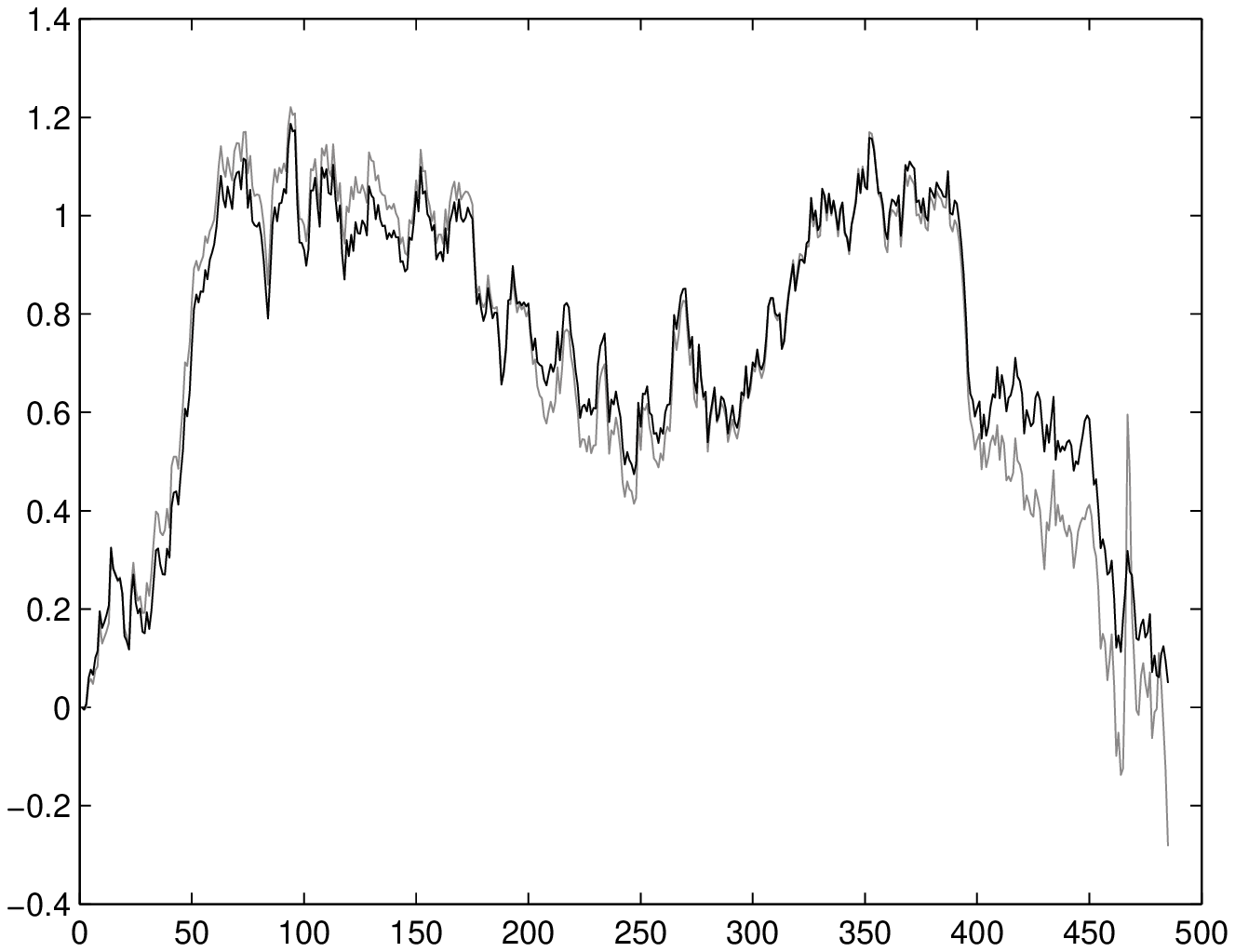}
	\caption{The left column shows the resulting filter distributions for the $x$-, $y$-, and
	$b$-coordinate respectively. The right column shows the resulting estimates from
	the linearization.}
\label{fig:FilterExample2}
\end{figure}

\begin{figure}[!ht]
	\centering
	\includegraphics[scale=.35]{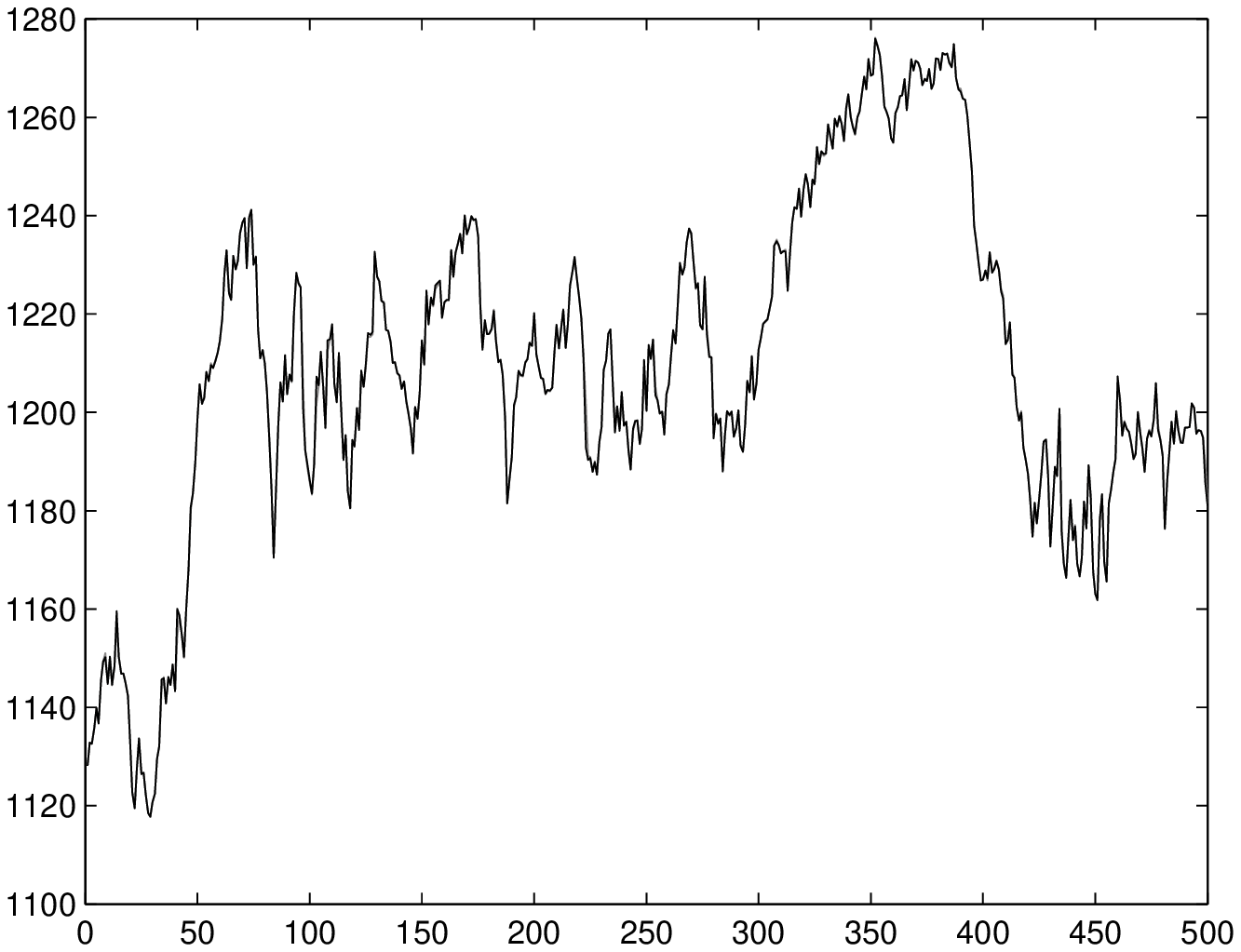}
	\includegraphics[scale=.35]{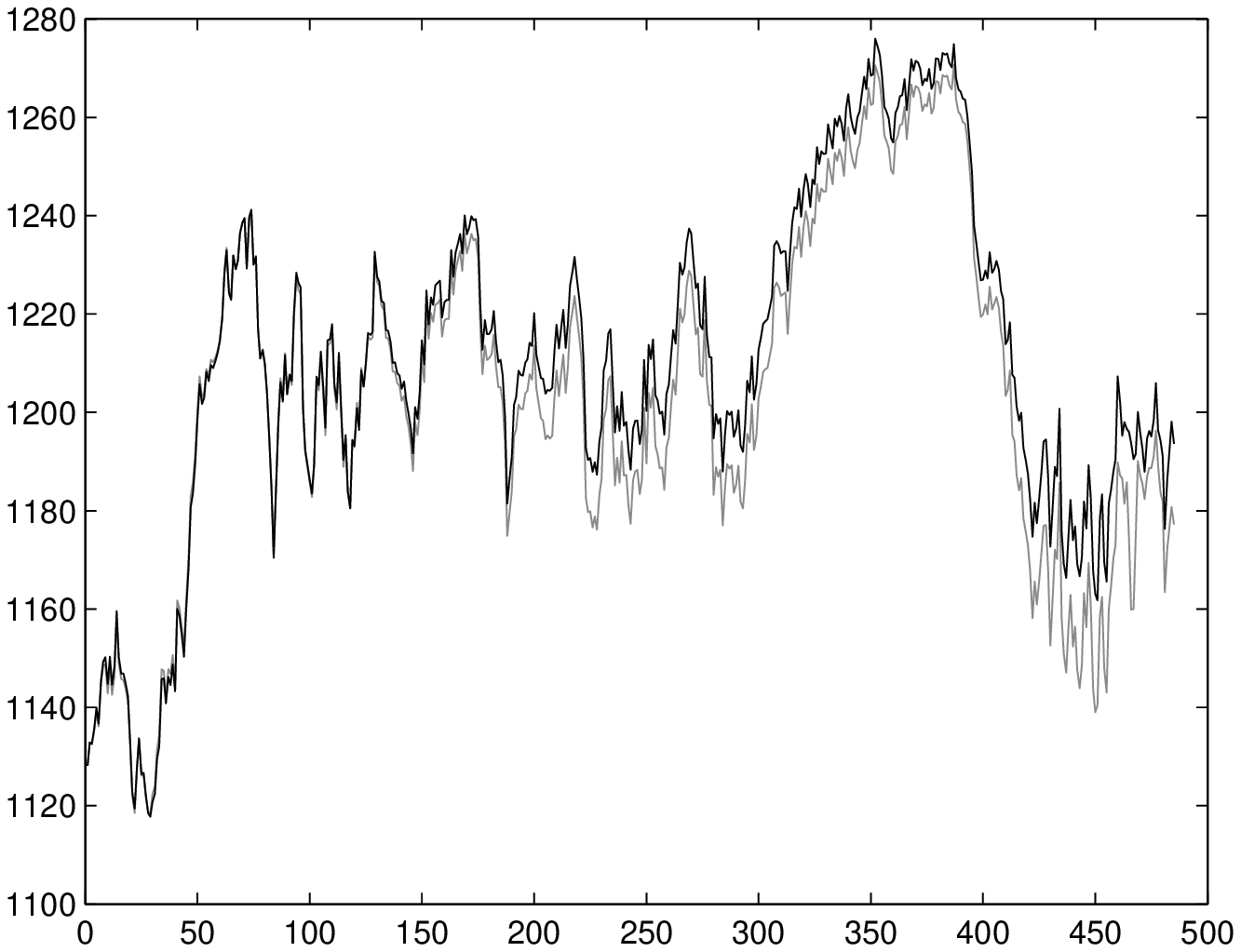}
	\includegraphics[scale=.35]{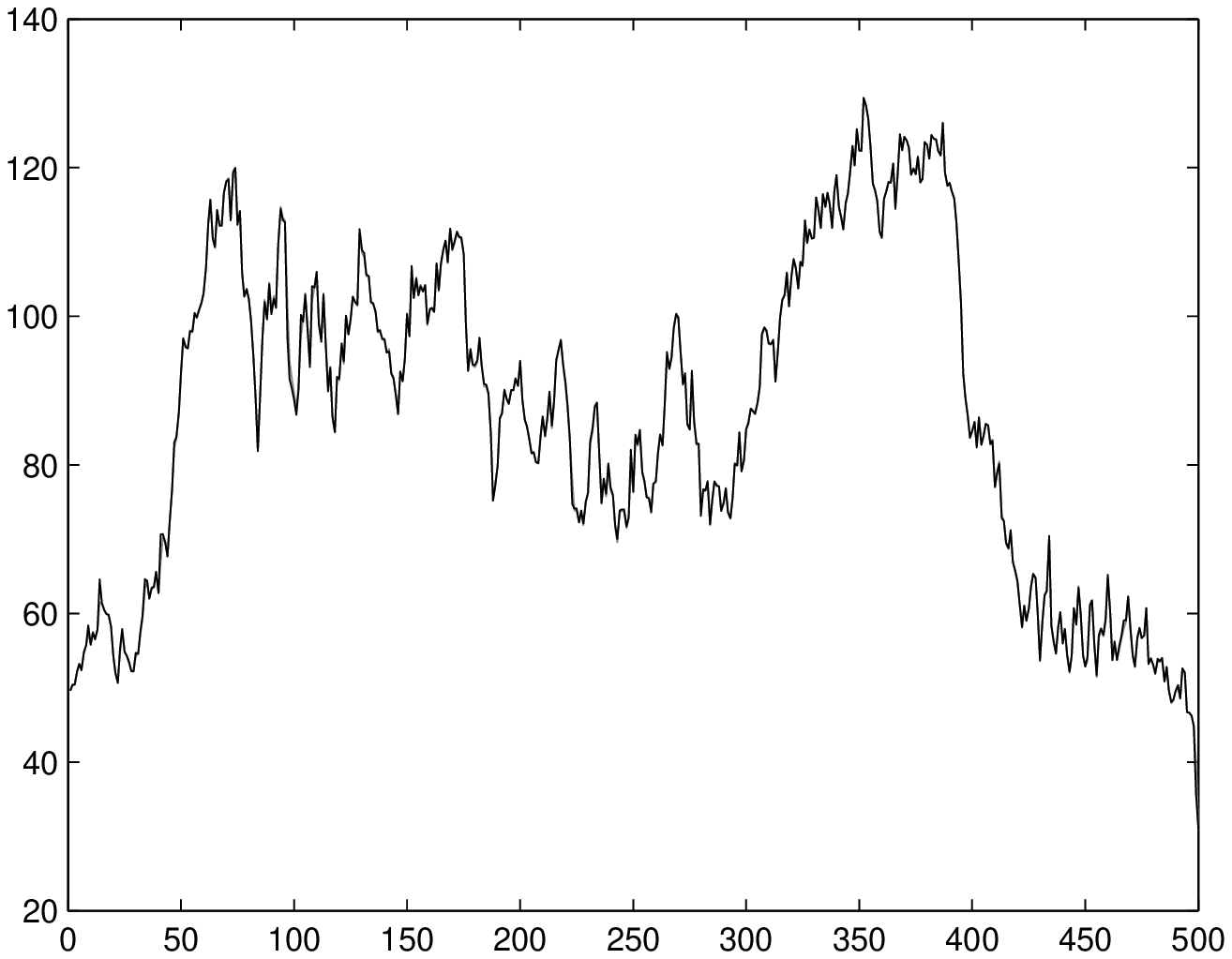}
	\includegraphics[scale=.35]{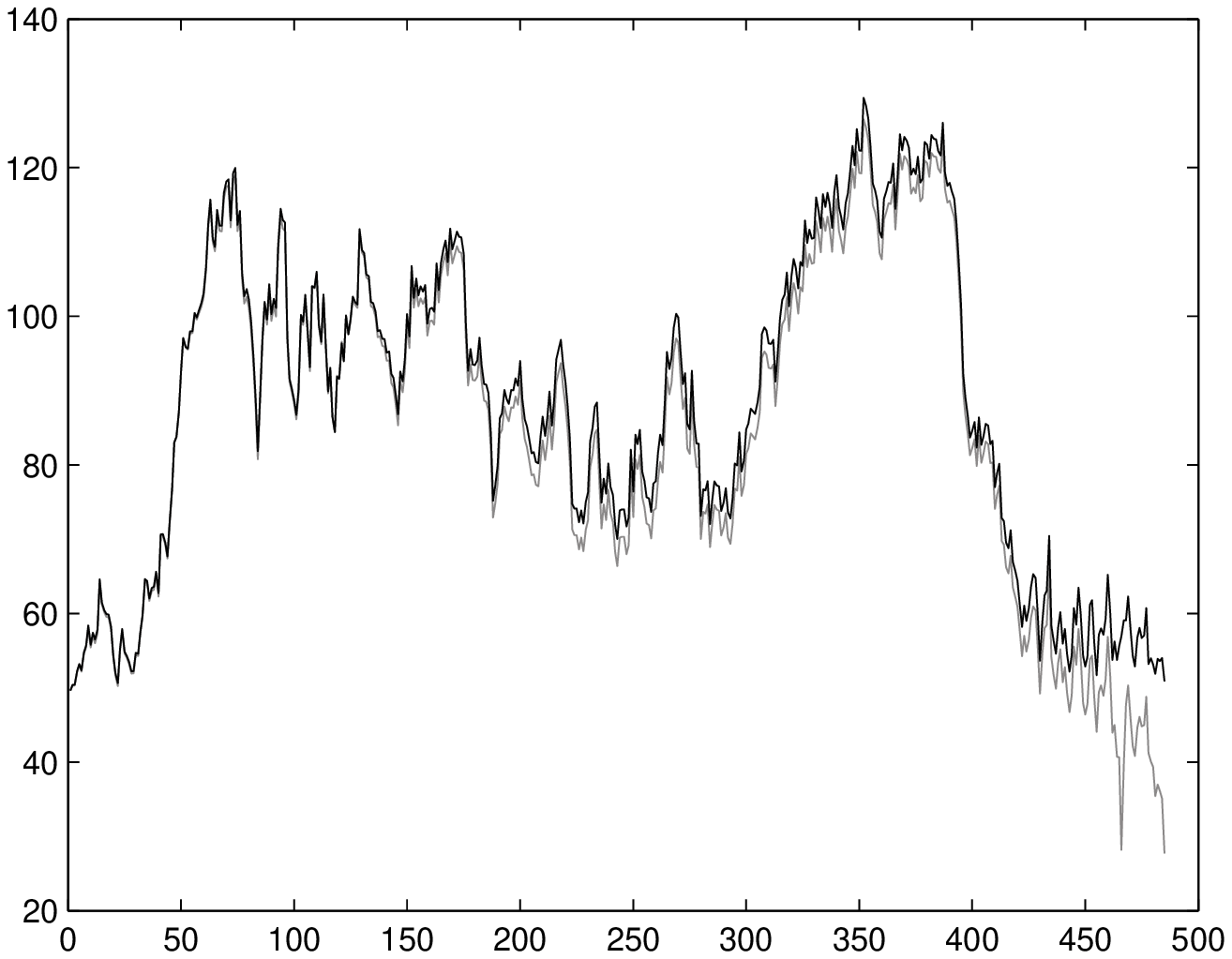}
	\includegraphics[scale=.35]{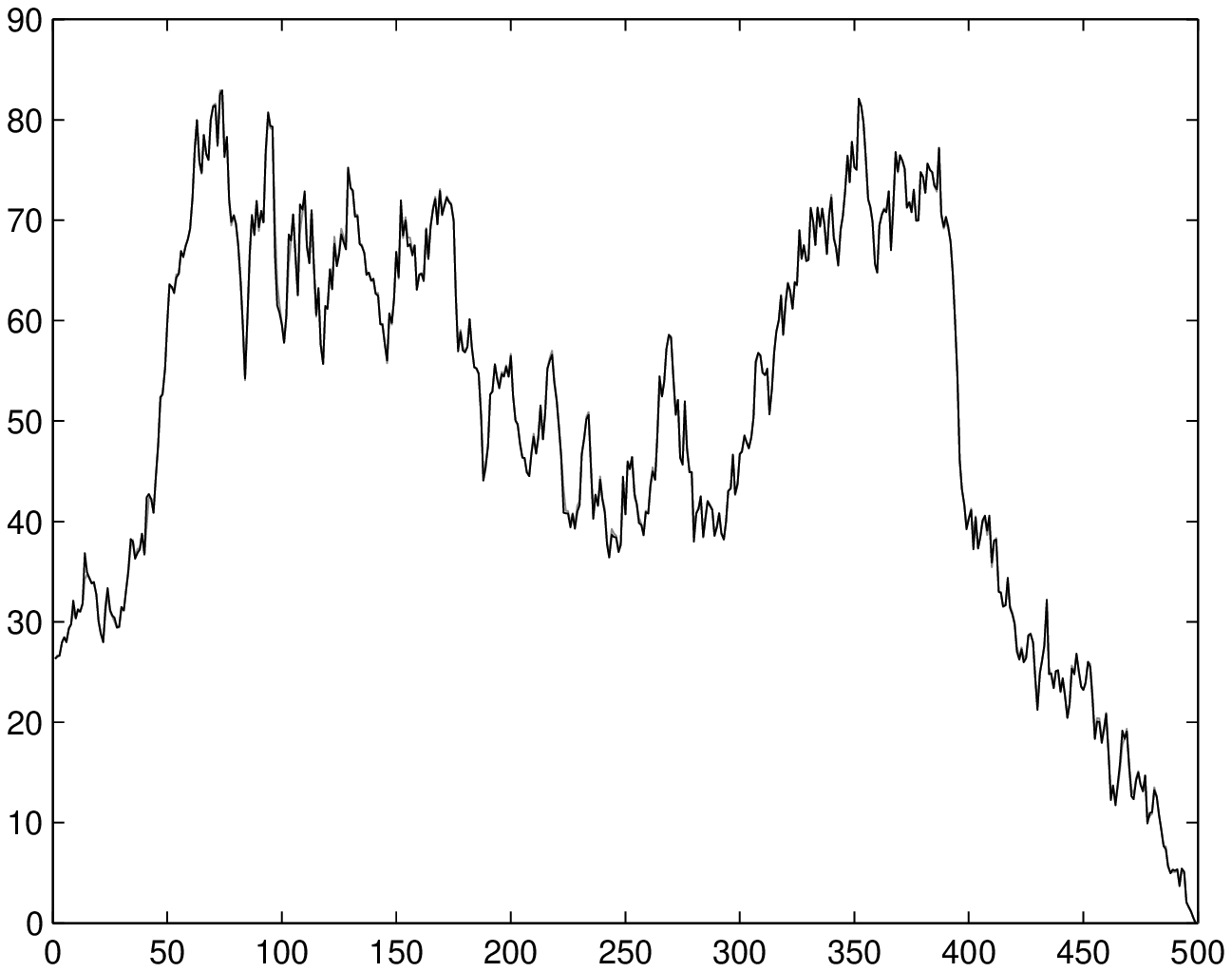}
	\includegraphics[scale=.35]{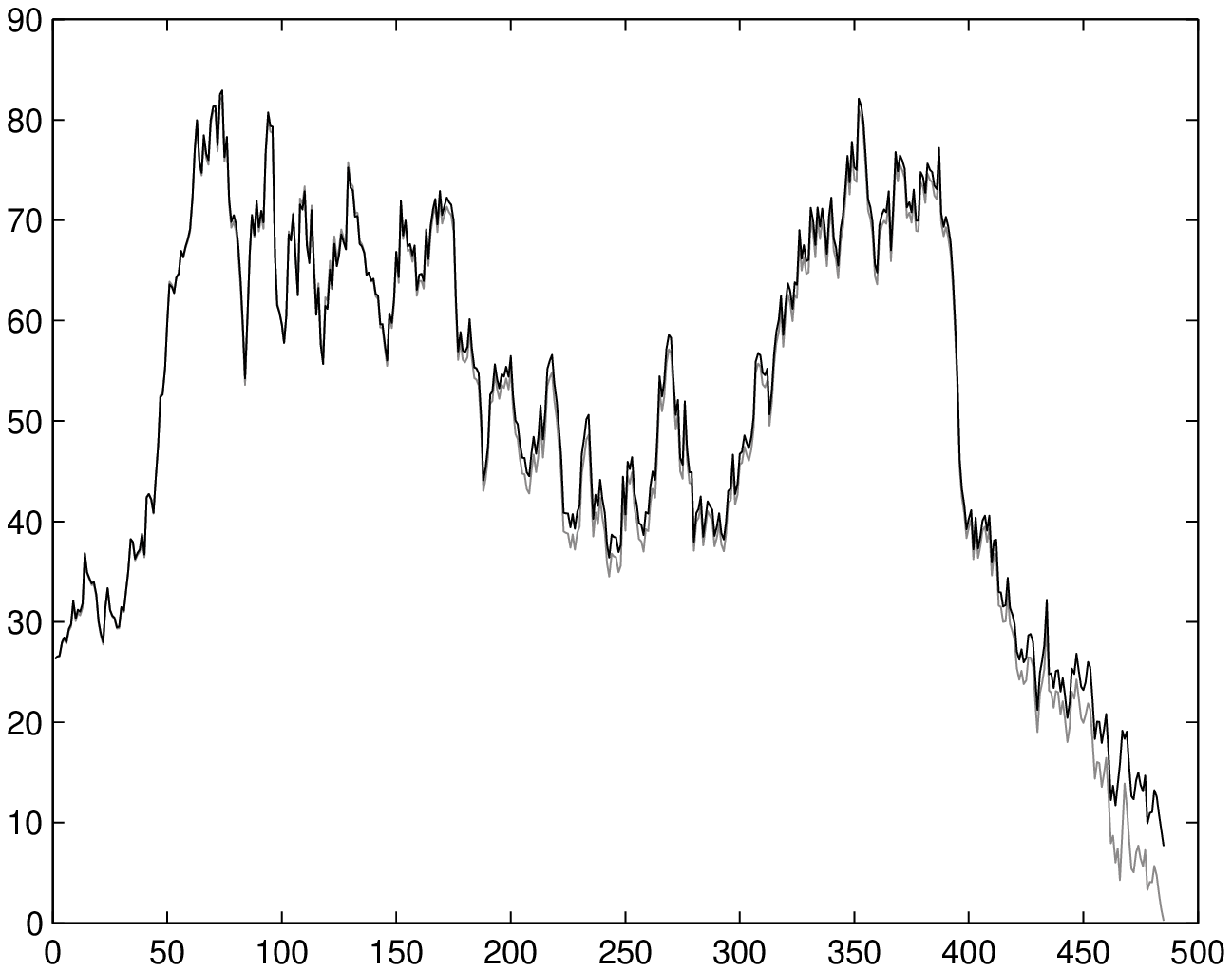}	
 	\caption{The left column shows the true price trajectories (black) and the resulting
	filter prices (gray) for $G$, $G_1$ and $G_2$ respectively. The right column shows
	the corresponding price trajectories from the linearization.}
\label{fig:FilterPricesExample2}
\end{figure}
\subsubsection{A simulation study}

In this section the performance of the particle filter and the local linear approximation will be illustrated in a small simulation study. 

Consider the model for $n = 2$ with parameters given by \eqref{eq:paramsneq2}.
Take $T=1$ and $\Delta t=1/500$ and simulate $\{(G_{k\Delta t}, G_{k\Delta t}^1, G_{k\Delta t}^2):k=1,\dots,500\}$ by feeding the model with a Gaussian random walk whose increment distribution is the 3-dimensional Normal distribution $\Normal_3(0,\Delta t I)$, where $I$ denotes the identity matrix. The problem we consider here is to estimate the location of the Gaussian random walk from the simulated price data $\{(G_{k\Delta t}, G_{k\Delta t}^1, G_{k\Delta t}^2):k=1,\dots,500\}$.

The particle filter parameters are selected as follows. The number of particles is $R=250$. The matrix $\Sigma_1$ is chosen as the sample covariance matrix of the simulated increments for the $3$-dimensional forward price process. The matrix $\Sigma_2$ needed to assign the particles' second-stage weights is chosen as $\Sigma_2=\Sigma_1$.


The output of the particle filter is a distribution of the location of the three-dimensional Gaussian random walk. The three components of the true simulated random walk 
and  the corresponding estimates from the particle filter are displayed in the left plots in 
 Figure \ref{fig:FilterExample2}. The empirical distributions of the particles estimating the location of the random walk are displayed in grey on top of the true simulated trajectories. The estimates from the local linear approximation of the random walk trajectories are 
displayed in the right plots in Figure \ref{fig:FilterExample2}. The particle filter approach is reasonably good at tracking the underlying Gaussian walk, whereas the performance of the local linear approximation is clearly worse. 

In addition, the forward prices of the index and the two options are recalculated using the corresponding  particle filter estimates and linearization estimates, respectively. For the particle filter, at any given time each particle (an estimate of the location of the Gaussian random walk) gives rise to a forward price and the weighted sum of the prices corresponding to different particles is the value of the gray price trajectory in the left plots in Figure \ref{fig:FilterPricesExample2}. The true price trajectory is plotted in black (the one corresponding to the simulated Gaussian random walk). The two price trajectories are essentially indistinguishable. The plots to the right in Figure \ref{fig:FilterPricesExample2} show the price trajectories computed from the linearization estimates of the Gaussian random walk (in gray) and the true price trajectories (in black). The linearization estimates of the Gaussian random walk do not reproduce the simulated price trajectories as accurately as the particle filter estimates.

\subsubsection{Tracking the Brownian particle for S\&P 500 option data}

\begin{figure}
	\centering
	\includegraphics[scale=.35]{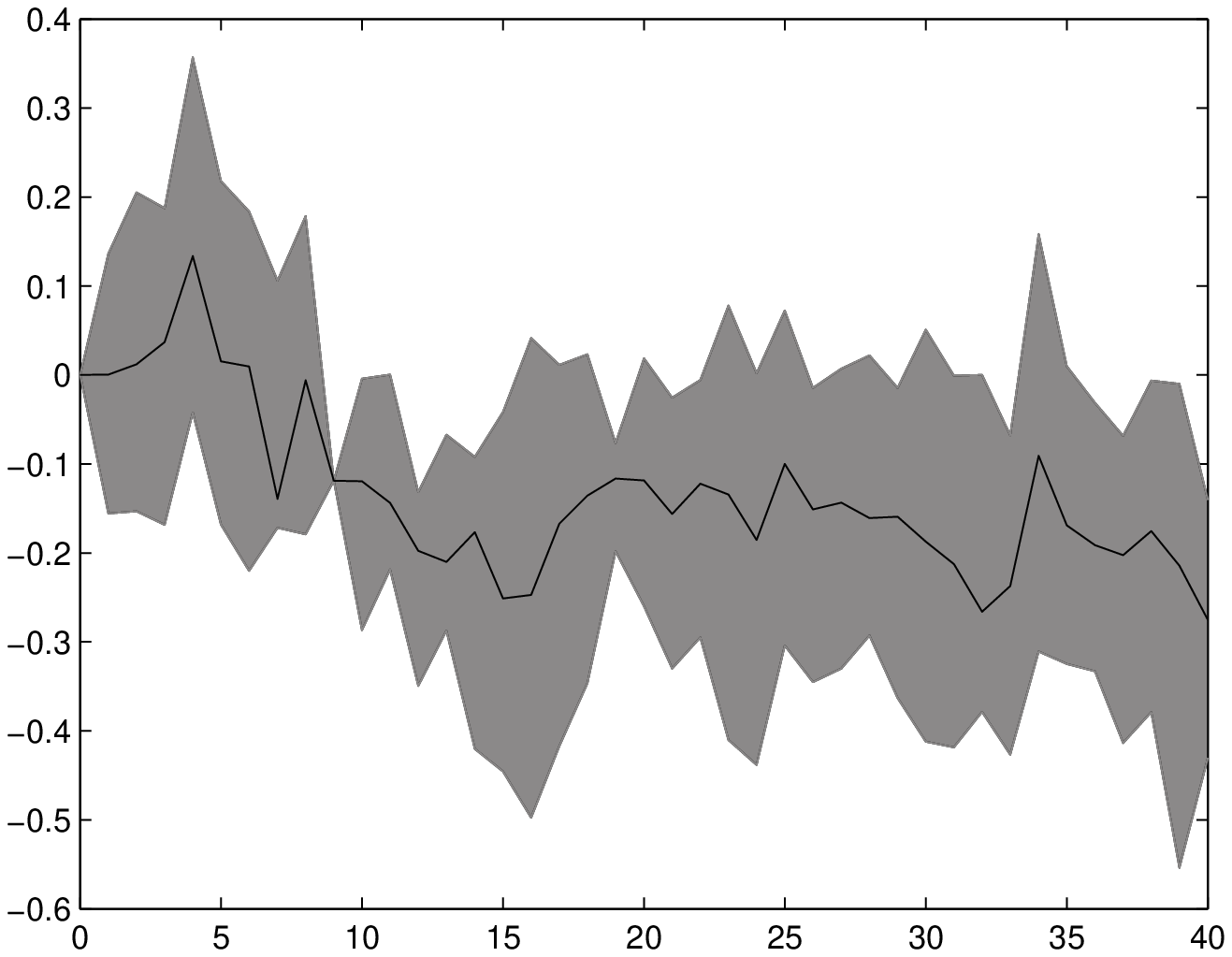}
	\includegraphics[scale=.35]{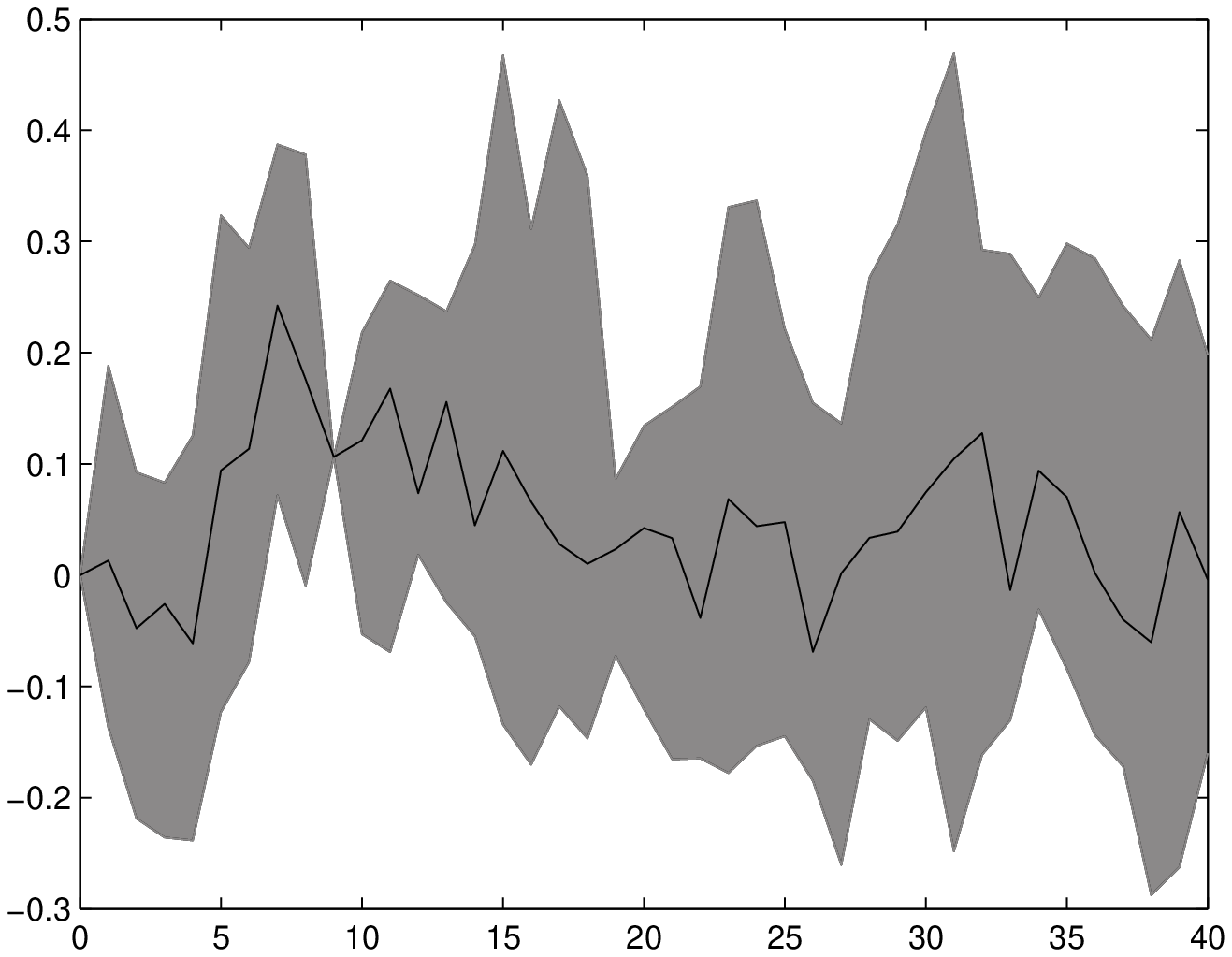}
	\includegraphics[scale=.35]{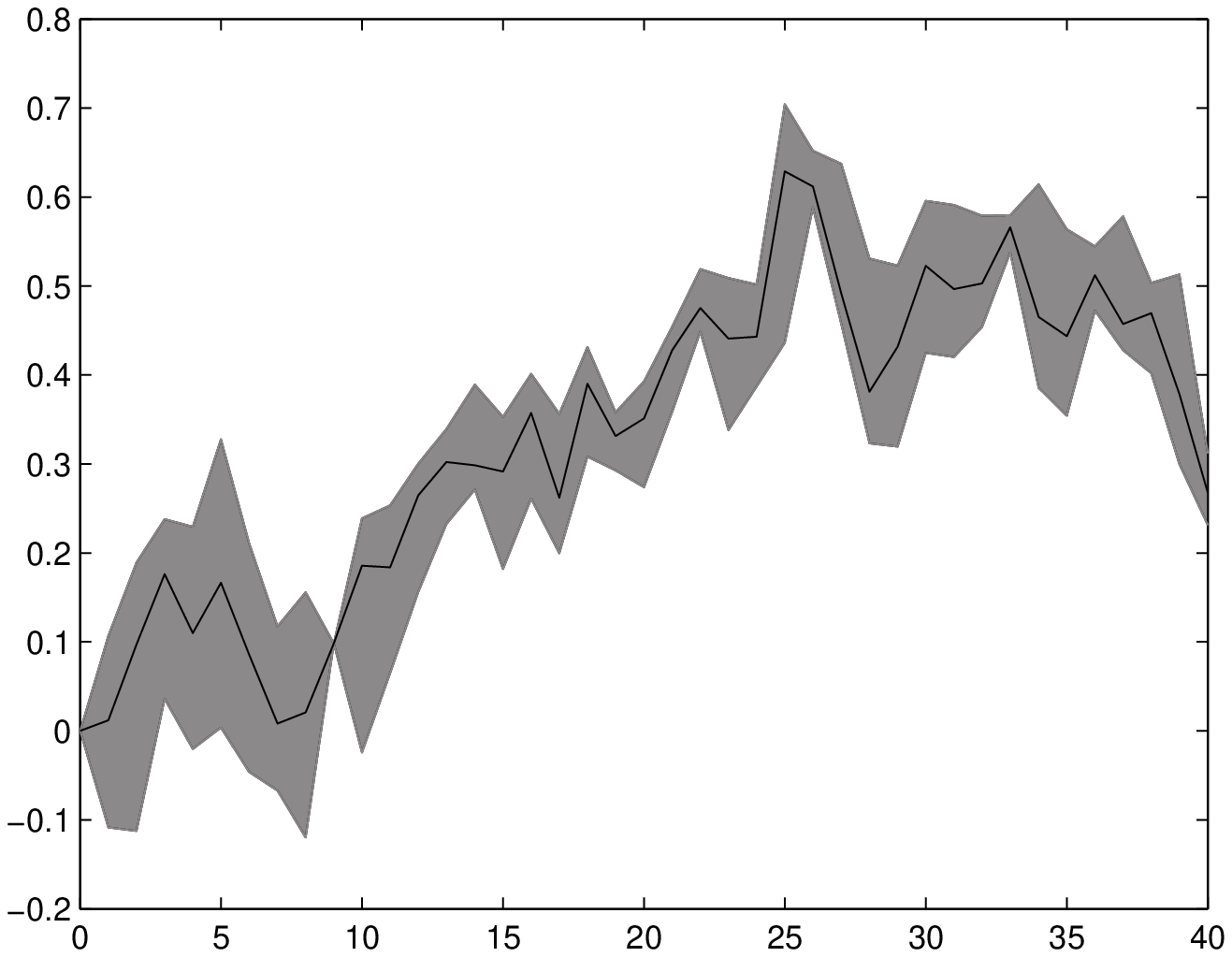}	
	\includegraphics[scale=.35]{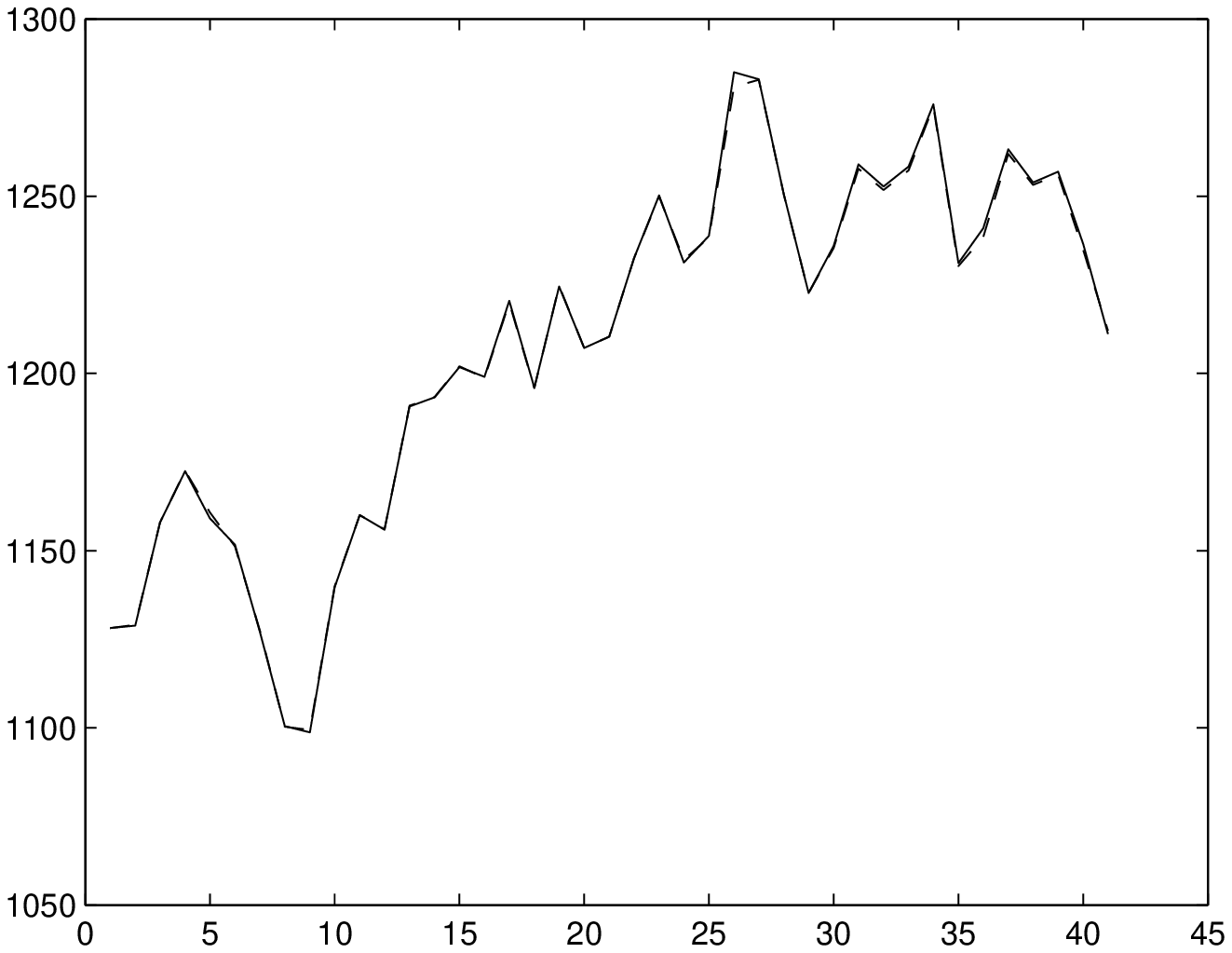}
	\includegraphics[scale=.35]{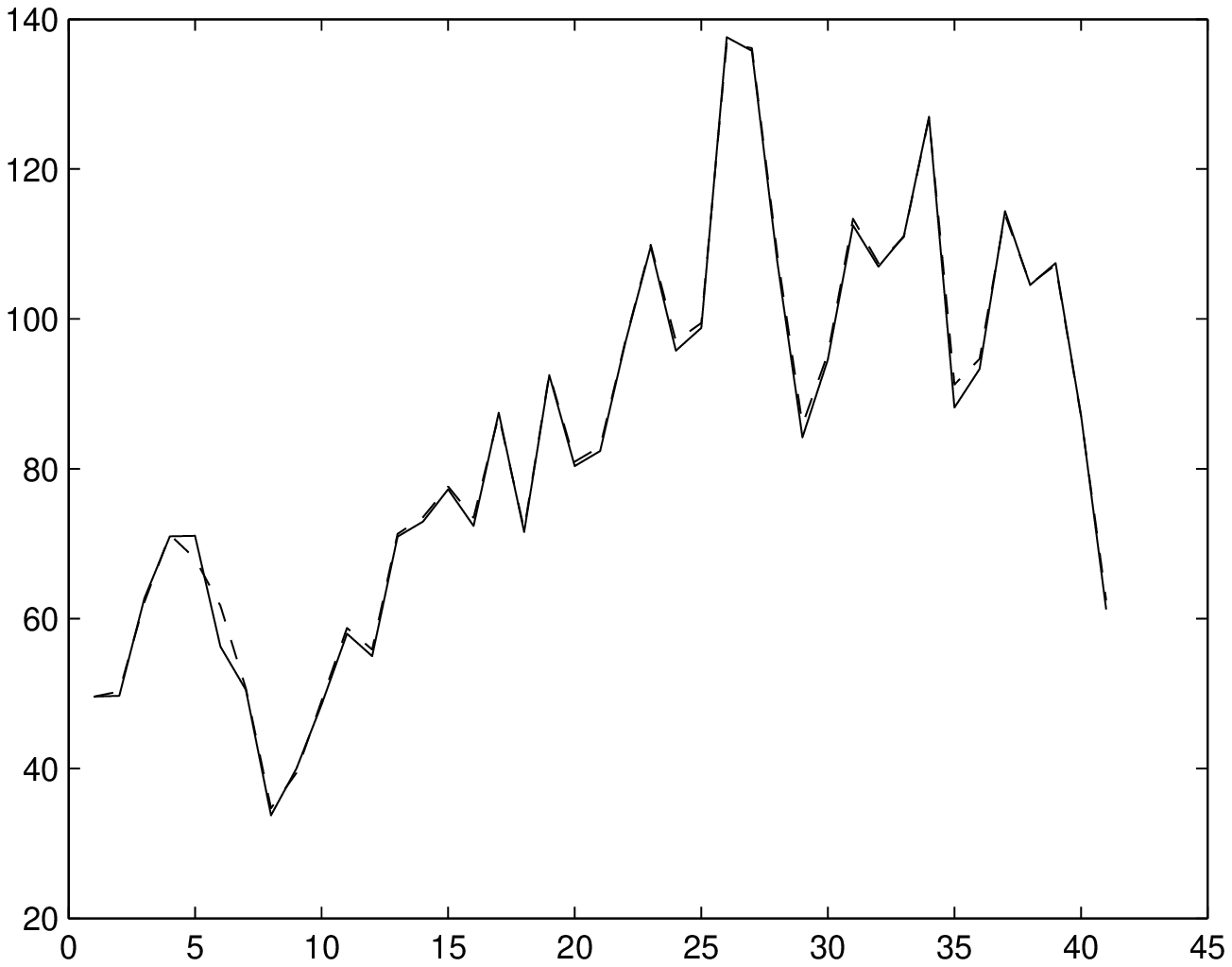}
	\includegraphics[scale=.35]{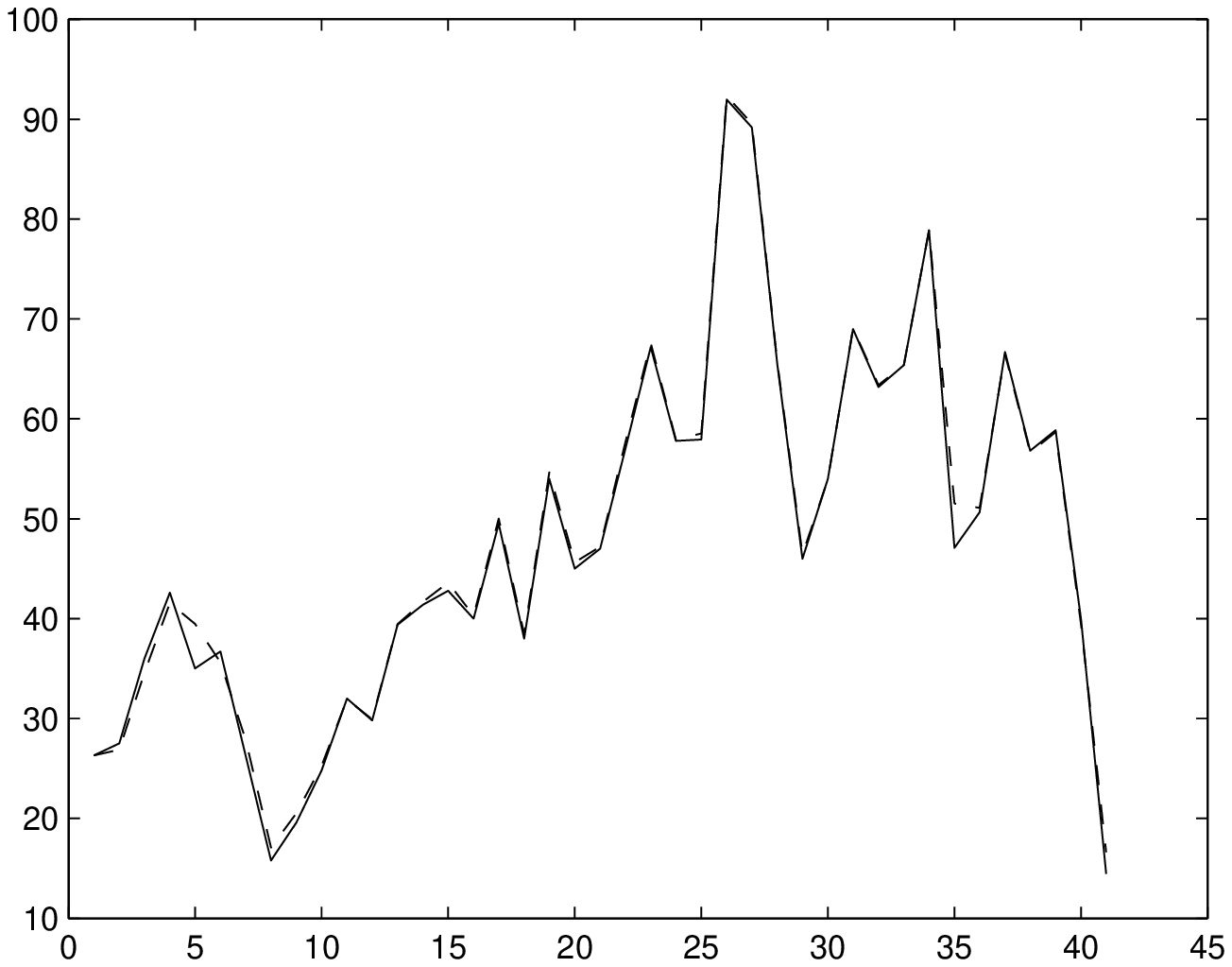}
 	\caption{The upper plots and the left middle plot show the empirical distributions (in gray) of the particles of the particle filter applied to real price data, and the mean of the empirical distributions (in black) for the $x$-, $y$- and $b$-coordinate, respectively. The right middle plot and the lower plots show the real price trajectories (in black) for $G^{0}$, $G^{1}$ and $G^{2}$ and the price trajectories computed from the filter estimates (in gray).}
\label{fig:realfilter}
\end{figure}

Now that the initial calibration of the model and the particle filter is well understood, 
the particle filter is applied to the S\&P 500 option price data; 41 vectors of daily closing prices for the index forward and two call options. The results with $n=2$ are shown in Figure \ref{fig:realfilter}. The particle filter distribution of the underlying Gaussian random walk is rather wide but nevertheless are the computed prices based on the filter estimates very close to the real prices. The plots in Figure \ref{fig:realfilter} demonstrate that the model is very good at reproducing the true price trajectories. 

\begin{figure}
	\centering
	\includegraphics[scale=.35]{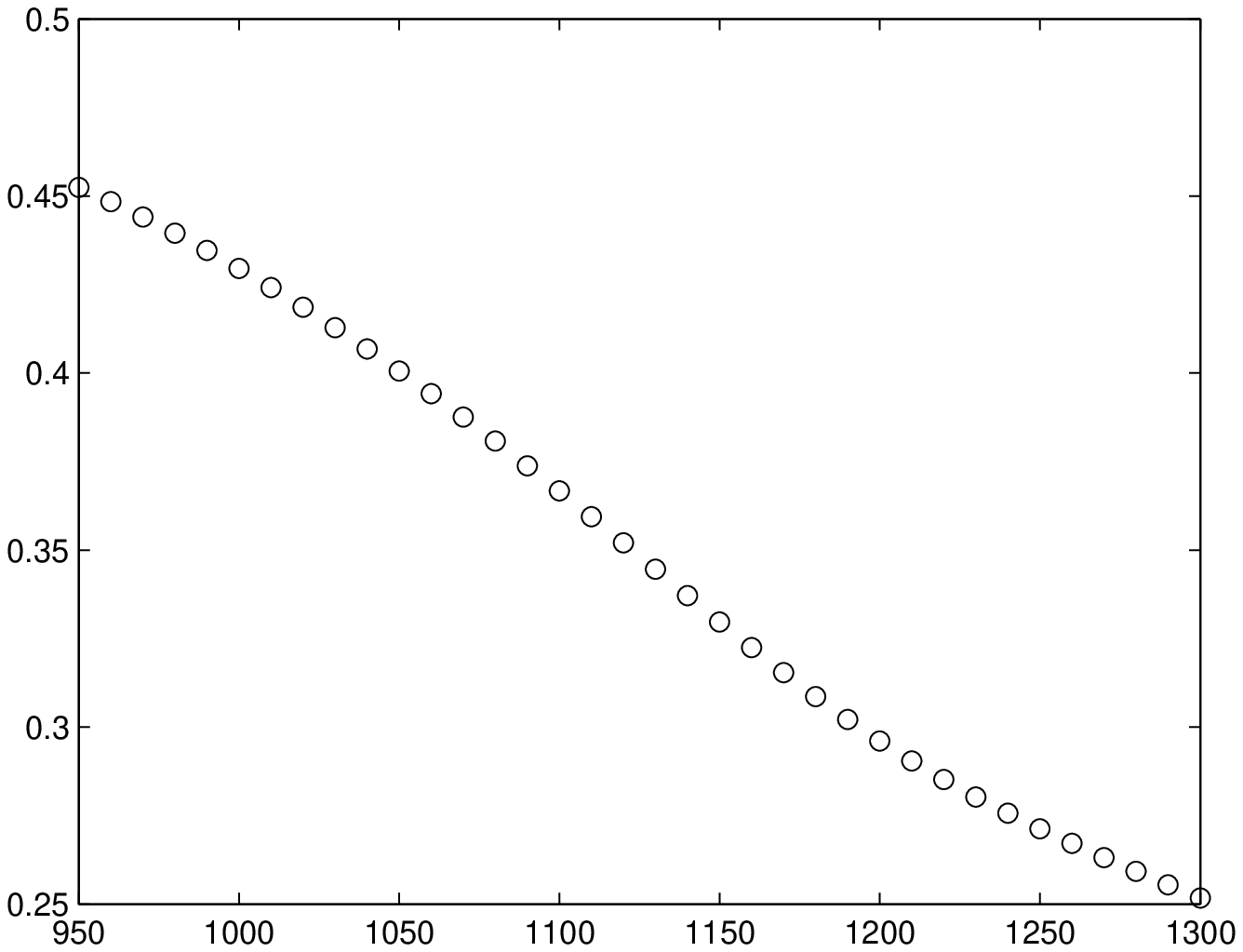}
	\includegraphics[scale=.35]{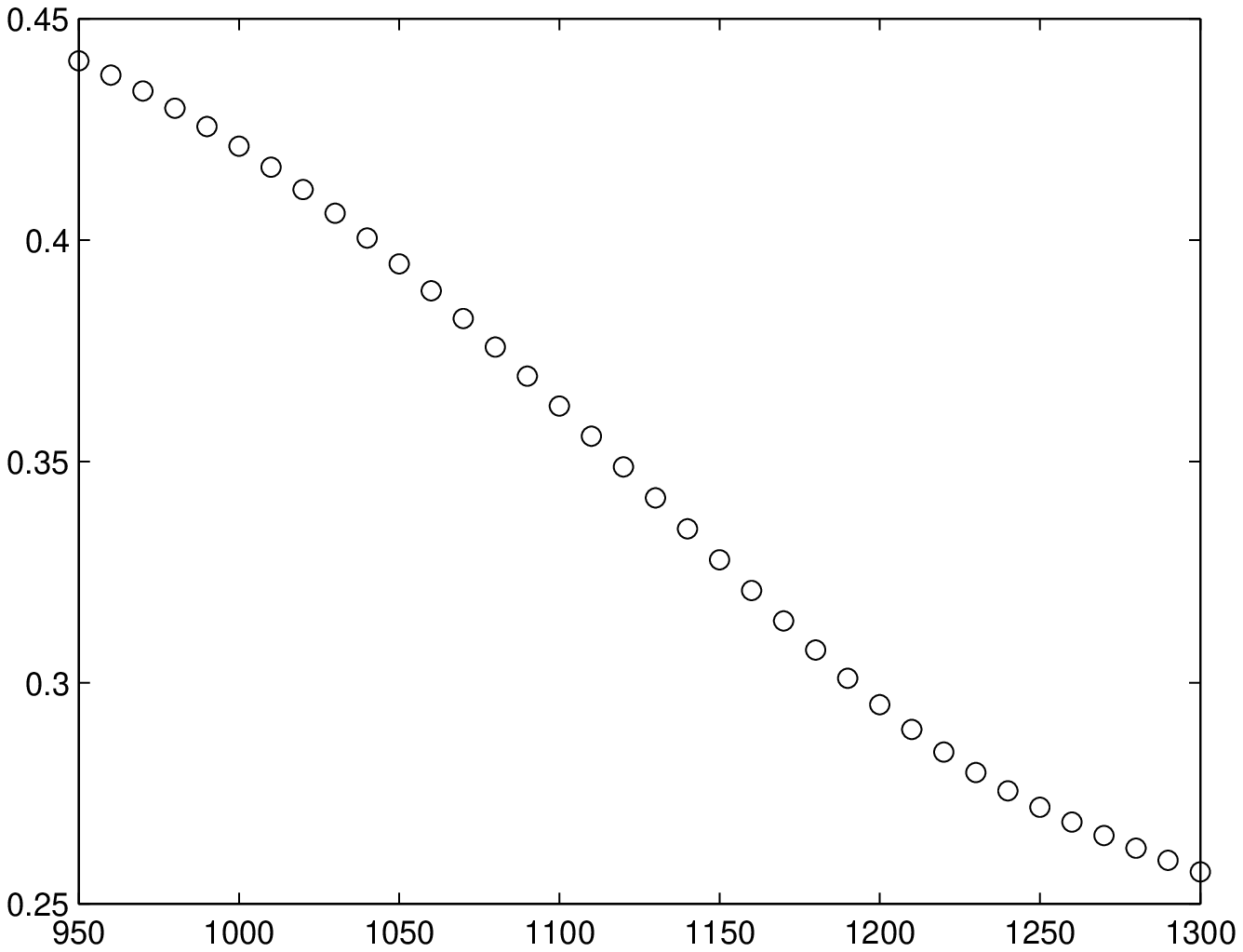}
	\includegraphics[scale=.35]{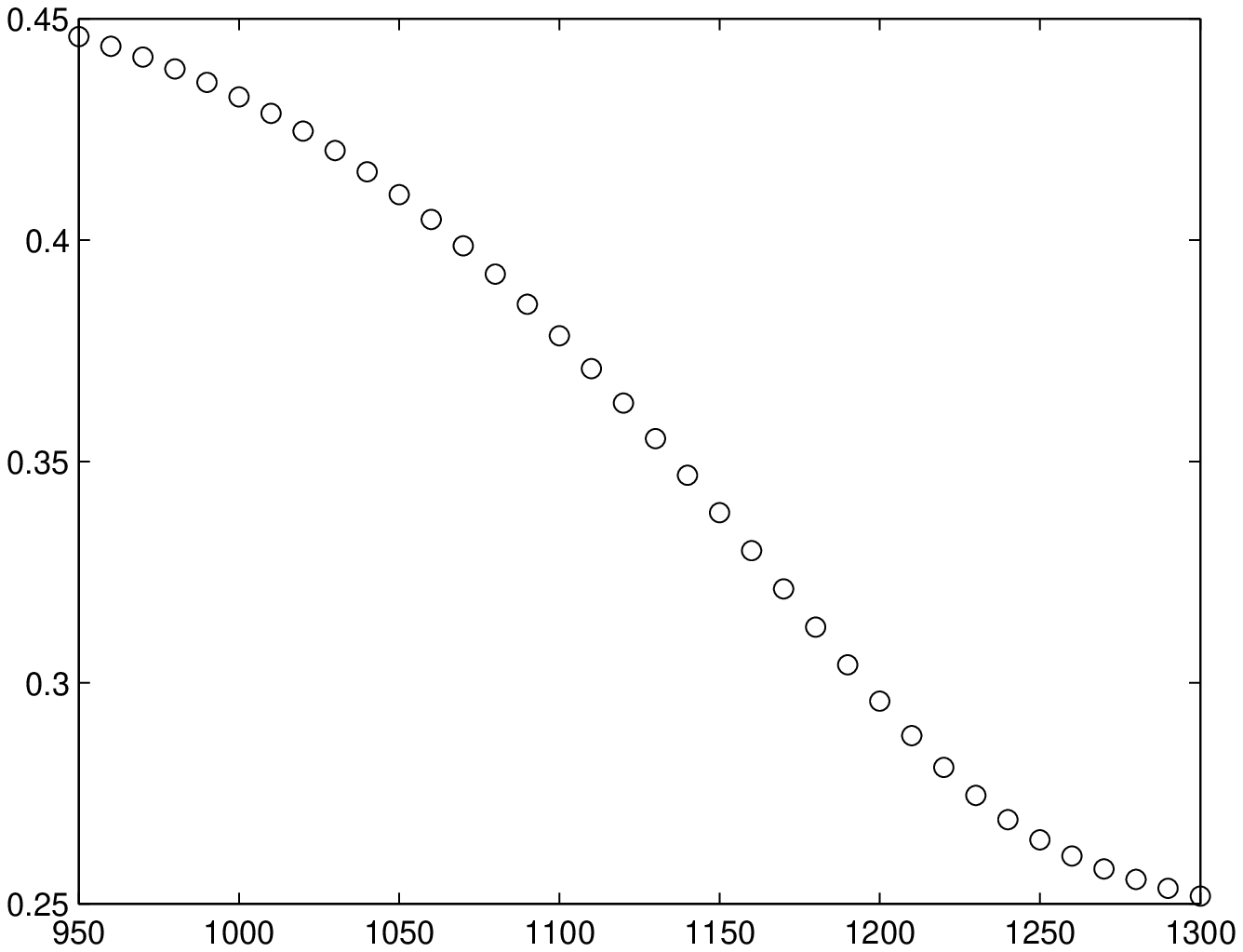}
	\includegraphics[scale=.35]{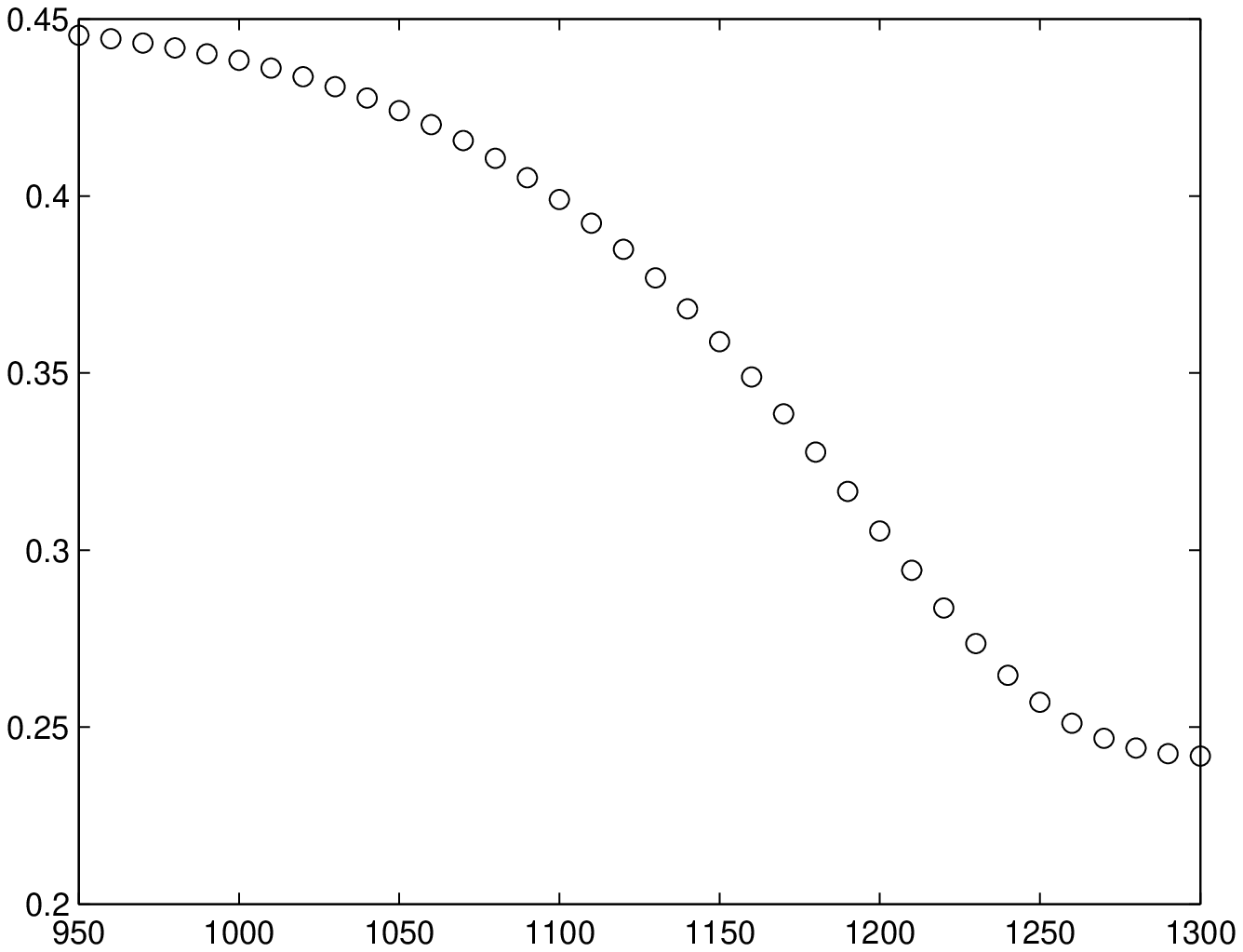}
 	\caption{The plots display volatility smiles produced by the particle filter and the model fitted to the S\&P 500 option data, at times $t=0$ days (upper left), $t = 10$ days (upper right), $t = 20$ days (lower left), and $t=30$ days (lower right).}
\label{fig:VolSmilesSP500}
\end{figure}

For each of the times $t \in \{0,10,20,30\}$ days from today, the set of particles from the particle filter is used to compute model prices for a fine grid of strikes. For each of these strikes a call option price is computed as a weighted average (second stage weights) of the model prices corresponding to different particles. Then, the produced prices are transformed into implied volatilities using Black's formula. For each of the four times, the procedure thus produces a volatility smile (a set of implied volatilities), and we observe how the volatility smile varies over time.
Figure \ref{fig:VolSmilesSP500} shows that the model and particle filter produce volatility smiles at all times that appear to be reasonable.

\appendix

\section{Proof of Proposition \ref{prop:simplemodel}}

Consider the equation $Ap=b$ in \eqref{eq:lineqsyst} with the choice of $x_k$s according to the statement of the proposition. Clearly, it has a unique solution. We need to determine this 
solution $p$ and verify that $p\in [0,1]^{n+2}$.
Using backward substitution, we solve for the three last 
probabilities $p_{n+2},p_{n+1},p_n$ to obtain
\begin{align*}
  p_{n+2} &= \frac{G_0^n}{x_{n+2}-K_n},\\ 
  p_{n+1} &= \frac{G_0^{n-1}}{K_n-K_{n-1}}
  -\frac{G_0^n(x_{n+2}-K_{n-1})}{(K_n-K_{n-1})(x_{n+2}-K_n)},\\
  p_n &= \frac{G_0^{n-2}}{K_{n-1}-K_{n-2}} 
  -\frac{G_0^{n-1}(K_n-K_{n-2})}{(K_{n-1}-K_{n-2})(K_n-K_{n-1})} 
  +\frac{G_0^n}{K_n-K_{n-1}}.
\end{align*}
We begin by showing, by a standard induction argument, that $p_k$ can be 
written as in \eqref{pk} for $k=3,\dots,n$. 
We know that this holds for $k=n$. We now assume that it holds for
$k=n-j,n-j+1,...,n$ and show that it holds for $k=n-j-1$. 
We know that
\begin{align*}
  G_0^{n-j-1}=\sum_{k=n-j+1}^{n+1} p_k(K_{k-1}-K_{n-j-1}) 
  + p_{n+2}(x_{n+2}-K_{n-j-1})
\end{align*}
which means that
\begin{align*}
  p_{n-j-1}=\frac{G_0^{n-j-3}-(K_{n-j-1}-K_{n-j-3})p_{n-j}
    -\dots-p_{n+2}(x_{n+2}-K_{n-j-3})}{K_{n-j-2}-K_{n-j-3}}.
\end{align*}
Inserting the expressions for $p_{n+1},p_{n+2}$ and the expression for 
$p_k$ for $k=n-j,n-j+1,...,n$, and collecting the terms we obtain
\begin{align*}
  p_{n-j-1}&=\frac{G_0^{n-j-3}}{K_{n-j-2}-K_{n-j-3}}
  -\frac{G_0^{n-j-2}(K_{n-j-1}-K_{n-j-3})}
  {(K_{n-j-2}-K_{n-j-3})(K_{n-j-1}-K_{n-j-2})}\\
  &\quad +\frac{G_0^{n-j-1}}{K_{n-j-1}-K_{n-j-2}}\\
  &\quad -\frac{\sum_{k=1}^j G_0^{n-k}\gamma(n-k)}{K_{n-j-2}-K_{n-j-3}}
  -\frac{G_0^n\gamma_n}{K_{n-j-2}-K_{n-j-3}},
\end{align*}
where
\begin{align*}
  \gamma_{n-k}&=\frac{K_{n-k-1}-K_{n-k-3}}{K_{n-k}-K_{n-k-1}}
  -\frac{(K_{n-k}-K_{n-k-3})(K_{n-k+1}-K_{n-k-1})}
  {(K_{n-k}-K_{n-k-1})(K_{n-k+1}-K_{n-k})}\\
  &\quad +\frac{K_{n-k+1}-K_{n-k-3}}{K_{n-k+1}-K_{n-k}}
\end{align*}
for $k=1,\dots,j$, and 
\begin{align*}
  \gamma_n&=\frac{K_{n-1}-K_{n-j-3}}{K_n-K_{n-1}} 
  -\frac{(K_n-K_{n-j-3})(x_{n+2}-K_{n-1})}{(K_n-K_{n-1})(x_{n+2}-K_n)}\\
  &\quad +\frac{x_{n+2}-K_{n-j-3}}{x_{n+2}-K_n}.
\end{align*}
Straightforward calculations show that $\gamma_{n-k}=0$ and $\gamma_n=0$.
By induction we have therefore shown that \eqref{pk} holds for $k=3,\dots,n$ 
and it remains to solve for $p_1$ and $p_2$. We have
\begin{align*}
  p_1+p_2 &= 1-\sum_{k=3}^{n+2}p_k\\
  p_1x_1 + p_2K_1 &= G_0-\sum_{k=3}^{n+1}p_kK_{k-1}-p_{n+2}x_{n+2}.
\end{align*}
Using \eqref{pk} we obtain
\begin{align*}
  p_1+p_2 &= 1-\frac{G_0^1-G_0^2}{K_2-K_1}\\
  p_1x_1+p_2K_1 &= G^0_0-\frac{G_0^1K_2-G_0^2K_1}{K_2-K_1}
\end{align*} 
Solving for $p_1$ and $p_2$ gives
\begin{align*}
  p_1 &= \frac{K_1+G_0^1-G^0_0}{K_1-x_1},\\
  p_2 &= \frac{x_1[(G_0^1-G_0^2)-(K_2-K_1)]+G^0_0(K_2-K_1)-G_0^1K_2+G_0^2K_1}
  {(K_1-x_1)(K_2-K_1)}
\end{align*}
We must show that $(p_1,\dots,p_{n+2})$ corresponds to a probability 
distribution, i.e.~that $p_k \geq 0$ for all $k$ 
(and $p_1+\dots+p_{n+2}=1$).

\smallskip
\noindent
$\mathbf{p_1 \geq 0:}$ 
Notice that $p_1\geq 0$ is equivalent to $K_1+G_0^1-G^0_0 \geq 0$ which 
follows from \eqref{eq:noarbcon1}.

\smallskip
\noindent
$\mathbf{p_k \geq 0}$ {\bf for }$\mathbf{k=3,...,n}:$ 
We have shown that $p_k$ is given by \eqref{pk} for $k=3,\dots,n$, i.e.~that
\begin{align*}
  p_k &= \frac{1}{K_{k-1}-K_{k-2}}\Big(G_0^{k-2}
  -\frac{K_k-K_{k-2}}{K_k-K_{k-1}}G_0^{k-1}
  +\frac{K_{k-1}-K_{k-2}}{K_k-K_{k-1}}G_0^k\Big).
\end{align*}
The non-negativity of $p_k$ is therefore an immediate consequence of 
\eqref{eq:noarbcon2}.

\smallskip
\noindent
$\mathbf{p_{n+1},p_{n+2}\geq 0:}$ Notice that
\begin{align*}
  p_{n+1}=\frac{G_0^{n-1}(x_{n+2}-K_n)-G_0^n(x_{n+2}-K_{n-1})}
  {(K_n-K_{n-1})(x_{n+2}-K_n)} \geq 0 
\end{align*}
is equivalent to  
$G_0^{n-1}(x_{n+2}-K_n)-G_0^n(x_{n+2}-K_{n-1}) \geq 0$.
Solving for $x_{n+2}$ shows that the latter is equivalent to
\begin{align}\label{x5boundary}
  x_{n+2} \geq \frac{G_0^{n-1}K_n - G_0^nK_{n-1}}{G_0^{n-1}-G_0^n}.
\end{align}
Moreover, since $x_{n+2}>K_n$ it obviously holds that 
$p_{n+2}=G_0^n/(x_{n+2}-K_n) \geq 0$.

\smallskip
\noindent
$\mathbf{p_2 \geq 0:}$
First note that $p_2\geq 0$ is equivalent to
\begin{align*}   
  x_1(K_2-K_1+G_0^2-G_0^1)\leq G^0_0(K_2-K_1)+G_0^2K_1-G_0^1K_2.
\end{align*}
Moreover, from \eqref{eq:noarbcon1} we know that 
$K_2-K_1+G_0^2-G_0^1\geq 0$ and therefore $p_2\geq 0$ is equivalent to
\begin{align}\label{x1boundary}
  x_1\leq \frac{G^0_0(K_2-K_1)+G_0^2K_1-G_0^1K_2}{K_2-K_1+G_0^2-G_0^1}.
\end{align}
\hfill{$\square$}


\begin{thebibliography}{100}
\bibitem{B76}
  F.~Black (1976),
  The pricing of commodity contracts,
  \emph{Journal of Financial Economics}, 3, 167-179.

\bibitem{BL78}
  D.~T.~Breeden and R.~H.~Litzenberger (1978), 
  Prices of state-contingent claims implicit in option prices,
  \emph{Journal of Business}, 51, 621-51.

\bibitem{BM02}
D.~Brigo and F.~Mercurio (2002), 
Lognormal-mixture dynamics and calibration to market volatility smiles, 
\emph{International Journal of Theoretical and Applied Finance}, 5(4), 427-446.

\bibitem{B06}
H.~Buehler (2006),
Expensive martingales,
\emph{Quantitative Finance}, 6, 207-218.

\bibitem{CN09}
R.~Carmona and S.~Nadtochiy (2009),
Local volatility dynamic models,
\emph{Finance and Stochastics}, 13, 1-48.

\bibitem{CM05}
  P.~Carr and D.~B.~Madan (2005),
  A note of sufficient conditions for no arbitrage,
  \emph{Finance Research Letters}, 2, 125-130. 

\bibitem{C05}
L.~Cousot (2005)
When can given European call prices be met by a martingale? An answer based on the
building of a Markov chain model, June 2005, 
\verb|http://ssrn.com/abstract=754544|

\bibitem{DH07}
M.~H.~A.~Davis and D.~G.~Hobson (2007),
The range of traded option prices,
\emph{Mathematical Finance}, 17, 1-14.

\bibitem{D94}
B.~Dupire (1994),
Pricing with a smile, 
\emph{Risk}, 7, 18-20.

\bibitem{FHM11}
D. Filipovic, L.~P. Hughston, and A. Macrina (2012),
 Conditional density models for asset pricing,
 \emph{International Journal of Theoretical and Applied Finance}, 15(1)

\bibitem{G06}
J.~Gatheral,
\emph{The volatility surface},
Wiley, 2006.

\bibitem{G05}
P.~Giot (2005),
Relationships between implied volatility indexes and stock index returns,
\emph{The Journal of Portfolio Management}, 31(3), 92-100.

\bibitem{KK10}
J.~Kallsen and P.~Kr\"uhner (2010),
On a Heath-Jarrow-Morton approach for stock options. \emph{Preprint}.
 
\bibitem{LL00}
J.~P.~Laurent and D.~Leisen,
Building a consistent pricing model from observed option prices. 
In: Avellaneda, M. (ed.) Collected papers of the New York University Mathematical Finance Seminar, vol. II, pp. 216-238. World Scientific, Singapore, 2000.
  
\bibitem{P04}
P.~Protter,
\emph{Stochastic Integration and Differential Equations}, second edition,
Springer, 2004

\bibitem{R76}
  W.~Rudin,
  \emph{Principles of Mathematical Analysis}, third edition,
  McGraw-Hill, 1976.

\bibitem{SW08}
M.~Schweizer and J.~Wissel (2008),
Arbitrage-free market models for option prices: the multi-strike case,
\emph{Finance and Stochastics}, 12, 469-505.

\bibitem{S93}
  D.~Shimko (1993), 
  Bounds of probability. 
  \emph{Risk}, 6, 33-37.

\end{thebibliography}
\end{document}